\documentclass[a4paper, 11pt]{article}
\usepackage{jcappub} 

\usepackage{amssymb}
\usepackage{amsmath}
\usepackage{amsfonts}
\usepackage{amssymb}
\usepackage{float}
\usepackage{soul}
\usepackage{bm}
\usepackage[normalem]{ulem}
\usepackage{multirow}
\usepackage[dvipsnames]{xcolor}
\newcommand\Tstrut{\rule{0pt}{2.9ex}} 

\usepackage[sorting=none,style=numeric-comp]{biblatex}
\usepackage{academicons}
\usepackage{fontawesome5}
\definecolor{orcidlogocol}{rgb}{0.65, 0.807, 0.223}

\newcommand{\orcid}[1]{$\,$\href{https://orcid.org/#1}{\textcolor{orcidlogocol}{\faOrcid}}}

\def\beq{\begin{equation}}
\def\eeq{\end{equation}}
\def\ber{\begin{eqnarray}}
\def\eer{\end{eqnarray}}
\def\benu{\begin{enumerate}}
\def\eenu{\end{enumerate}}

\def\l{\left}
\def\r{\right}
\def\d{{\rm d}}
\newcommand{\sq}{\lower.25ex\hbox{\large$\Box$}}

\def\f{\frac}

\def \blue {\color{blue}}

\usepackage{cancel}

\usepackage{framed}

\colorlet{shadecolor}{CadetBlue!5}

\title{Braneworld Dark Energy in light of DESI~DR2}
\author[a,b]{Swagat~S.~Mishra\,\orcid{0000-0003-4057-145X},}
\author[c]{William~L.~Matthewson\,\orcid{0000-0001-6957-772X},}
\author[b]{Varun~Sahni\,\orcid{0000-0002-9470-9939},}
\author[c,d]{Arman~Shafieloo\,\orcid{0000-0001-6815-0337}}
\author[e]{and Yuri~Shtanov\,\orcid{0000-0002-4891-7059}}
\affiliation[a]{School of Physics and Astronomy,  University of Nottingham, Nottingham, NG7 2RD, United Kingdom.}
\affiliation[b]{Inter-University Centre for Astronomy and Astrophysics,
Post Bag 4, Ganeshkhind, Pune 411~007, India}
\affiliation[c]{Korea Astronomy and Space Science Institute (KASI),
776 Daedeok-daero, Yuseong-gu, Daejeon 34055, Korea}
\affiliation[d]{KASI Campus, University of Science and Technology,
217 Gajeong-ro, Yuseong-gu, Daejeon 34113, Korea}
\affiliation[e]{Bogolyubov Institute for Theoretical Physics, Metrologichna St. 14-b, Kiev 03143, Ukraine}

\emailAdd{swagat.mishra@nottingham.ac.uk}
\emailAdd{willmatt4th@kasi.re.kr}
\emailAdd{varun@iucaa.in} 
\emailAdd{shafieloo@kasi.re.kr}
\emailAdd{shtanov@bitp.kyiv.ua}

\abstract{Recent observational results from the DESI collaboration reveal tensions with the standard $\Lambda$CDM model and favour a scenario in which dark energy (DE) decays over time. The DESI DR2 data also suggest that the DE equation of state (EoS) may have been phantom-like ($w < - 1$) in the past, evolving to $w > - 1$ at present\,---\,implying a recent crossing of the phantom divide at $w = - 1$. 

Scalar field models of DE naturally emerge in ultraviolet-complete theories such as string theory, which is typically formulated in higher dimensions.  In this work, we investigate a broad class of {\em thawing scalar field models\/}\,---\,including the simple quadratic, quartic, exponential, symmetry-breaking and axion potentials\,---\,propagating on a (4+1)-dimensional  ghost-free phantom braneworld,  
and demonstrate that their effective EoS exhibits a  phantom-divide crossing. Alongside the Hubble parameter and EoS of DE, we also analyse the evolution of the {\em Om\/} diagnostic, and demonstrate that the time dependence of these quantities is  in excellent agreement with the DESI DR2 observations. Furthermore, we perform a comprehensive parameter estimation using Markov Chain Monte Carlo sampling, and find that the $\chi^2$ values for all our models are remarkably close to that of the widely used CPL parametrisation\,---\,indicating that our models fit the data very well.
} 
\arxivnumber{arXiv:2507.07193}

\begin{document}

\maketitle
\flushbottom

\section{Introduction}
\label{sec:Intro}
Ever since the remarkable serendipitous discovery of cosmic acceleration \cite{SupernovaSearchTeam:1998fmf,SupernovaCosmologyProject:1998vns}, the physical nature of dark energy (DE) has remained elusive and somewhat of an enigma. Although a cosmological constant (CC) with $T_{ik} = \Lambda\,g_{ik} \, \Rightarrow \, p = -\rho$ provides a good fit to most of the data, if CC is to be associated with the vacuum energy then its extremely small value $\rho_{\rm vac} \sim 10^{-47}\,{\rm GeV}^4$, suggested by observations, appears to be difficult to reconcile with expectations from fundamental physics and quantum field theory \cite{Zeldovich:1968ehl,Weinberg:1988cp}.
Alternatives to the cosmological constant that can induce late-time acceleration have been actively explored in the literature. These usually involve modifications either to the right-hand side or the left-hand side of the Einstein equations $G_{ik} = 8\pi G \, T_{ik}$. Early candidates for the former include scalar fields with appropriately chosen potentials \cite{Ratra:1987rm,Peebles:1987ek}, while to the latter belong late-time modifications to Einstein's general relativity (GR); see \cite{Sahni:1999gb,Peebles:2002gy,Padmanabhan_2003,Sahni_2004,Copeland:2006wr,Bousso_2007,Frieman_2008,durrer2008darkenergymodifiedgravity,Li_2011,Nojiri_2011,Clifton_2012,Amendola:2015ksp,Padilla:2015aaa} for extensive reviews on the subject.

In the absence of terrestrial (laboratory) tests of DE, the key to understanding its properties lies in a comprehensive analysis of cosmological data sets. In this connection, model-independent techniques \cite{Sahni:2006pa} applied to observational data provide an attractive route to discerning the properties of DE at a deeper level. To this category belong the popular CPL ansatz \cite{CHEVALLIER_2001, Linder_2003} for the  DE equation of state (EoS), defined as
\beq
w(a) = w_0 + w_a(1-a) = w_0 + w_a \l(\f{z}{1+z}\r) \, ,
\label{eq:CPL}
\eeq
and the $Om$ diagnostic \cite{Sahni:2008xx, Zunckel_2008, Sahni_2014} which provides a null test of $\Lambda{\rm CDM}$\,\footnote{We use the notation $h(z) \equiv H(z)/H_0$, as defined in Eqs.~(\ref{eq:Om}) and (\ref{eq:h_PhBrane}), throughout this paper, where $h(z)$ is a variable that changes with redshift. However, in order to avoid confusion, it is worth pointing out that often the letter $h$ stands for $h \equiv H_0\, \l(100~\text{km/(s\,Mpc)}\r)^{-1}$ in the cosmology literature, where $h$ is a constant, and does not vary with redshift.} 
\begin{equation} \label{eq:Om}
Om(z) = \frac{h^2(z) - 1}{(1 + z)^3-1}\,, \qquad h (z) \equiv \frac{H(z)}{H_0}\,.
\end{equation}
Interestingly, when applied to the DESI DR2 data \cite{DESI:2024uvr,DESI:2024lzq,DESI:2025fxa}, both methods suggest that cosmic acceleration may be slowing down at the present epoch \cite{lodha2024desi2024constraintsphysicsfocused,DESI:2024mwx,DESI:2024aqx,DESI:2024hhd,DESI:2025fii}. What is even more surprising is the fact that a model-independent analysis of several data sets indicates that DE passed through a recent {\em phantom\/} phase when its EoS dropped below $-1$~\cite{lodha2024desi2024constraintsphysicsfocused}. If these results stand up to further scrutiny~\cite{Cortes:2024lgw,Park:2024pew,Park:2025azv,wang2025diddesidr2truly}, then they appear to imply that DE crossed the phantom divide in its recent past, since $w(z) < -1$ at $z \gtrsim 1$ and $w(z) > -1$ at $z \lesssim 1$. (Tension with the $\Lambda$CDM model stands at between $2\sigma$ and  $3\sigma$  confidence levels \cite{ormondroyd2025comparisondynamicaldarkenergy, cortês2025desisdr2exclusionlambdacdm, wang2025diddesidr2truly, RoyChoudhury:2024wri,Park:2024vrw,RoyChoudhury:2025dhe,Braglia:2025gdo}.)

Since the DESI results point towards decaying DE whose EoS increases with time, it is instructive to work with {\em thawing models\,}\footnote{Thawing models complement  {\em freezing models\/} in which the DE EoS decreases with time and tends towards a more negative value in the asymptotic future~\cite{Caldwell:2005tm}.} of dynamical DE~\cite{Caldwell:2005tm}. Although these models cannot accommodate phantom-crossing in GR, they do so quite easily when propagating on a higher dimensional braneworld. 
Indeed, it is well known that a {\em phantom brane} with a brane tension has an effective EoS which is lower than $-1$~\cite{Sahni:2002dx,Lue:2004za,Alam_2017,Bag_2021}. In fact, the lowering of the dark energy EoS relative to its GR value is an important generic feature of phantom brane cosmology. In the present analysis, we assume that the brane tension (vacuum energy) is zero, and that any effective vacuum energy at a given epoch arises from an ultra-light scalar field that has not yet relaxed to its vacuum state. (This is in line with the customary assumption  made in most scalar-field (quintessence) models of DE in the literature~\cite{Frieman:1995pm}.)  Such a scalar field propagating on the brane will initially have an effective DE EoS $w < -1$, when the field $\phi(t)$ is frozen to its initial value. At late times, when $\phi$ begins rolling down its potential,\footnote{An ultra-light scalar field remains frozen to its initial value as long as the `friction term' $3H{\dot\phi}$ is larger than the `force' term $V_{,\phi}$ in the equation of motion ${\ddot\phi} + 3H{\dot\phi} + V_{,\phi} = 0$. As the universe expands, $H$ decreases  allowing $V_{,\phi}$ to influence the motion of $\phi$, which leads to growth in the EoS $w(t)$ at late times. This is also true if the effective DE EoS  drops below $w = -1$ in the recent past, because the total effective density of the not-yet-DE-dominated universe, nevertheless, keeps dropping, and therefore $H$ keeps decreasing, allowing $V_{,\phi}$ to influence the late-time motion of $\phi$. } the effective DE EoS crosses the phantom divide and increases to $w > -1$. (Dynamical DE models have also been discussed in \cite{Wolf:2025jed,you2025dynamicaldarkenergyimplies,shah2025interactingdarksectorslight,akrami2025desidetectedexponentialquintessence,ye2025necviolationbeyondhorndeski,tyagi2025constraintsgeneralizedgravitythermodynamiccosmology,kumar2025evidencenoncolddarkmatter,ahlen2025positiveneutrinomassesdesi,Gialamas:2025pwv,Andriot:2025los,Scherer:2025esj,Silva:2025hxw,Wolf:2024stt,Ye:2024ywg,Wolf:2025jlc} in the context of DESI DR2.)

\begin{figure}[htp]
    \centering
    \includegraphics[width = 0.495\textwidth]{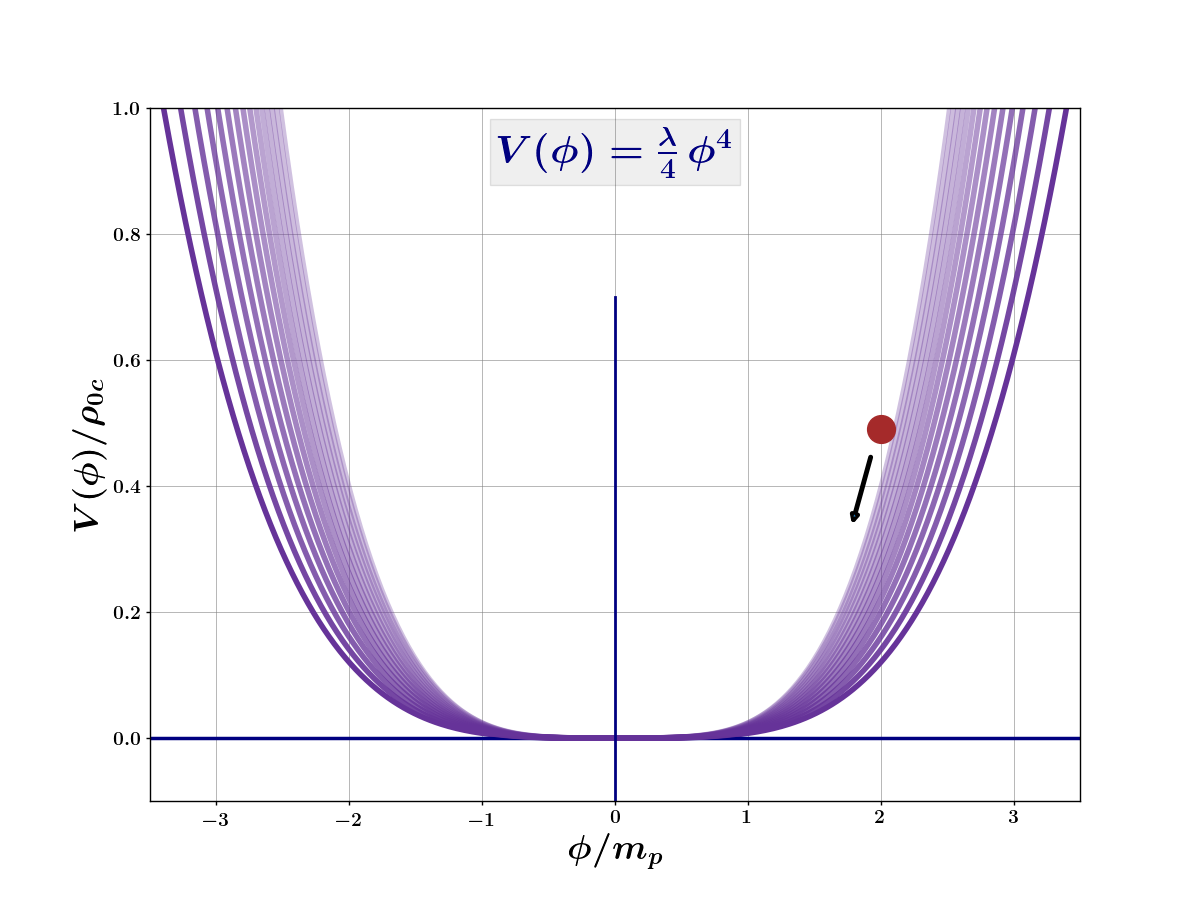}
     \includegraphics[width = 0.495\textwidth]{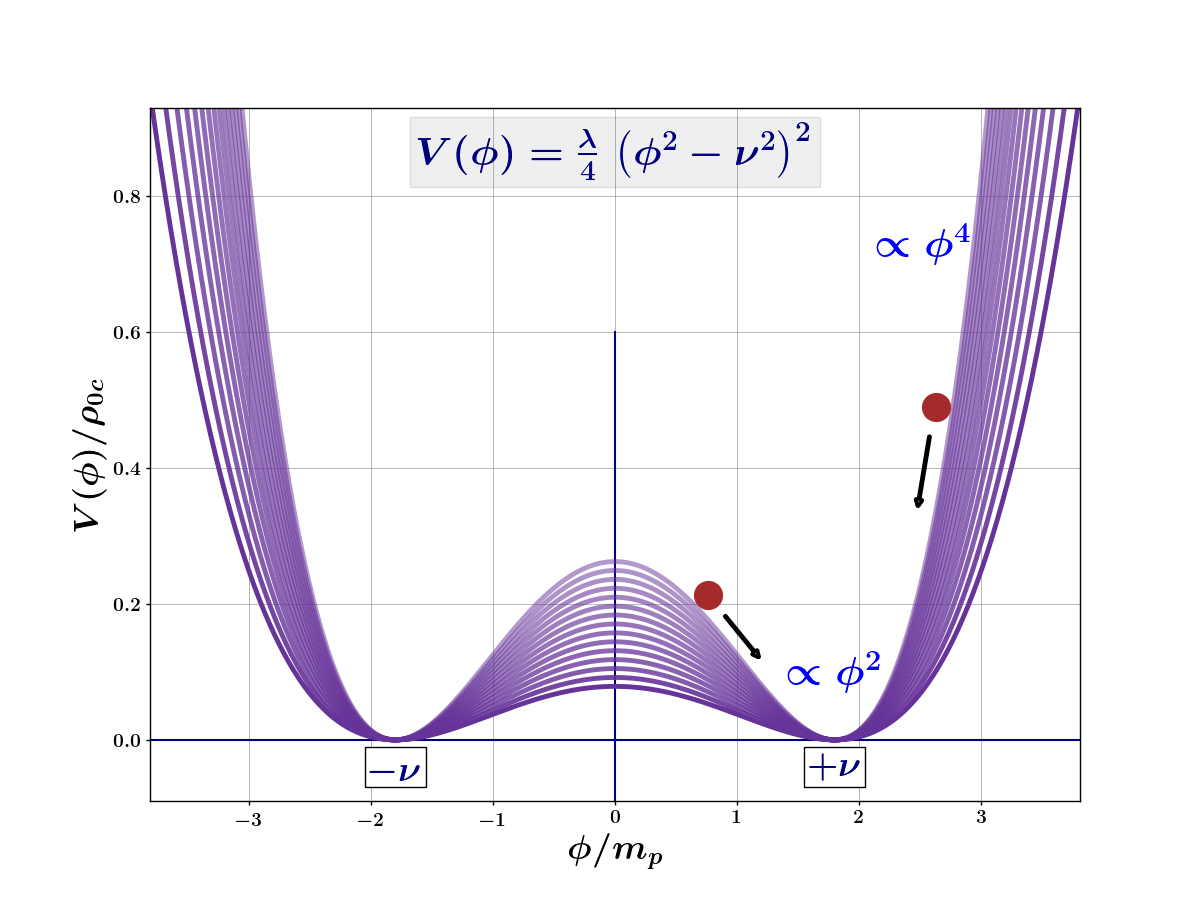}
     \includegraphics[width = 0.495\textwidth]{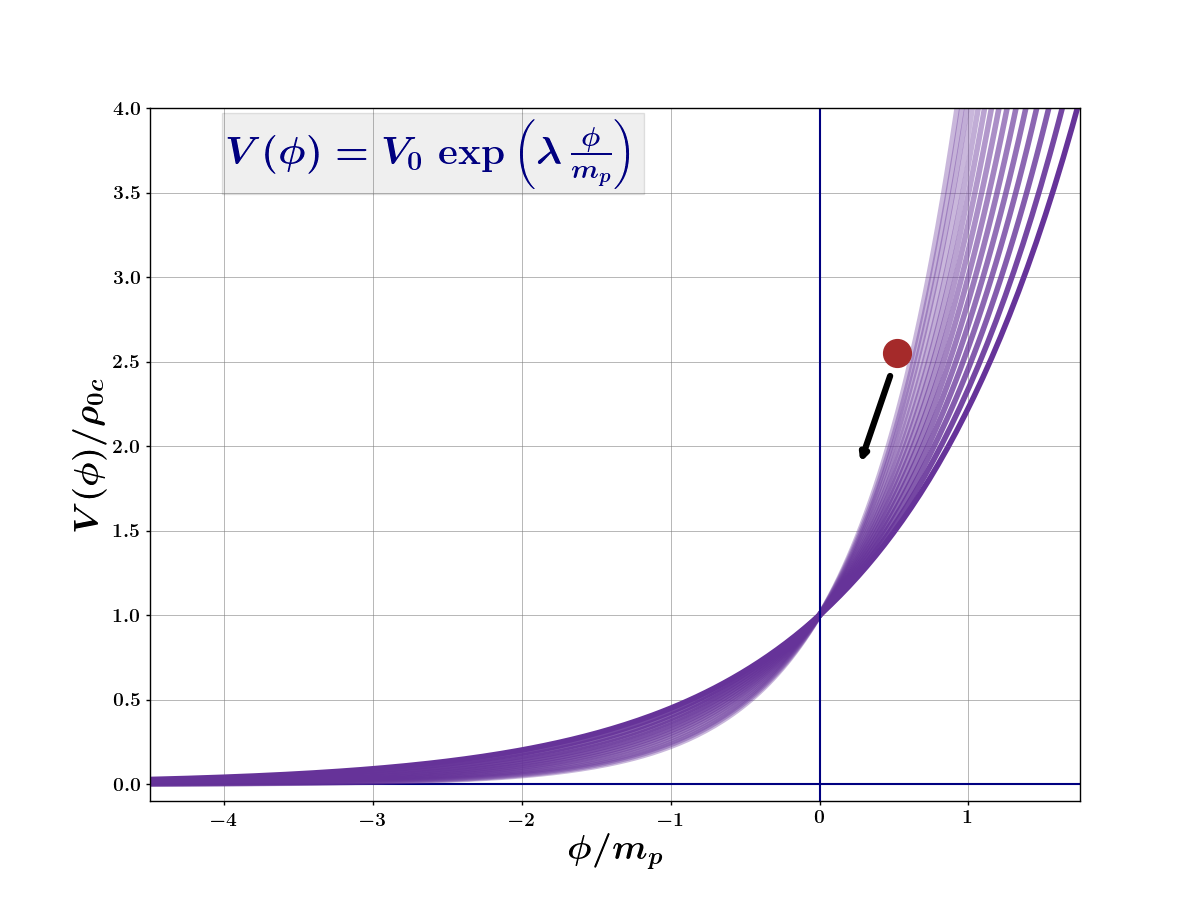}
     \includegraphics[width = 0.495\textwidth]{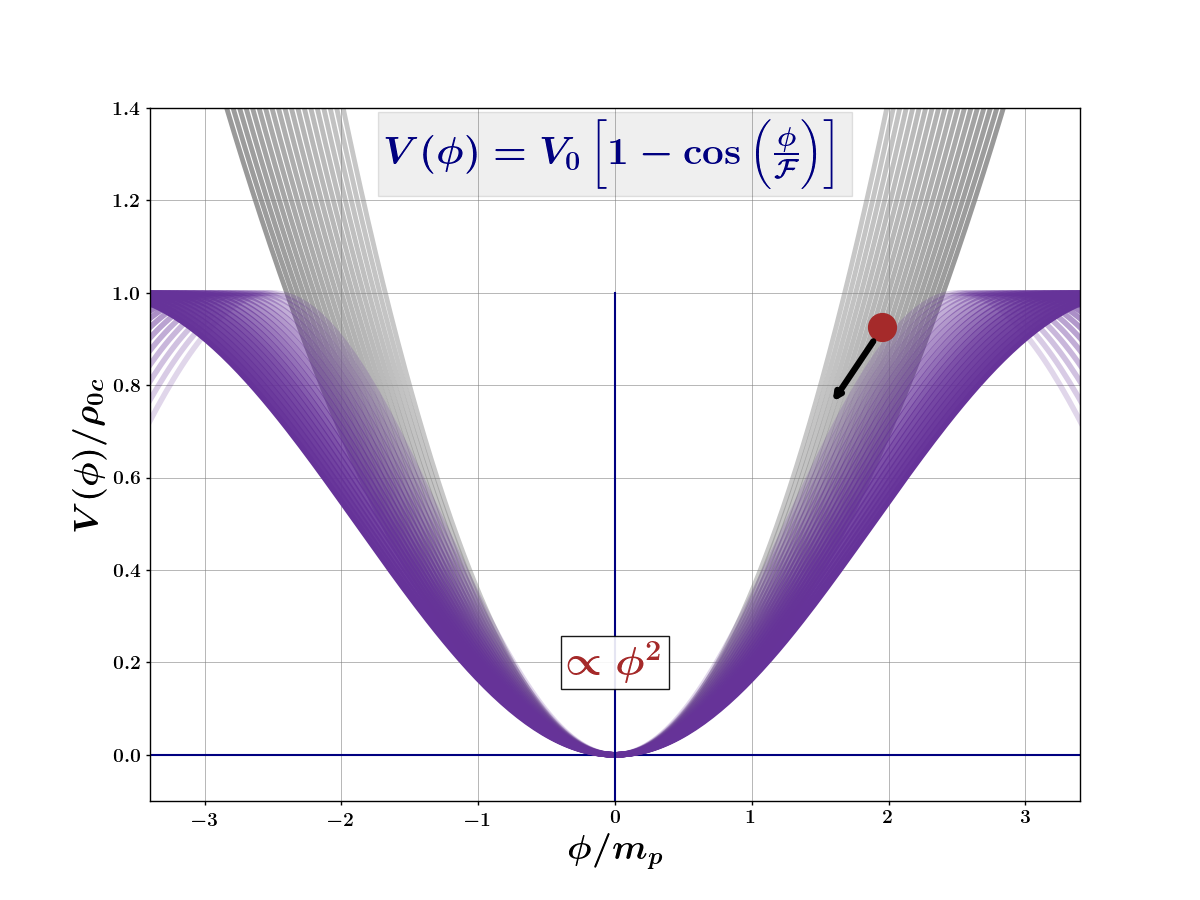}
     \vspace{-0.2in}
    \caption{Schematic plots of different potentials considered in this work. {\bf Top-left panel} shows the quartic potential~(\ref{eq:pot_quartic}) for different values of $\lambda$. {\bf Top-right panel} shows the symmetry-breaking potential~(\ref{eq:pot_SSB})  for fixed value of $\nu$, and different values of $\lambda$.  {\bf Bottom-left panel} shows the exponential potential~(\ref{eq:pot_Exp}) for fixed value of $V_0$, and different values of $\lambda$. {\bf Bottom-right panel} shows the pseudo Nambu-Goldstone boson/axion potential~(\ref{eq:pot_PNGB}) for different values of the axion mass~(\ref{eq:pot_Axion_quad_mass}) (the quadratic approximation~(\ref{eq:pot_Axion_quad}) is displayed in light-gray curves).} 
    \label{fig:DE_pots}
\end{figure}

In this paper, we demonstrate that the following scalar-field potentials residing on the brane give rise to dynamical dark energy with an EoS which crosses the phantom divide, in broad agreement with the DESI DR2 results. The potentials are illustrated in Fig.~\ref{fig:DE_pots}.

\begin{enumerate}

\item An  ultra-light scalar field with negligible self-interaction, described by the quadratic potential\,\footnote{Note that the quadratic potential can be realised as a  limiting case of the symmetry-breaking potential (\ref{eq:pot_SSB}) and the axion potential~(\ref{eq:pot_PNGB}).}  
\begin{equation}
V(\phi) = \frac{1}{2}m^2\phi^2\, .
\label{eq:pot_free}
\end{equation}
\item A massless scalar field with a quartic self-interaction
\begin{equation}
V(\phi) = \frac{\lambda}{4}\,\phi^4\,.
\label{eq:pot_quartic}
\end{equation}
\item The symmetry-breaking potential 
\begin{equation}
V(\phi) = \frac{\lambda}{4} \, \l( \phi^2 - \nu^2 \r)^2 \, .
\label{eq:pot_SSB}
\end{equation}
\item An exponential potential
\begin{equation}
V(\phi) = V_0 \, \exp{\l(\lambda\, \f{\phi}{m_p}\r)} \, .
\label{eq:pot_Exp}
\end{equation}
\item A pseudo Nambu-Goldstone boson with an axion potential \cite{Frieman:1995pm}
\begin{equation}
V(\phi) = V_0 \l[ 1 -  \cos{\l( \frac{\phi}{{\cal F}}\r)} \r] \, .
\label{eq:pot_PNGB}
\end{equation}
A similar potential $V(\phi) = V_0\cos{\l( \frac{\phi}{{\cal M}}\r)}$ has been discussed\,\footnote{An interesting property of this potential is that it becomes negative at late times which can cause the universe to stop expanding and begin to contract \cite{Alam_2003}. More general axion-like potentials have also been discussed in \cite{chiang2025observationalconstraintsgeneralisedaxionlike} and references therein.  However, in our analysis, the axion potential is always positive semi-definite, \textit{i.e.\@} $V(\phi)\geq 0$, as can be inferred from Eq.~(\ref{eq:pot_PNGB}).} in the context of the string theory axion \cite{Choi_2000} and tested against observations in \cite{Alam_2003}.  
\item
And finally, an asymptotically flat  (plateau) potential
\begin{equation}
V(\phi) = V_0 \, \tanh^2{\l(\lambda \, \frac{\phi}{m_p}\r)}  \, .
\label{eq:pot_plateau}
\end{equation}
\end{enumerate}

Our paper is organised as follows. In Section~\ref{sec:Brane_dynamics}, we introduce the braneworld model employed in our study and present the cosmological equations that are solved numerically.  Section~\ref{sec:Dynamics_Models_Brane} details the numerical solutions and the resulting behaviour of the effective dark energy for various scalar-field potentials, while Section~\ref{sec:MCMC} is dedicated to a comprehensive parameter estimation for these potentials using  Markov Chain Monte Carlo (MCMC) sampling against the  publicly available DESI DR2 data. Sections~\ref{sec:Dynamics_Models_Brane} and \ref{sec:MCMC} constitute the core of this work. We discuss the implications of our results in Section~\ref{sec:discussion}. Appendix~\ref{app:Analytics} is dedicated to a discussion on the existence of a pole in the effective EoS of DE on the phantom brane at a higher redshift. Appendix~\ref{app:PhBrane_others} discusses the evolution of other physical quantities in our models. Finally,  details for a specific asymptotically flat scalar field potential are provided in Appendix~\ref{sec:PhBrane_Tmodel}.

\section{Cosmological equations on the phantom brane}
\label{sec:Brane_dynamics}

The phantom braneworld model, detailed in \cite{Sahni:2002dx}, features a phantom-like effective EoS for dark energy, characterised by $w_\text{eff} < -1$~\cite{Lue:2004za}. The generic action for this  cosmological braneworld model,  originally suggested by Collins and Holdom~\cite{Collins:2000yb} (and independently by Shtanov~\cite{Shtanov:2000vr}),  also known as the Dvali--Gabadadze--Porrati (DGP) model~\cite{Dvali:2000hr},  reads\,\footnote{We adopt the mostly-plus metric signature and work in natural units, setting $\hbar = c = 1$. In \eqref{action}, we use the standard general-relativistic normalisation  of the action on the brane, which differs by a factor of 1/2 from the normalisation used in \cite{Dvali:2000hr, Shtanov:2000vr, Sahni:2002dx}.} 
\cite{Collins:2000yb, Dvali:2000hr, Shtanov:2000vr}
\begin{equation}\label{action}
S = M_p^3 \int_{\rm bulk} \left( {\cal R} - 2 \Lambda_{\rm b} \right) + \frac{m_p^2}{2} \int_{\rm brane} \left( R - 2 \Lambda \right) + \int_{\rm brane} L_{\rm m} \, . 
\end{equation} 
This model represents a simple general-relativistic action in the five-dimensional bulk (with scalar curvature ${\cal R}$) and on the four-dimensional brane (with scalar curvature $R$), with matter confined only to the brane and described by the Lagrangian $L_{\rm m}$. Integrations over the bulk and
brane are taken with the corresponding natural volume elements. The universal constants $m_p = 1 / \sqrt{ 8 \pi G}$ and $M_p$ play the role of the Planck masses on the brane and in the bulk space, respectively. The symbols $\Lambda$ and  $\Lambda_{\rm b}$ denote, respectively, the cosmological constants on the brane and in the bulk, so that $m_p^2 \Lambda$ is the brane tension from the five-dimensional bulk perspective.

This paper examines a braneworld model in which the cosmological constants associated with both the brane and the bulk are excluded, \textit{i.e.\@} $\Lambda_\text{b} = \Lambda = 0$. It is postulated that there exists a mechanism that nullifies these constants in vacuum. Instead, any effective pseudo-vacuum energy present at a specific epoch is attributed to an ultra-light scalar field that has not yet reached its vacuum state. Cosmologically, this implies that the effective dark energy in the universe evolves due to two distinct effects: the braneworld dynamics and the time-dependent evolution of the ultra-light scalar field.

We consider the universe at relatively late times, that is, $z \ll 10^2$, for which the density of radiation can be safely ignored. The expansion rate on a phantom brane  (which corresponds to the normal, ghost-free \cite{Charmousis:2006pn, Gorbunov:2005zk, Koyama:2007za}, branch of the DGP model \cite{Shtanov:2000vr, Deffayet:2000uy, Deffayet:2001pu}) filled with matter and a scalar field describing dynamical dark energy is then expressed as \cite{Sahni:2002dx, Bag:2018jle} 
\begin{equation}
h (z) \equiv \frac{H (z)}{H_0} = \sqrt{ \Omega_{0m} (1 + z)^3 + \frac{\rho_\phi}{3 m_p^2 H_0^2} + \Omega_{0\ell}} - \sqrt{\Omega_{0\ell}} \, ,   
\label{eq:h_PhBrane}
\end{equation}
where 
\begin{equation}
\rho_\phi = \frac12 \dot \phi^2 + V (\phi) 
\label{eq:rho_phi}
\end{equation}
is the energy density of the scalar field (regarded as a function of redshift), and
\begin{equation}\label{omegas}
\Omega_{0m} = \frac{\rho_{0m}}{3 m_p^2 H_0^2} \, , \qquad \Omega_{0\ell} = \frac{1}{\ell^2 H_0^2} \, , \qquad \ell = \frac{m_p^2}{M_p^3} \, .    
\end{equation}
The Omega parameters obey the constraint that follows from the condition $h (0) = 1$:
\begin{equation}\label{constraint}
\Omega_{0m} + \frac{\rho_{0\phi}}{3 m_p^2 H_0^2} - 2 \sqrt{\Omega_{0\ell}} = 1 \, , 
\end{equation}
where $\rho_{0\phi}$ is the current value of $\rho_\phi$.
Note that $\Omega_{0\ell}$ is the new fundamental parameter in our model. It is associated with the presence of a fifth dimension (the bulk). Setting $\Omega_{0\ell} = 0$ in (\ref{eq:h_PhBrane}), one recovers the expansion rate in GR.

Constraints on the fundamental parameter $\ell$ from local gravitational physics are significantly weaker than those derived from cosmological observations, including those presented in this work. The presence of extra-dimensional gravity induces corrections to Newton's laws, potentially leading to observable anomalies within the Solar System. Notably, all planets would experience a uniform anomalous perihelion precession with a rate given by $\dot \phi = - 3 / 4 \ell$ \cite{Lue:2002sw}. The non-detection of this effect in observations imposes a lower bound on the fundamental length parameter, $\ell \gtrsim 0.26$~Gpc \cite{Battat:2008bu}. Another constraint, $\ell \gtrsim 0.32$~Gpc, arises from Lunar Laser Ranging experiments \cite{Dvali:2002vf, Koyama:2015vza}.\footnote{Our constant $\ell$ is equal to twice the length scale $r_0$ introduced and constrained in \cite{Lue:2002sw, Battat:2008bu, Dvali:2002vf, Koyama:2015vza}.} In terms of the parameter $\Omega_{0\ell}$ defined in \eqref{omegas}, this translates into an upper bound
\begin{equation}
\Omega_{0\ell} \lesssim 180\, h_{70}^{-2} \, ,
\end{equation}
where $h_{70} = H_0 / \left( 70~\text{km/(s\,Mpc)} \right)$. This is several orders of magnitude larger than the typical best-fit values $\Omega_{0\ell} \simeq 0.013$ obtained in this work (see below).

From the perspective of general relativity, the effective dark energy density in our model is defined as
\begin{equation} \label{eq:de}
\rho_{\rm DE} = 3 m_p^2 H_0^2 h^2 - \rho_m = \rho_\phi - 6 m_p^2 H_0^2 \sqrt{\Omega_{0\ell}}\, h \, .
\end{equation}
The second equality follows from Eq.~\eqref{eq:h_PhBrane} by rearranging its terms and squaring both sides.

The scalar field respects the usual equation 
\begin{equation}
\ddot \phi + 3 H \dot \phi + V_{,\phi}(\phi) = 0 \, ,
\label{eq:EoM_phi}
\end{equation}
 where $V_{,\phi}={\d V}/{\d \phi}$. According to our assumptions, the vacuum value for the scalar field potential is zero.

In this work, we integrate the above equations starting from an initial redshift of $z = 99$. This choice corresponds to a sufficiently early evolutionary epoch, while, at the same time, allowing us to neglect the contribution from radiation, which is subdominant at this and later epochs. The initial time derivative $\dot \phi$ of the scalar field is set to zero, and the initial value of $\phi$ is chosen so that the constraint Eq.~\eqref{constraint} is satisfied with $\Omega_{0m} = 0.315$. This value of $\Omega_{0m}$ is held fixed across all our simulations, while the remaining parameters are varied.

After the solutions are obtained,  we can evaluate  the deceleration parameter $q (z)$,  the effective DE density relative to its present-epoch value $f_{\rm DE}(z)$  and the effective equation-of-state parameter $w_\text{DE} (z)$ of DE using the following expressions:
\begin{align}
q(z) &\equiv - \frac{\ddot{a}}{a\,H^2} = \frac{h' (z)}{h (z)} \, (1 + z) - 1  \, , \label{eq:Brane_q} \\
 f_{\rm DE}(z) &\equiv  \f{\rho_{\rm DE}}{\rho_{0\rm DE}} =  h^2(z) \l(\f{1 - \Omega_m(z)}{1 - \Omega_{0m}}\r)  \, ,  \label{eq:Brane_fDE} \\
w_\text{DE}(z) &= \frac{2\,q(z) - 1}{3\left( 1 - \Omega_m(z) \right)} \, , \label{eq:Brane_wDE}
\end{align}
with 
\beq
  \Omega_{m}(z) = \Omega_{0m} \,\frac{(1 + z)^3}{h^2(z)} \, .
\label{eq:Omega_m}
\eeq

It is worth noting that the effective EoS of DE on the phantom brane is known to exhibit a pole~\cite{Sahni:2002dx} at a redshift $z=z_p$ where $\Omega_m(z_p) =1$, as can be seen from Eq.~(\ref{eq:Brane_wDE}). The existence of such a pole is purely a result of the effective description of DE, and does not correspond to an actual singularity in the system. See Appendix~\ref{app:Analytics} for a discussion on the location of the pole.\footnote{See Ref.~\cite{Andriot:2025los} for the analysis of a coupled matter--quintessence model in GR which exhibits two poles in the effective EoS of DE.}

By combining Eqs.~(\ref{eq:h_PhBrane}), (\ref{eq:rho_phi}), (\ref{eq:EoM_phi}), (\ref{eq:Brane_q}), and (\ref{eq:Omega_m}), the effective EoS of DE (\ref{eq:Brane_wDE}) can be written as 
\beq
w_\text{DE}(z) = -1 - \l[ \f{\sqrt{\Omega_{0\ell}}\,\Omega_{m}(z) - \f{1}{h(z)} \, \f{\dot{\phi}^2}{3m_p^2H_0^2}}{\bigl( 1- \Omega_{m}(z)\bigr) \l(h(z)+\sqrt{\Omega_{0\ell}}\r)} \r] \, ,
\label{eq:Brane_wDE_ext}
\eeq
which indicates that, at higher redshifts, when the scalar field is frozen at  its initial value, \textit{i.e.\@} $\dot{\phi} \simeq 0$, the EoS of DE is phantom like, satisfying $w_\text{DE}(z) < -1$ (provided $\Omega_m \neq 1$, which is a pole, as noted above). The expression for the EoS of DE at the present epoch, $z=0$, is given by
\beq
w_\text{DE}(0) = -1 + \l[ \f{ \Omega_{0\phi,{\rm KE}} - \Omega_{0m}\,\sqrt{\Omega_{0\ell}}}{\bigl(1-\Omega_{0m}\bigr)\l(1+\sqrt{\Omega_{0\ell}}\r)} \r] \, ,
\label{eq:Brane_wDE_present}
\eeq
where $\Omega_{0\phi,{\rm KE}} \equiv \dot{\phi}_0^2/\l(3m_p^2H_0^2\r)$.  Importantly, Eq.~(\ref{eq:Brane_wDE_present}) demonstrates that, at the present epoch, if $\Omega_{0\phi,{\rm KE}} >  \Omega_{0m}\,\sqrt{\Omega_{0\ell}}$, then $w_\text{DE} > -1$, ensuring the phantom-divide crossing at some intermediate epoch $z > 0$.

\section{Dynamics of different DE models on the phantom brane}
\label{sec:Dynamics_Models_Brane}

  In this section, we discuss the dynamics of DE for the quadratic~(\ref{eq:pot_free}), quartic~(\ref{eq:pot_quartic}), symmetry-breaking~(\ref{eq:pot_SSB}), exponential~(\ref{eq:pot_Exp}) and axion~(\ref{eq:pot_PNGB}) potentials as representatives of a broad class of simple thawing models in our braneworld scenario. While the plateau potential~(\ref{eq:pot_plateau}) also fits the data quite well, its best-fit parameters lead to an EoS of DE that resembles that of the quadratic potential (as well as the axion potential). Therefore we discuss it in the App.~\ref{sec:PhBrane_Tmodel}, in order to maintain a simpler presentation in the main text of the manuscript.
     
As discussed above, we set both the brane tension and the bulk cosmological constant to zero. In the limit $\Omega_{0\ell} \to 0$, our model reduces to that of a massive scalar field in General Relativity.  In all plots appearing in this section, we fix the present-day dark energy density parameter to 
$\Omega_{0{\rm DE}} \equiv 1 - \Omega_{0m} = 0.685$, and investigate the effects of varying $\Omega_{0\ell}$
along with other parameters specific to the chosen scalar-field potentials. 
\subsection{Quadratic potential}
\label{sec:PhBrane_Quad}

We first turn our attention to the simple quadratic scalar-field potential (gray colour curves in bottom-right panel of Fig.~\ref{fig:DE_pots}), given by
\begin{equation}
V(\phi) = \frac{1}{2}\, m^2 \phi^2 \, ,
\label{eq:pot_Quad}
\end{equation}
where $m$ is mass of the scalar field. 

\begin{figure}[htp]
\vspace{-0.1in}
\begin{center}
\vspace{-0.1in}
\includegraphics[width=0.495\textwidth]{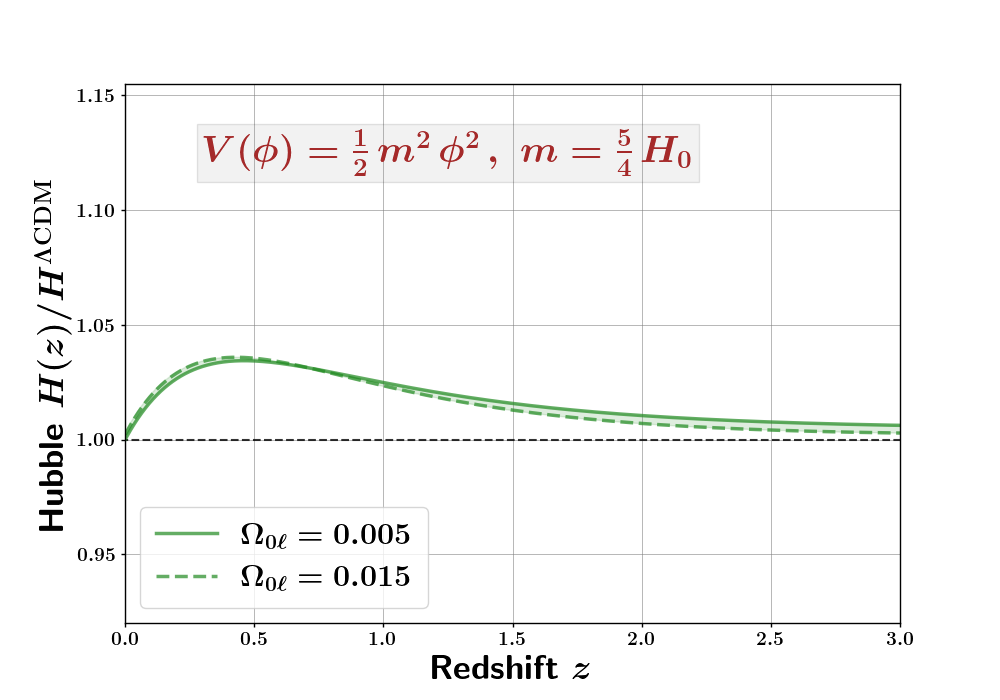}
\includegraphics[width=0.495\textwidth]{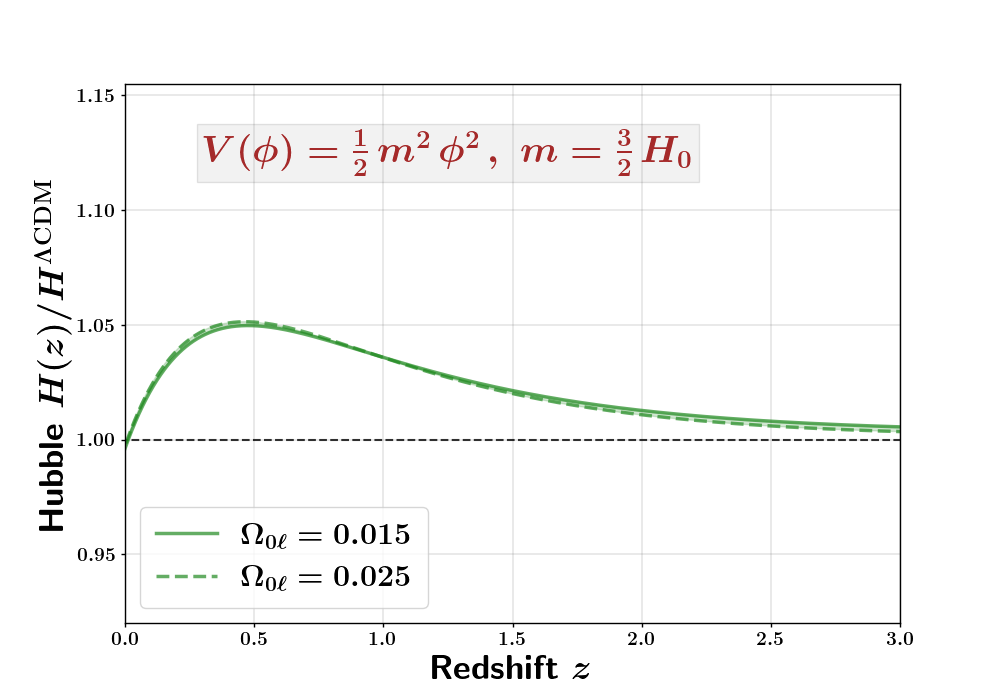}
\vspace{-0.2in}
\caption{The Hubble parameter relative to $\Lambda$CDM is shown for the quadratic potential~(\ref{eq:pot_Quad}) with $m = \frac{5}{4} H_0$ ({\bf left panel}), and $m = \frac{3}{2} H_0$ ({\bf right panel}).}
\label{fig:DE_PhBrane_Quad_Hubble}
\end{center}
\end{figure}

\begin{figure}[htp]
\vspace{-0.15in}
\begin{center}
\vspace{-0.15in}
\includegraphics[width=0.495\textwidth]{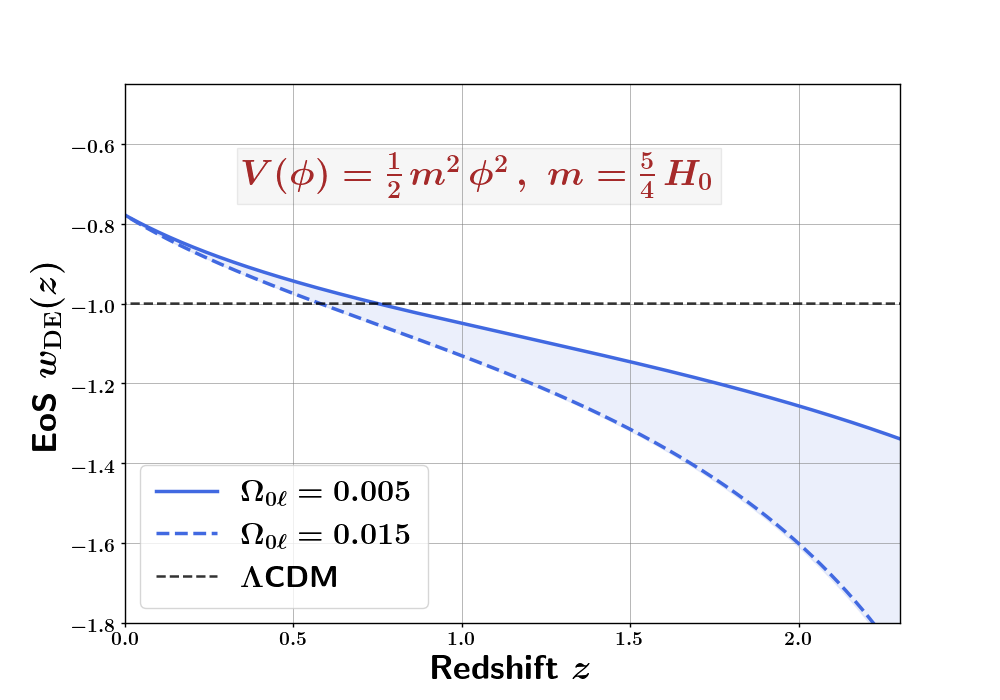}
\includegraphics[width=0.495\textwidth]{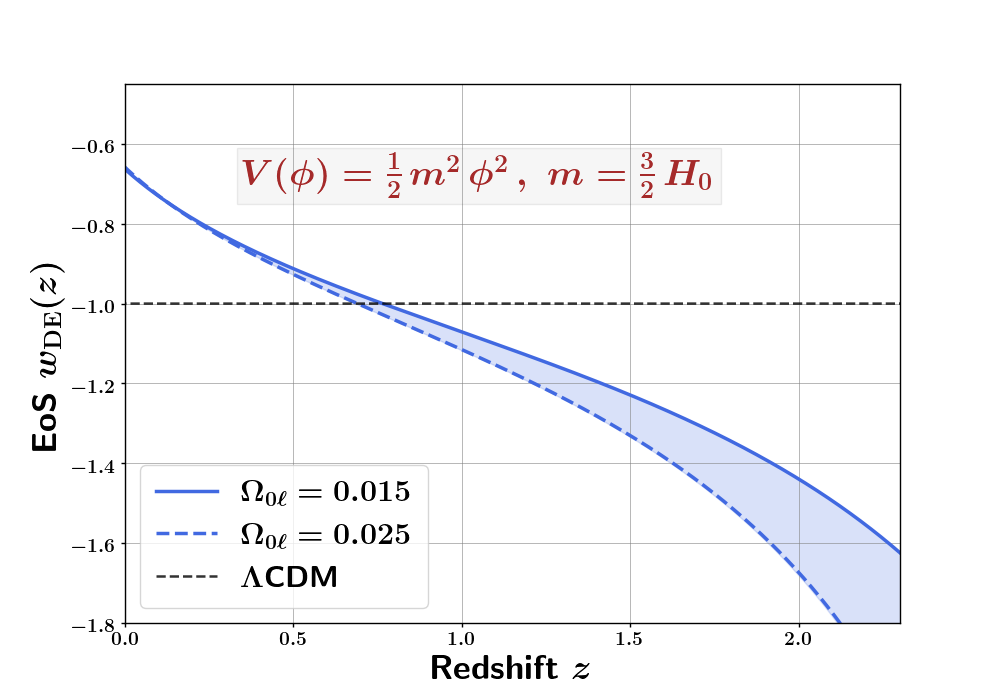}
\vspace{-0.2in}
\caption{The DE equation-of-state parameter corresponding to the quadratic potential (\ref{eq:pot_Quad}) is shown for $m=\frac54 H_0$ ({\bf left panel}), and  for  $m=\frac32 H_0$ ({\bf right panel}). Note that a larger value of the higher dimensional parameter, $\Omega_{0\ell}$, results in a lower redshift of phantom crossing.}
\label{fig:DE_PhBrane_Quad_EoS}
\end{center}
\end{figure}

 The results of our simulations are illustrated in Figs.~\Ref{fig:DE_PhBrane_Quad_Hubble} and \ref{fig:DE_PhBrane_Quad_EoS} for the evolution of the Hubble parameter (\ref{eq:h_PhBrane}) (relative to its value in $\Lambda$CDM) and the effective EoS of DE (\ref{eq:Brane_wDE}), respectively. Plots for the evolution of the density fraction of DE~(\ref{eq:Brane_fDE}), the deceleration parameter~(\ref{eq:Brane_q}) and the $Om$ diagnostic~(\ref{eq:Om}) are shown in App.~\ref{app:Quadratic}.  Note that the plots are generated for two distinct values of the scalar field mass, namely, $m=\frac54 H_0$ and $\frac32 H_0$. 


The blue curves in Fig.~\ref{fig:DE_PhBrane_Quad_EoS} demonstrate that the EoS of  DE undergoes a phantom crossing in the recent past, leading to $w_{\rm DE} > -1$ at the present epoch. For a fixed mass of the scalar field, comparing the solid and dashed blue curves in the left and right panels of Fig.~\ref{fig:DE_PhBrane_Quad_EoS}, we notice that the epoch of phantom crossing gets shifted towards a lower redshift, while the DE EoS falls off sharply towards higher redshifts, upon increasing the value of $\Omega_{0\ell}$. In fact, this appears to be a universal feature of the braneworld parameter $\Omega_{0\ell}$  across all the potentials used in this work.

On the other hand, for a given $\Omega_{0\ell}$, an increase in the mass of the scalar field results in a steeper phantom crossing, and consequently, a higher DE EoS at the present epoch. However, if the mass is exceedingly high, \textit{i.e.\@} $m/H_0 \gtrsim 1$, then the scalar field begins to oscillate prior to the present epoch, leading to {\em phantom oscillations\/} \cite{Zhao:2017cud}, which is not consistent with the DESI DR2 results. Similarly, if the mass is very small, \textit{i.e.\@} $m \ll H_0$, then the scalar field continues to remain frozen at its initial value, leading to $w_{\rm DE} \lesssim -1$ at the present epoch, with no phantom crossing. 

A visual comparison of our plots in  Figs.~\Ref{fig:DE_PhBrane_Quad_Hubble}--\ref{fig:DE_PhBrane_Quad_EoS}, with those of Ref.~\cite{DESI:2025fii}, indicates that the DESI DR2 constraints  appear to be consistent with a massive scalar field (on the brane) with parameters $ m \sim H_0$, $\Omega_{0{\ell}} \sim 0.01$.  As discussed before, for $\Omega_{0\ell} \to 0$, our model reduces to that of a scalar field with a quadratic potential in GR\@. A more systematic and detailed investigation of parameter estimation, involving MCMC analysis, is carried out in Sec.~\ref{sec:MCMC},  where results for the quadratic potential in GR ($\Omega_{0\ell} =0$) is compared with that of the braneworld scenario ($\Omega_{0\ell} \neq 0$).

\subsection{Quartic potential}
\label{sec:PhBrane_Quartic}

We next turn our attention to the simple quartic potential, given by
\begin{equation}
V(\phi) = \frac{\lambda}{4}\,\phi^4 \, .
\label{eq:pot_Quartic}
\end{equation}
The potential is schematically illustrated in the top-left panel of Fig.~\ref{fig:DE_pots}. 

\begin{figure}[htp]
\begin{center}
\includegraphics[width=0.495\textwidth]{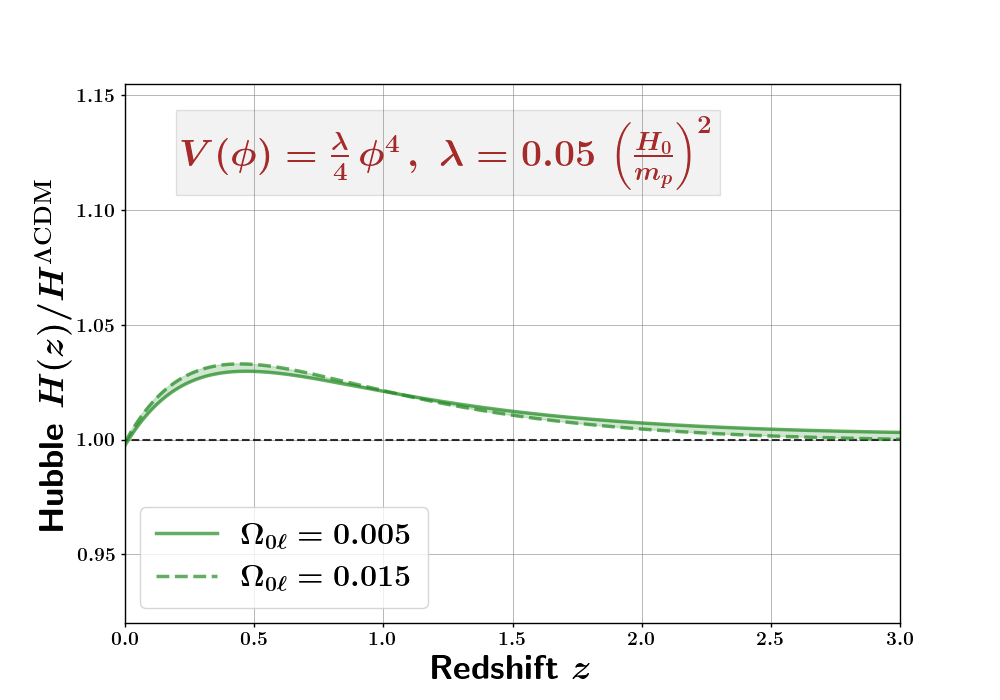}
\includegraphics[width=0.495\textwidth]{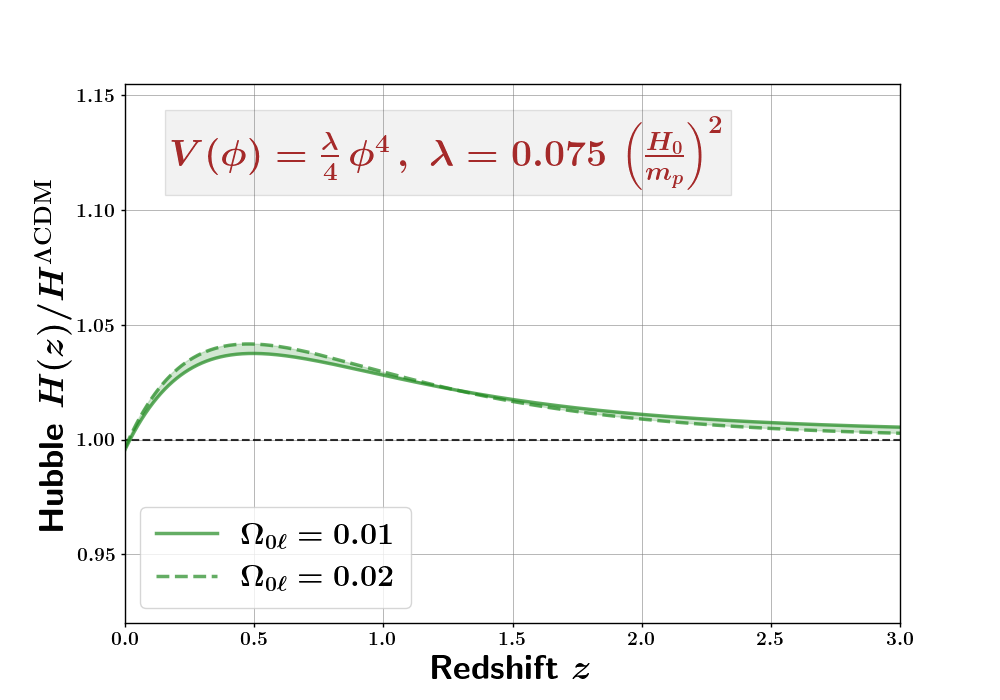}
\vspace{-0.2in}
\caption{The Hubble parameter corresponding to the quartic potential~(\ref{eq:pot_Quartic}) is shown for $\lambda = 0.05 \left( H_0/m_p \right)^2$ ({\bf left panel}), and $\lambda = 0.075 \left( H_0/m_p \right)^2$ ({\bf right panel}).}
\label{fig:DE_PhBrane_Quartic_Hubble}
\end{center}
\end{figure}

\begin{figure}[htp]
\begin{center}
\includegraphics[width=0.495\textwidth]{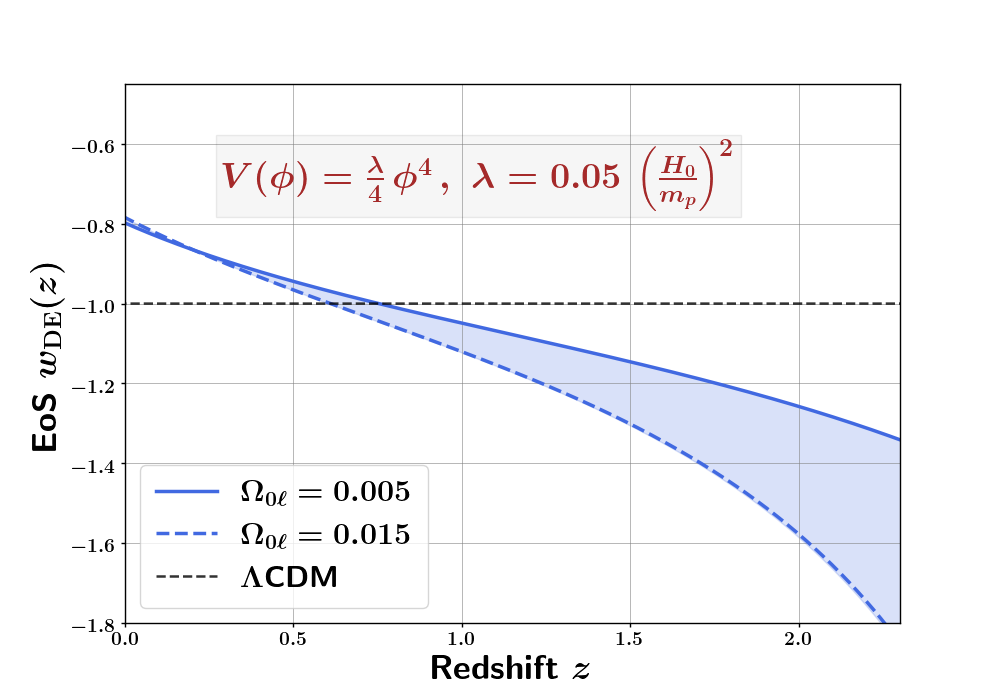}
\includegraphics[width=0.495\textwidth]{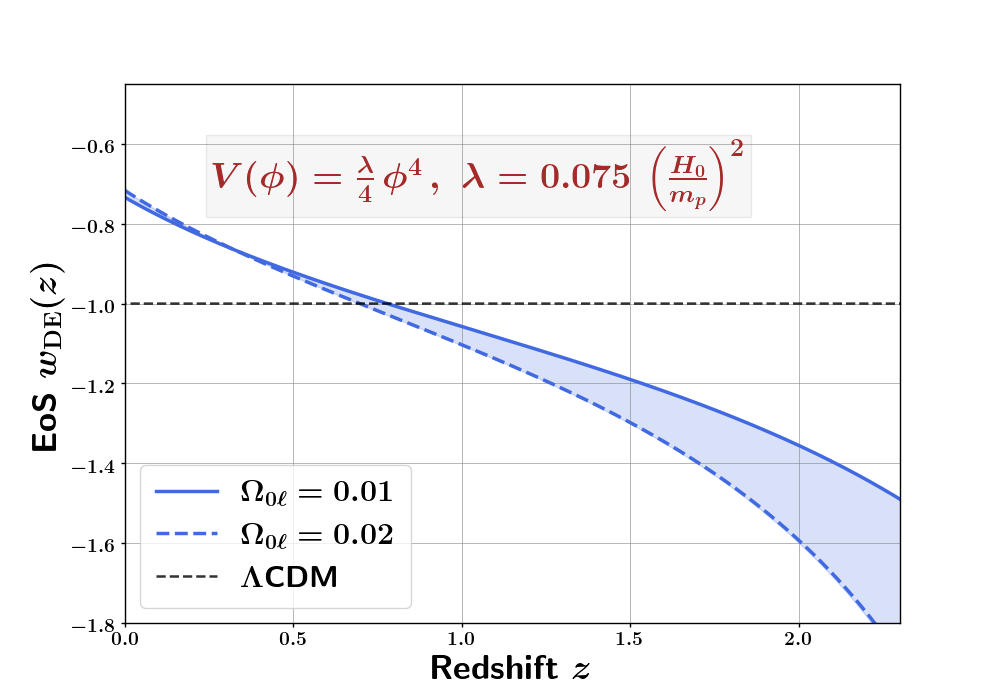}
\vspace{-0.2in}
\caption{The DE EoS corresponding to the quartic potential (\ref{eq:pot_Quartic}) is shown for $\lambda=0.05 \left( H_0/m_p \right)^2$ ({\bf left panel}), and  for  $\lambda = 0.075 \left( H_0/m_p \right)^2$ ({\bf right panel}). Note that a larger value of the higher dimensional parameter, $\Omega_{0\ell}$, results in a lower redshift of phantom crossing.}
\label{fig:DE_PhBrane_Quartic_EoS}
\end{center}
\end{figure}

We perform numerical simulations for a range of (fixed) values of $\lambda$.  The results of our simulations are illustrated in Figs.~\Ref{fig:DE_PhBrane_Quartic_Hubble} and \ref{fig:DE_PhBrane_Quartic_EoS} for the evolution of the Hubble parameter (\ref{eq:h_PhBrane}) (relative to its value in $\Lambda$CDM) and the effective  EoS of DE (\ref{eq:Brane_wDE}), respectively. Plots for the density fraction of DE~(\ref{eq:Brane_fDE}), the deceleration parameter~(\ref{eq:Brane_q}) and the $Om$ diagnostic~(\ref{eq:Om}) are shown in App.~\ref{app:Quartic}.  Note that these plots are generated for two distinct values of $\lambda$, namely, $\lambda = 0.05 \left( H_0/m_p \right)^2$ and $\lambda = 0.075 \left( H_0/m_p \right)^2$.  For a given $\Omega_{0\ell}$, an increase in the value $\lambda$ results in a steeper phantom crossing, and consequently, a higher DE EoS at the present epoch. 

 A visual comparison of our plots in Figs.~\Ref{fig:DE_PhBrane_Quartic_Hubble}--\Ref{fig:DE_PhBrane_Quartic_EoS}, with those of Ref.~\cite{DESI:2025fii}, indicates that DESI DR2 constraints appear to be well explained by the quartic potential on the phantom brane, closer to the range of parameters $\lbrace \lambda,\,\Omega_{0{\ell}}\rbrace$ used in these figures. A much larger or smaller value of $\lambda$ will clearly not be compatible with the DESI constraints. A systematic parameter estimation for this model is carried out in Sec.~\ref{sec:MCMC}.
 
\subsection{Symmetry-breaking potential}
\label{sec:PnBrane_SB}

Consider the symmetry-breaking potential
\begin{equation}
V(\phi) = \frac{\lambda}{4} \, \l( \phi^2 - \nu^2 \r)^2 \, ,
\label{eq:pot_SB}
\end{equation}
where $\lambda$  is  the strength of  self-interaction, while $\pm\,\nu$ is the vacuum expectation value of the field $\phi$ in the right and left vacua, respectively. The theory with this potential, schematically illustrated in the top-right panel of Fig.~\ref{fig:DE_pots}, is renormalisable and often appears in models of unification of gauge theories, including the Standard Model of particle physics.

Without loss of generality, we assume that the scalar field evolves on the right side, that is, $\phi > 0$. We then perform the field redefinition 
\begin{equation}
\Phi = \phi - \nu \, . 
\end{equation}
The potential in terms of the new field takes the form 
\begin{equation}
V(\Phi) = \frac12 \left( 2\lambda\nu^2 \right) \Phi^2 + \lambda\nu\,\Phi^3 +  \frac{\lambda}{4} \, \Phi^4 \, ,
\label{eq:pot_SB_Right}
\end{equation}
which indicates that the mass of the scalar field around the vacuum $\Phi = 0$ is given by
\begin{equation}
m_\Phi = \sqrt{2\lambda}\, \nu  \quad \Rightarrow \quad \frac{m_\Phi}{H_0} = \sqrt{2\left(\frac{\lambda m_p^2}{H_0^2}\right)} \left( \frac{\nu}{m_p} \right) \, .
\label{eq:pot_SB_mass}
\end{equation}
Given that there are two free parameters in the potential, namely, $\lbrace \lambda,\,\nu\rbrace$, we carry out the simulations systematically for different (fixed) values of the mass parameter $m_\Phi$. Eq.~(\ref{eq:pot_SB_mass}) then relates $\nu$ with a given choice of $\lambda$. Around this right-side vacuum, the scalar field can begin its descent from its frozen state either along the steep right wing or the flat hilltop of the potential (as illustrated by the brown circles and arrows in the top-right panel of Fig.~\ref{fig:DE_pots}), depending on the initial conditions in the early universe.

To facilitate comparison between our results for the symmetry-breaking potential and those for the quadratic and quartic potentials, we explicitly generate the plots using the values of $m_\Phi$ and $\lambda$ which give results similar to those in Sec.~\ref{sec:PhBrane_Quad} and Sec.~\ref{sec:PhBrane_Quartic}, respectively.

\subsubsection{Descent from the steep wing}
\label{sec:DE_PhBrane_SB_R}

\begin{figure}[htp]
\begin{center}
\includegraphics[width=0.495\textwidth]{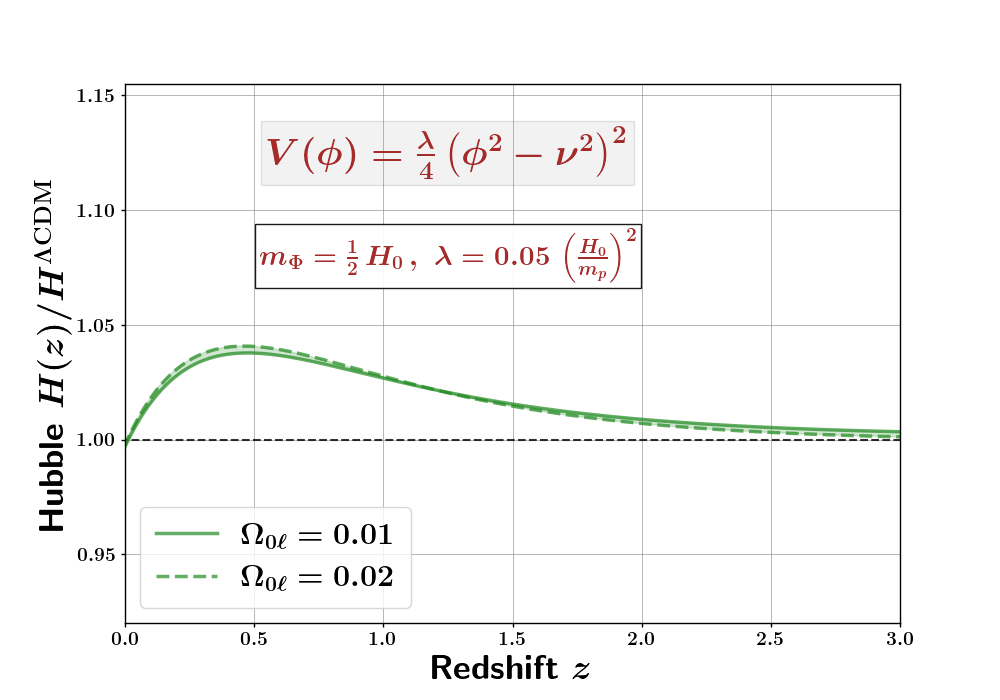}
\includegraphics[width=0.495\textwidth]{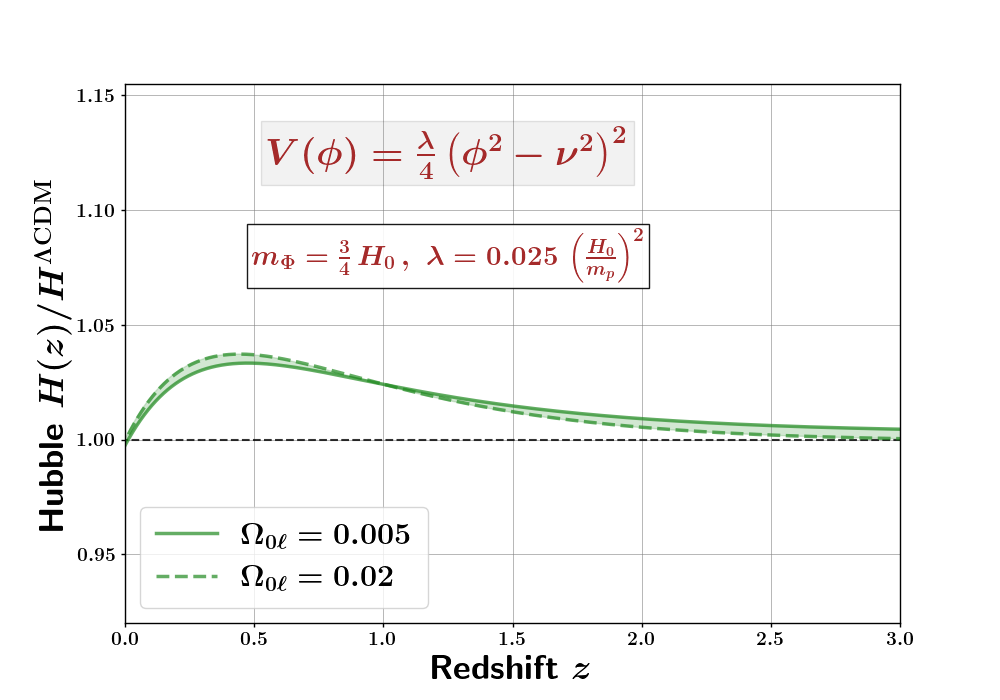}
\vspace{-0.2in}
\caption{The Hubble parameter corresponding to the steep right wing of the symmetry-breaking  potential~(\ref{eq:pot_SB}) is shown for $m_\Phi=\frac{1}{2}\,H_0\,,~\lambda = 0.05 \left( H_0/m_p \right)^2$ ({\bf left panel}), and $m_\Phi=\frac{3}{4}\,H_0\,,~\lambda = 0.025 \left( H_0/m_p \right)^2$ ({\bf right panel}).}
\label{fig:DE_PhBrane_SB_R_Hubble}
\end{center}
\end{figure}

\begin{figure}[htp]
\begin{center}
\includegraphics[width=0.495\textwidth]{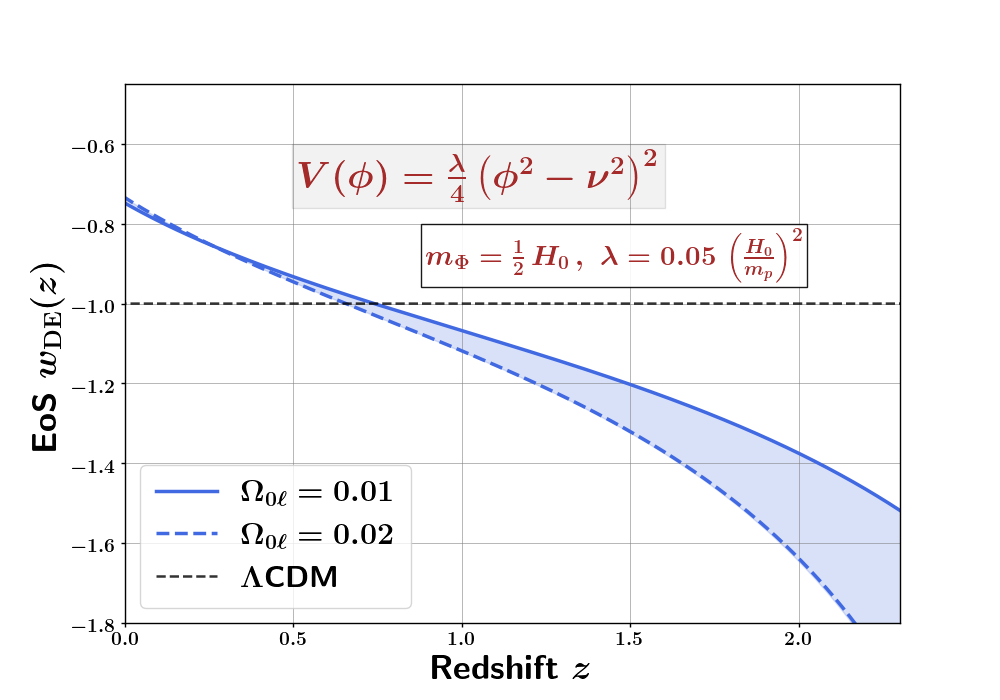}
\includegraphics[width=0.495\textwidth]{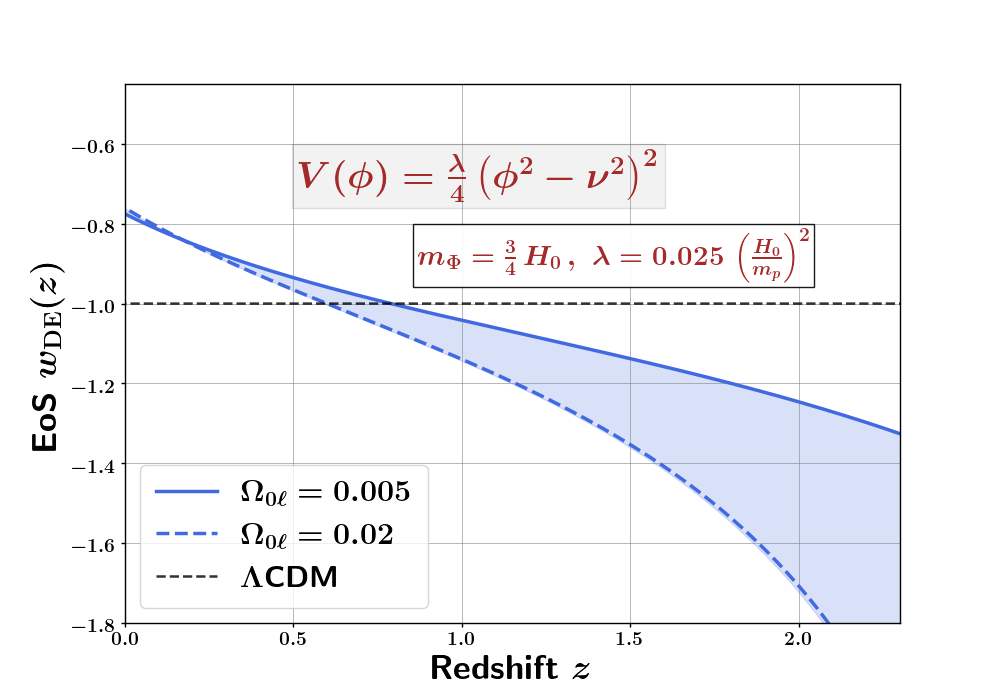}
\vspace{-0.2in}
\caption{The DE equation-of-state parameter  corresponding to the steep right wing of the symmetry-breaking  potential~(\ref{eq:pot_SB}) are shown for $m_\Phi=\frac{1}{2}\,H_0\,,~\lambda = 0.05 \left( H_0/m_p \right)^2$ ({\bf left panel}), and  for  $m_\Phi=\frac{3}{4}\,H_0\,,~\lambda = 0.025 \left( H_0/m_p \right)^2$ ({\bf right panel}). Note that a larger value of the higher dimensional parameter, $\Omega_{0\ell}$, results in a lower redshift of phantom crossing.  }
\label{fig:DE_PhBrane_SB_R_EoS}
\end{center}
\end{figure}

Our results for parameter values $\lbrace m_\Phi,\,\lambda \rbrace$ relevant for scalar-field  evolution along the steep right wing  are illustrated in Figs.~\Ref{fig:DE_PhBrane_SB_R_Hubble} and \ref{fig:DE_PhBrane_SB_R_EoS} for the evolution of the Hubble parameter~(\ref{eq:h_PhBrane}) (relative to its value in $\Lambda$CDM) and the effective  EoS of DE~(\ref{eq:Brane_wDE}), respectively. Plots for the density fraction of DE~(\ref{eq:Brane_fDE}), the deceleration parameter~(\ref{eq:Brane_q}) and the $Om$ diagnostic~(\ref{eq:Om}) are shown in App.~\ref{app:DE_PhBrane_SB_R}.  

 A visual comparison of our plots with those of Ref.~\cite{DESI:2025fii} indicates that the DESI constraints can be accommodated by the steep right wing of the symmetry-breaking potential on the phantom brane, provided that both $m_\Phi$ and $\lambda$ are close to the values in our plots. When these parameters are exceedingly large, the effective dynamical dark energy undergoes a phantom crossing that occurs too early and too abruptly, which is inconsistent with the DESI DR2 constraints. For a more robust and systematic estimation of the parameters, see Sec.~\ref{sec:MCMC}.
 
\subsubsection{Descent from the flat wing}
\label{sec:DE_PhBrane_SB_L}

\begin{figure}[htb]
\vspace{-0.2in}
\begin{center}
\includegraphics[width=0.495\textwidth]{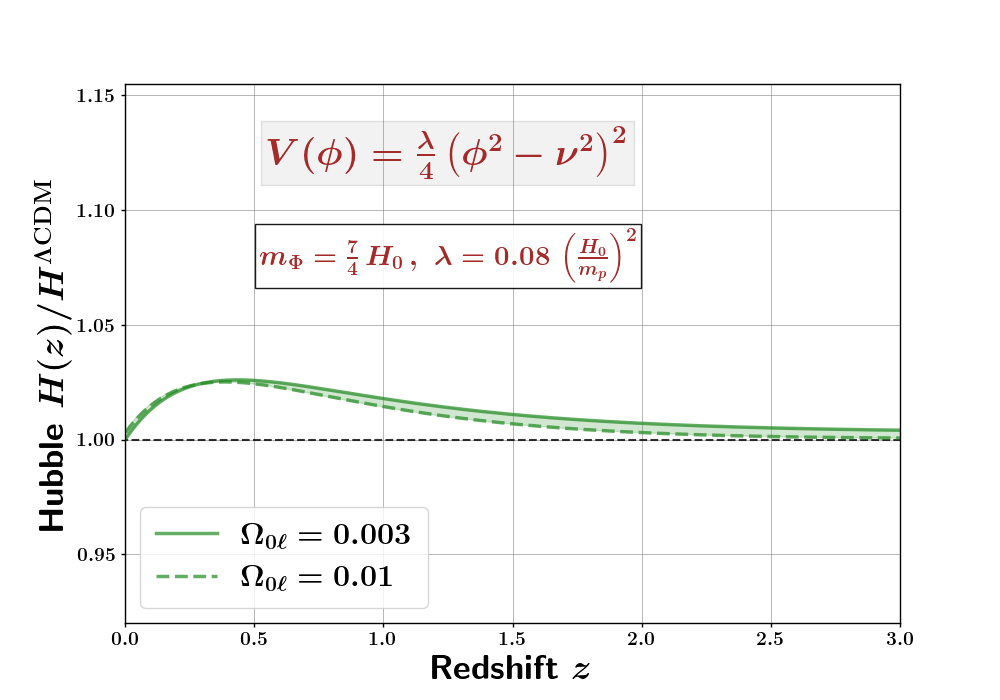}
\includegraphics[width=0.495\textwidth]{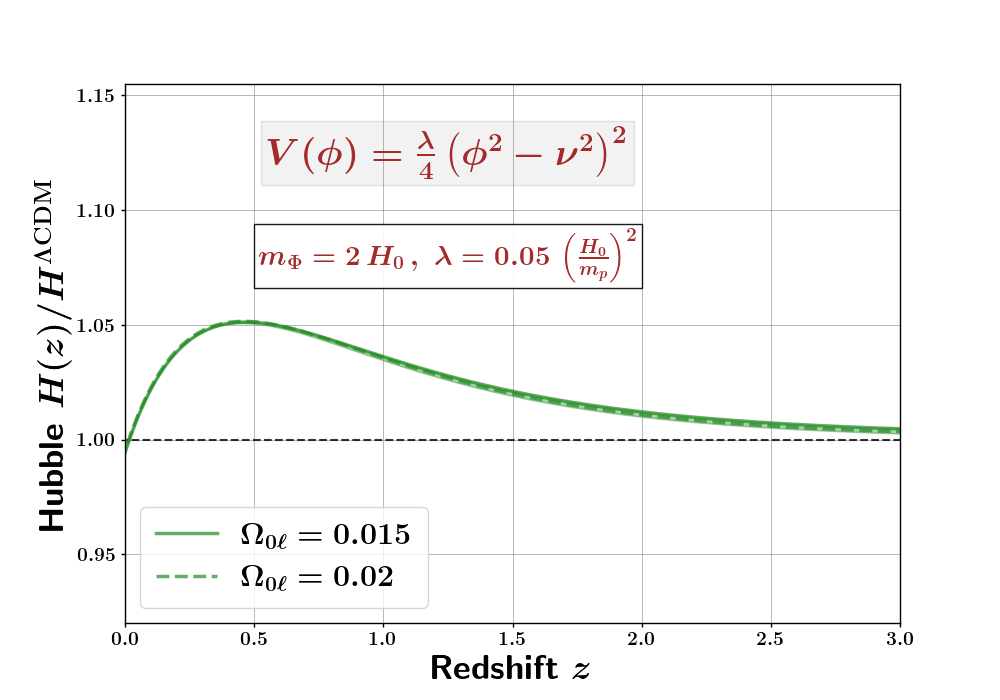}
\vspace{-0.2in}
\caption{The Hubble parameter corresponding to the flat left wing of the symmetry-breaking  potential~(\ref{eq:pot_SB}) is shown for $m_\Phi=\frac{7}{4}\,H_0\,$, $\lambda = 0.08 \left( H_0/m_p \right)^2$ ({\bf left panel}), and $m_\Phi=2\,H_0\,$, $\lambda = 0.05 \left( H_0/m_p \right)^2$ ({\bf right panel}).}
\label{fig:DE_PhBrane_SB_L_Hubble}
\end{center}
\end{figure}

\begin{figure}[htb]
\vspace{-0.2in}
\begin{center}
\includegraphics[width=0.495\textwidth]{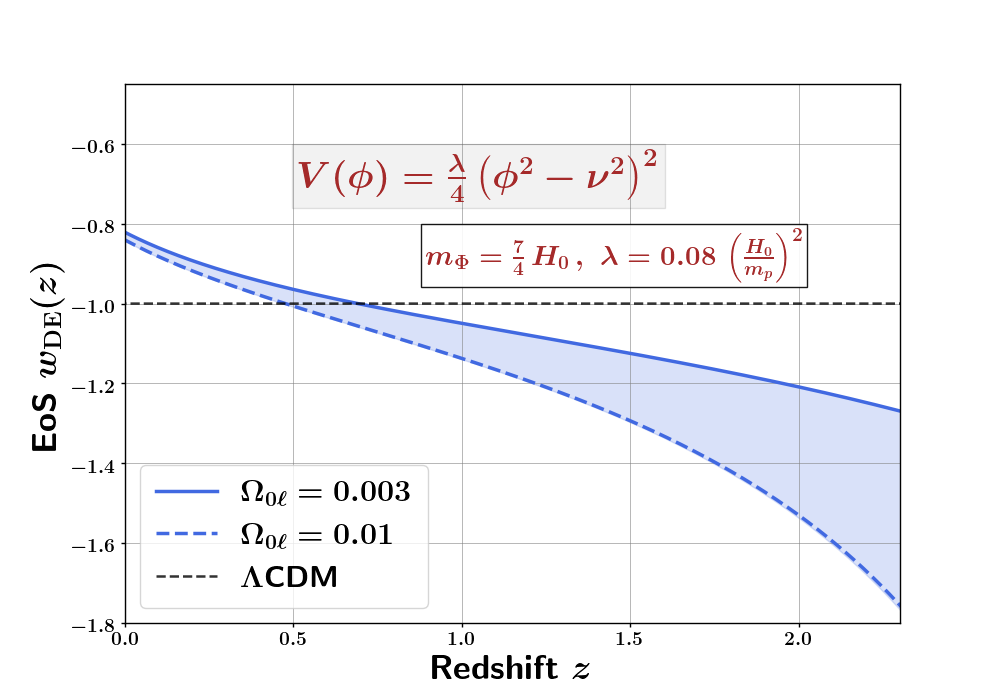}
\includegraphics[width=0.495\textwidth]{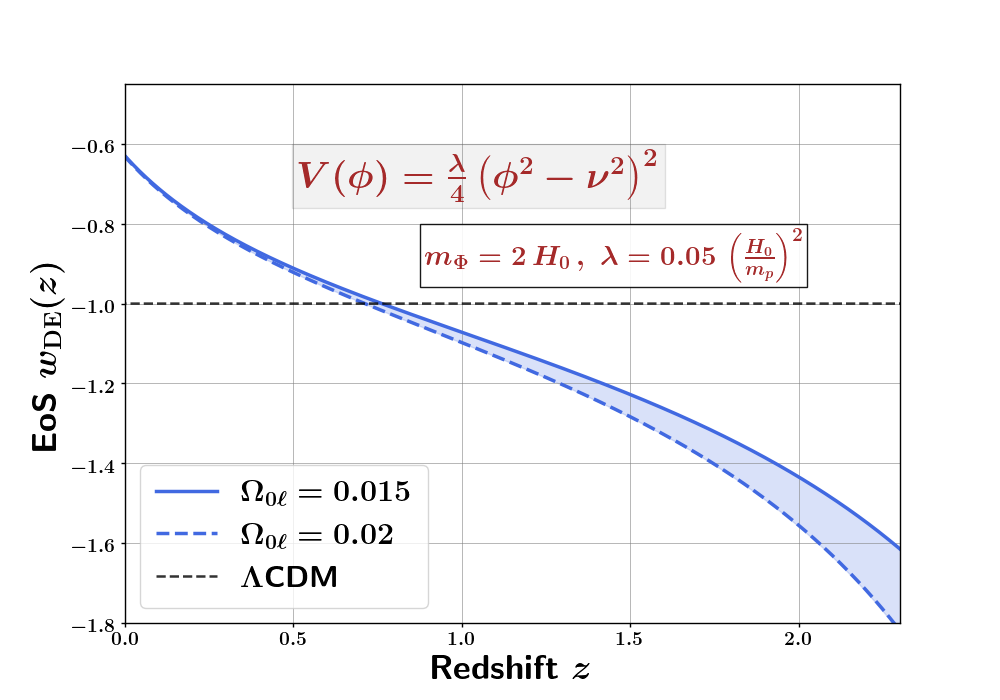}
\vspace{-0.2in}
\caption{The DE equation-of-state parameter corresponding to the flat left wing of the symmetry-breaking  potential~(\ref{eq:pot_SB}) is shown for $m_\Phi=\frac{7}{4}\,H_0\,$, $\lambda = 0.08 \left( H_0/m_p \right)^2$ ({\bf left panel}), and  for  $m_\Phi=2\,H_0\,$, $\lambda = 0.05 \left( H_0/m_p \right)^2$ ({\bf right panel}). Note that a larger value of the higher dimensional parameter, $\Omega_{0\ell}$, results in a lower redshift of phantom crossing. }
\label{fig:DE_PhBrane_SB_L_EoS}
\end{center}
\end{figure}

Our results for a range of parameter values of $\lbrace m_\Phi,\,\lambda \rbrace$ relevant for scalar-field  evolution along the flat left wing   are illustrated in Figs.~\Ref{fig:DE_PhBrane_SB_L_Hubble} and \ref{fig:DE_PhBrane_SB_L_EoS} for the evolution of the Hubble parameter~(\ref{eq:h_PhBrane}) (relative to its $\Lambda$CDM value) and the effective EoS of DE~(\ref{eq:Brane_wDE}) respectively. Plots for the density fraction of DE~(\ref{eq:Brane_fDE}), the deceleration parameter~(\ref{eq:Brane_q}) and the $Om$ diagnostic~(\ref{eq:Om}) are shown in App.~\ref{app:DE_PhBrane_SB_L}. A rudimentary comparison of our plots with those of Ref.~\cite{DESI:2025fii} indicates that the DESI constraints are reasonably well explained by the flat left wing of the symmetry-breaking potential on the phantom brane, provided that both $m_\Phi$ and $\lambda$  are relatively higher than those in the case of the steep right wing discussed in Sec.~\ref{sec:DE_PhBrane_SB_R}, as can be seen by comparing Fig.~\ref{fig:DE_PhBrane_SB_L_EoS} with  Fig.~\ref{fig:DE_PhBrane_SB_R_EoS}.  


We note that for a given value of the effective mass ($m_\Phi$), a flatter potential is relatively less effective in explaining the DESI DR2 constraints compared to a quadratic potential of the same mass, as discussed in Sec.~\ref{sec:compare}. This is due to the fact that flat potentials display shallower rate of phantom crossing as compared to steeper potentials. Therefore, they require  relatively larger values of the effective mass in order to be consistent with the data.  A systematic estimation of parameters is carried out in Sec.~\ref{sec:MCMC} using MCMC.

\subsection{Exponential potential}
\label{sec:PhBrane_Exp}
 The exponential potential
\beq
V(\phi) = V_0 \, \exp{\l(\lambda\, \f{\phi}{m_p}\r)} \, ,
\label{eq:pot_Exponential}
\eeq
is ubiquitous in both early-Universe \cite{Halliwell:1986ja,Lucchin:1984yf} and late-Universe \cite{Copeland:1997et,Ferreira:1997au} cosmology, and appears naturally in string theory \cite{ValeixoBento:2020ujr, Cicoli:2023opf, Apers:2024ffe}. In standard GR, a pure exponential potential with $\lambda < \sqrt{2}$ behaves like DE, while $\lambda > \sqrt{2}$ does not result in late-time acceleration.\footnote{In particular, an exponential potential with $\lambda^2 > 3(1+w_B)$ in a Universe dominated by a perfect fluid with EoS $w_B$ results in  a scaling attractor solution~\cite{Copeland:1997et}.  Therefore, modifications to the exponential potential is desired~\cite{Sen:2001xu,Barreiro:1999zs,Sahni:1999qe,Mishra:2017ehw,Bag:2017vjp} in order to realise DE in GR\@. In fact, our analysis of DE in the braneworld for the pure exponential potential will also carry over conveniently to modifications of the exponential function, such as $V(\phi) = V_0\, \l[\cosh\l(\lambda \phi/m_p\r) - 1\r]$ \cite{Sahni_2000}.} However, in the braneworld scenario, it is possible to obtain transient late-time acceleration from exponential potentials\,\footnote{In GR, a scalar field with steep exponential potential ($\lambda \gtrsim \sqrt{2}$) with a matter-coupling has been studied in Ref.~\cite{Andriot:2025los} in the context of  DESI DR2.} with $\lambda \gtrsim \sqrt{2}$.

\begin{figure}[htb]
\vspace{-0.1in}
\begin{center}
\includegraphics[width=0.495\textwidth]{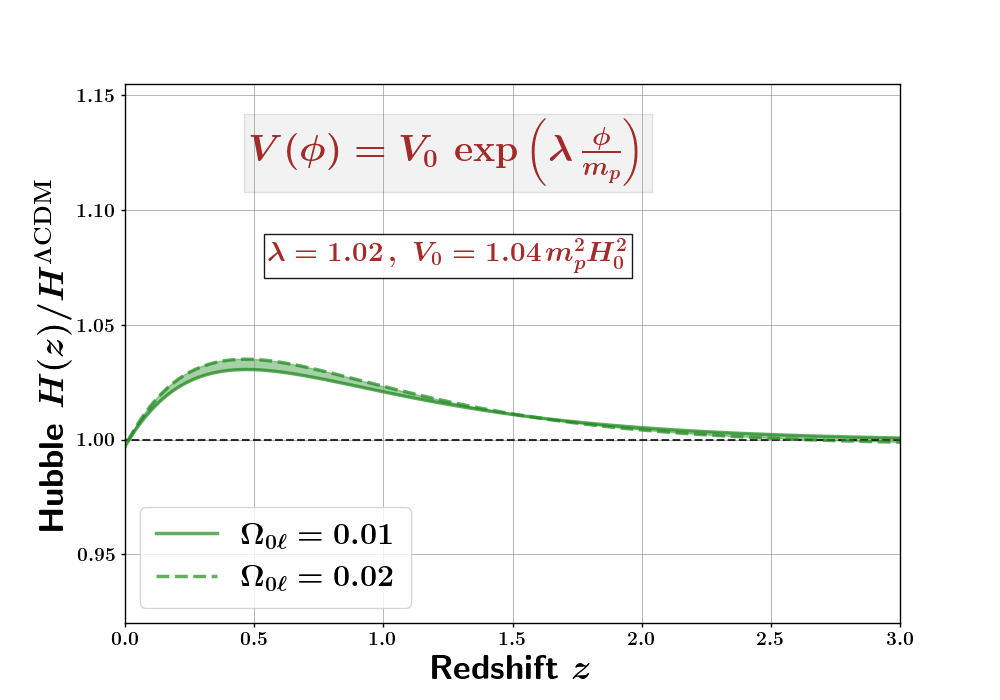}
\includegraphics[width=0.495\textwidth]{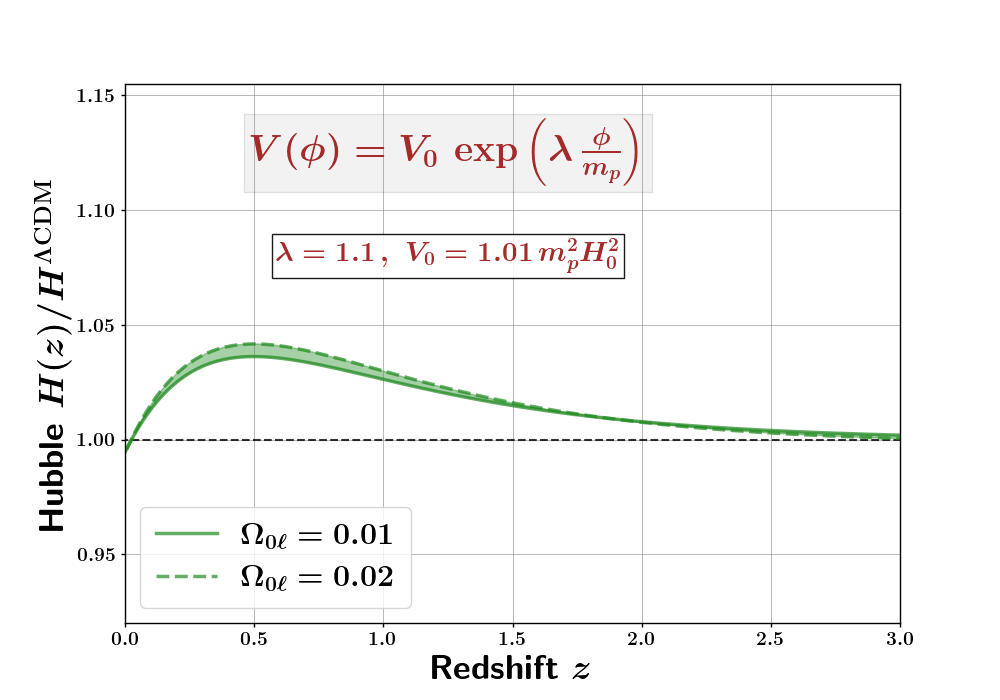}
\vspace{-0.2in}
\caption{The Hubble parameter corresponding to the exponential  potential~(\ref{eq:pot_Exponential}) is shown for $\lambda=1.02, \, V_0 =1.04\, m_p^2H_0^2$ ({\bf left panel}), and  for  $\lambda= 1.1, \, V_0 =1.01\, m_p^2H_0^2$ ({\bf right panel}).}
\label{fig:DE_PhBrane_Exp_Hubble}
\end{center}
\end{figure}

\begin{figure}[htb]
\vspace{-0.2in}
\begin{center}
\includegraphics[width=0.495\textwidth]{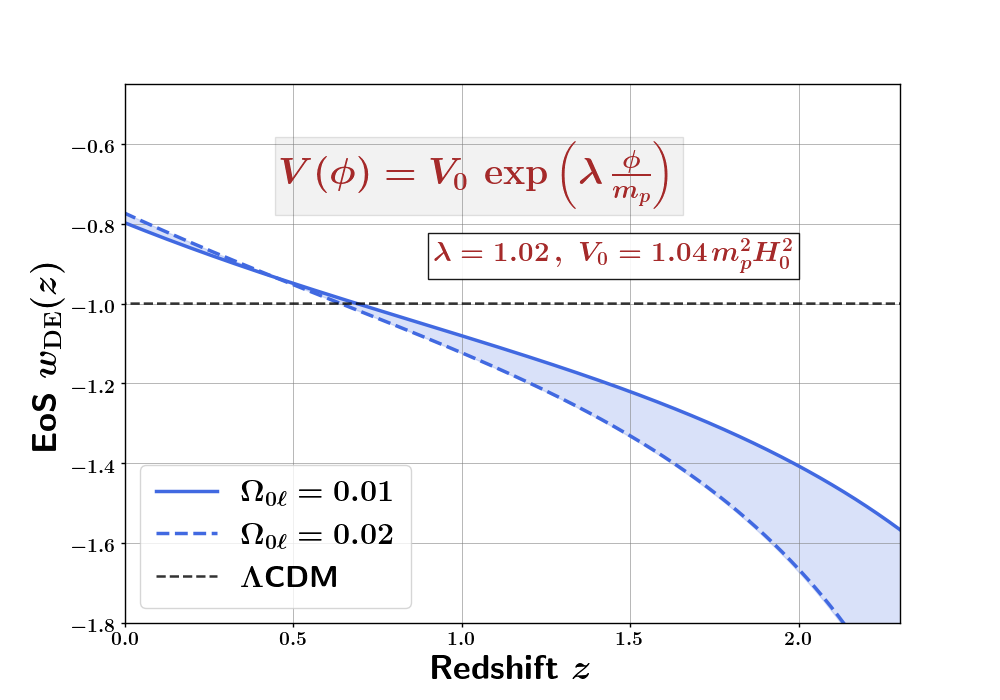}
\includegraphics[width=0.495\textwidth]
{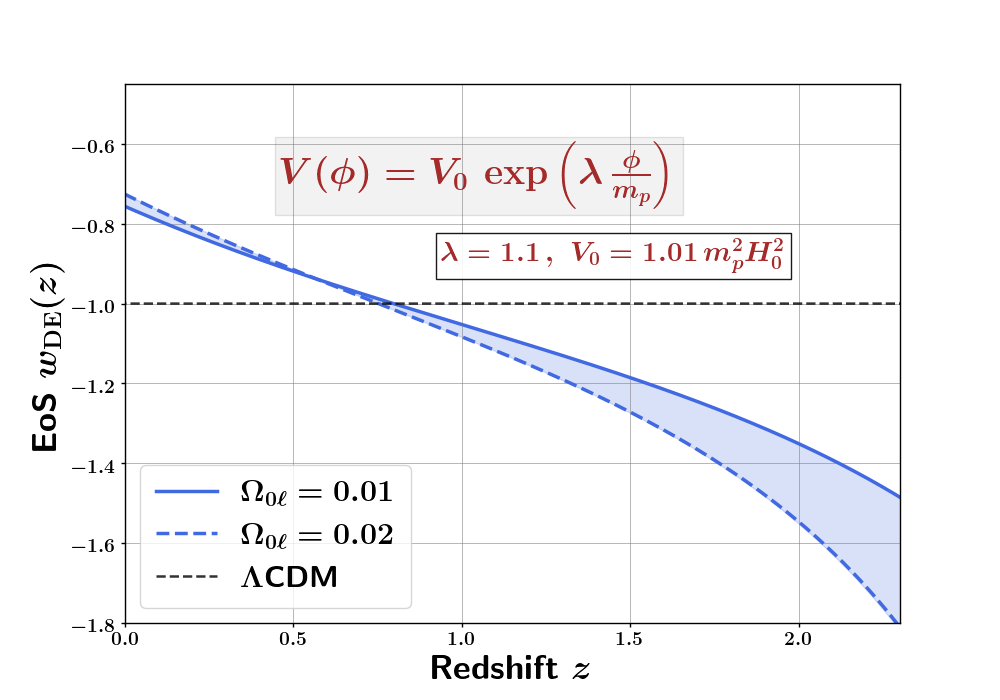}
\vspace{-0.2in}
\caption{The DE equation-of-state parameter corresponding to the exponential  potential~(\ref{eq:pot_Exponential}) is shown for $\lambda=1.02, \, V_0 =1.04\, m_p^2H_0^2$ ({\bf left panel}), and  for  $\lambda= 1.1, \, V_0 =1.01\, m_p^2H_0^2$ ({\bf right panel}). Note that a larger value of the higher dimensional parameter, $\Omega_{0\ell}$, results in a lower redshift of phantom crossing.}
\label{fig:DE_PhBrane_Exp_EoS}
\end{center}
\end{figure}

Our results for  a range of parameter values of $\lambda,\,V_0$  of the exponential potential   are illustrated in Figs.~\Ref{fig:DE_PhBrane_Exp_Hubble} and \ref{fig:DE_PhBrane_Exp_EoS} for the evolution of the Hubble parameter~(\ref{eq:h_PhBrane}) (relative to its $\Lambda$CDM value) and the effective EoS of DE~(\ref{eq:Brane_wDE}) respectively. Plots for the density fraction of DE~(\ref{eq:Brane_fDE}), the deceleration parameter~(\ref{eq:Brane_q}) and the $Om$ diagnostic~(\ref{eq:Om}) are shown in App.~\ref{app:Exp}. 

A visual comparison of our plots with those of Ref.~\cite{DESI:2025fii} indicates that the DESI DR2 constraints  can be accommodated by the exponential potential for $\lambda \approx \mathcal{O}(1)$. However, a more systematic parameter estimation using MCMC is discussed in Sec.~\ref{sec:MCMC}.

\subsection{Axion potential}
\label{sec:PhBrane_Axion}

The potential for a (pseudo) Nambu-Goldstone boson of the axion type is given by
\begin{equation}
V(\phi) = V_0 \l[ 1 -  \cos{\l( \frac{\phi}{{\cal F}}\r)} \r] \, ,
\label{eq:pot_Axion}
\end{equation}
where ${\cal F}$ is the axion decay constant. The potential is schematically illustrated in the bottom-left panel of Fig.~\ref{fig:DE_pots}  by purple curves, corresponding to a fixed value of $V_0$ and different values of ${\cal F}$.

 Note that, for $\phi \ll {\cal F}$, the potential can be approximated as a quadratic up to leading order in $\phi/{\cal F}$, that is,
\begin{equation}
V(\phi) \big\vert_{\,\phi \, \ll \, {\cal F}} \, \simeq \, \frac12 m_{\cal A}^2 \phi^2 - {\cal O}\l(\frac{\phi}{\cal F}\r)^4  \, ,
\label{eq:pot_Axion_quad}
\end{equation}
with 
\begin{equation}
m_{\cal A}^2 = \frac{V_0}{\mathcal{F}^2} \, ,
\label{eq:pot_Axion_quad_mass}
\end{equation}
which is illustrated by gray-colour plots in the bottom-left panel of Fig.~\ref{fig:DE_pots}.

\begin{figure}[htb]
\begin{center}
\includegraphics[width=0.495\textwidth]{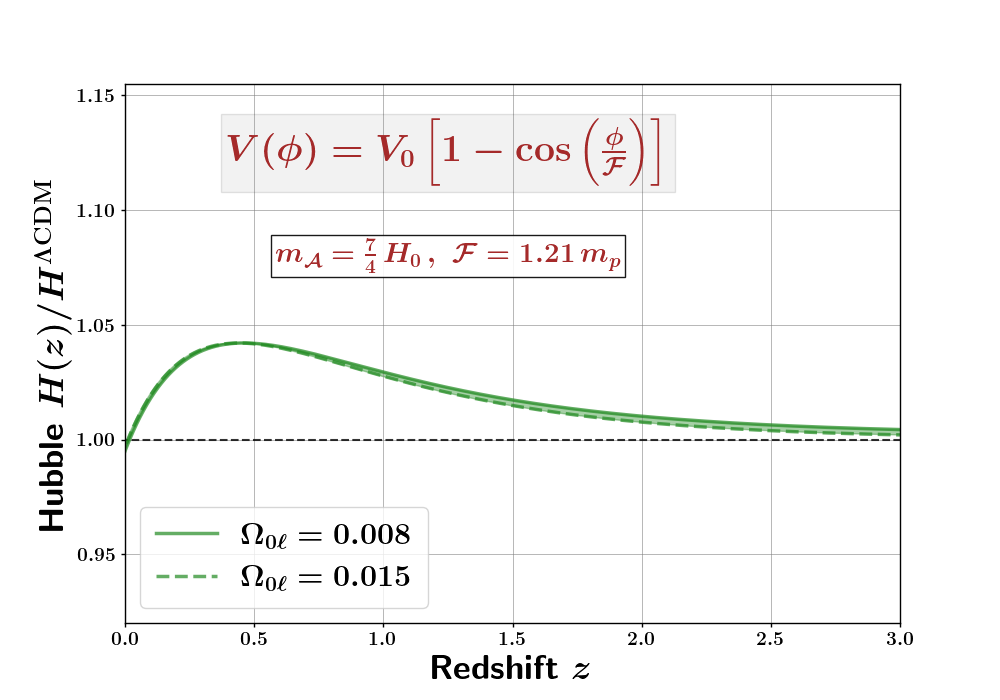}
\includegraphics[width=0.495\textwidth]{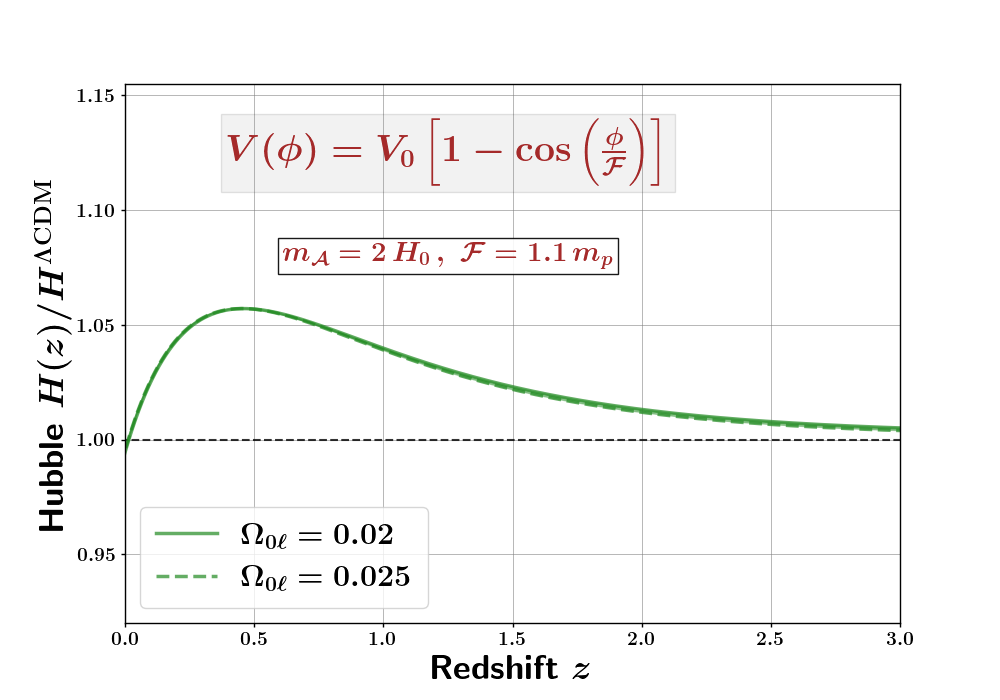}
\vspace{-0.2in}
\caption{The Hubble parameter corresponding to the Axion  potential~(\ref{eq:pot_Axion}) is shown for $m_{\cal A}=\frac{7}{4}\,H_0\,$, ${\cal F}=1.21\,m_p$ ({\bf left panel}), and $m_{\cal A}=2\,H_0\,$, ${\cal F}=1.1\,m_p$ ({\bf right panel}).}
\label{fig:DE_PhBrane_Axion_Hubble}
\end{center}
\end{figure}

\begin{figure}[htb]
\begin{center}
\includegraphics[width=0.495\textwidth]{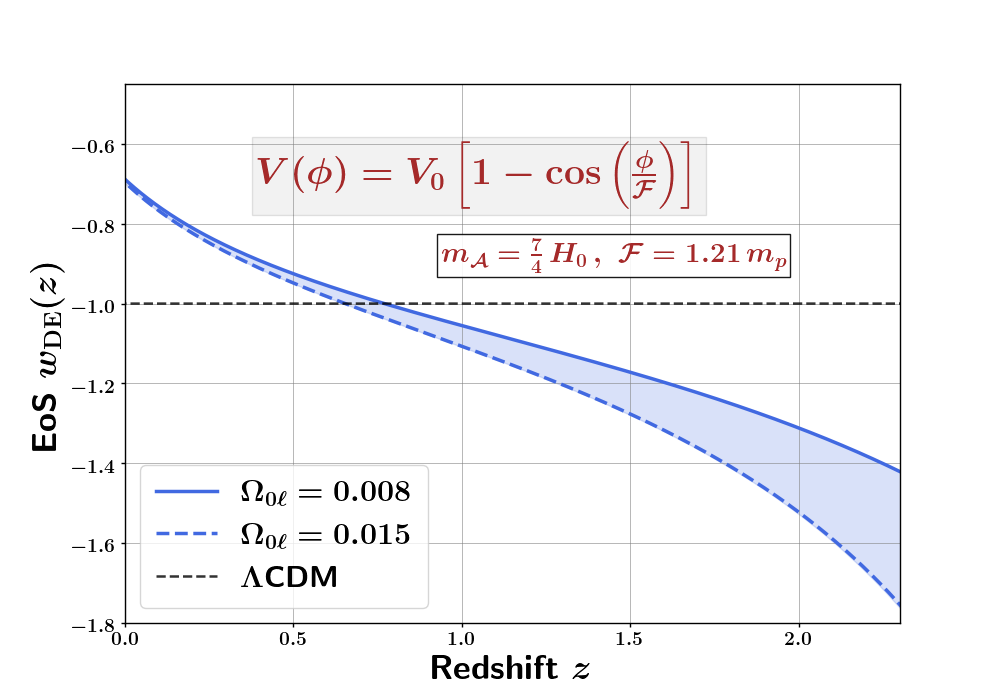}
\includegraphics[width=0.495\textwidth]{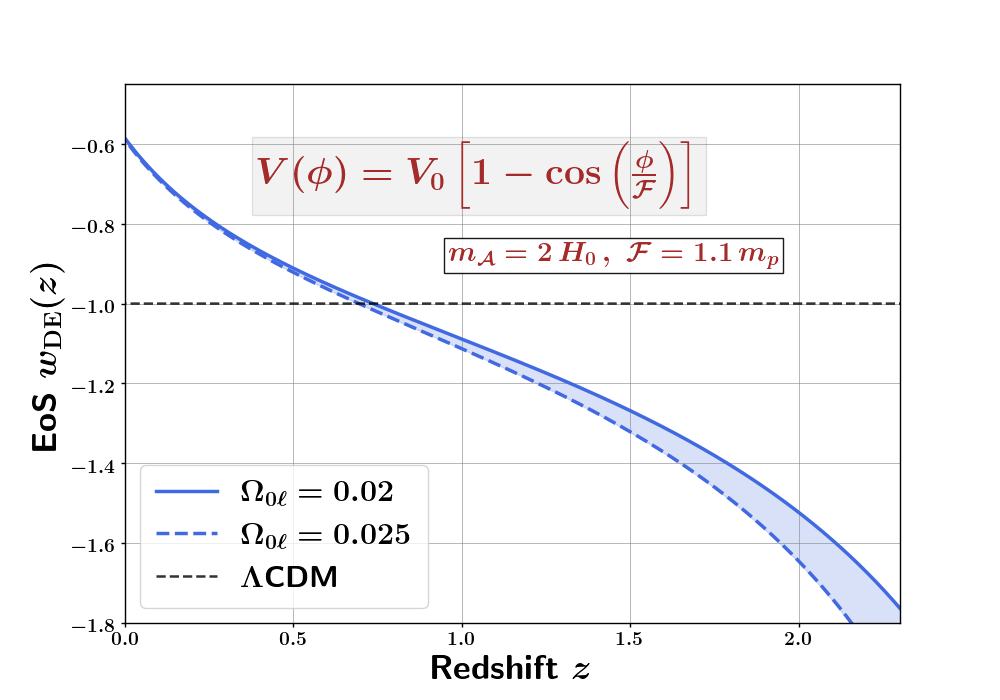}
\vspace{-0.2in}
\caption{The DE equation-of-state parameter corresponding to the Axion  potential~(\ref{eq:pot_Axion}) is shown for $m_{\cal A}=\frac{7}{4}\,H_0\,,~{\cal F}=1.21\,m_p$ ({\bf left panel}), and  for  $m_{\cal A}=2\,H_0\,,~{\cal F}=1.1\,m_p$ ({\bf right panel}). Note that a larger value of the higher dimensional parameter, $\Omega_{0\ell}$, results in a lower redshift of phantom crossing. }
\label{fig:DE_PhBrane_Axion_EoS}
\end{center}
\end{figure}

Our results for a range of parameter values of $\lbrace m_{\cal A},\,{\cal F} \rbrace$ for the Axion potential   are illustrated in Figs.~\Ref{fig:DE_PhBrane_Axion_Hubble} and \ref{fig:DE_PhBrane_Axion_EoS} for the evolution of the Hubble parameter~(\ref{eq:h_PhBrane}) (relative to its $\Lambda$CDM value) and the effective EoS of DE~(\ref{eq:Brane_wDE}) respectively. Plots for the density fraction of DE~(\ref{eq:Brane_fDE}), the deceleration parameter~(\ref{eq:Brane_q}) and the $Om$ diagnostic~(\ref{eq:Om}) are shown in App.~\ref{app:DE_PhBrane_Axion}.

A visual comparison of our plots with those of Ref.~\cite{DESI:2025fii} indicates that the DESI DR2 constraints  can be accommodated by the Axion potential, provided  $m_{\cal A}$ is somewhat larger than in the case of the quadratic potential discussed in Sec.~\ref{sec:PhBrane_Quad}. 

This is due to the fact that a flat potential, such as the Axion potential~(\ref{eq:pot_Axion}) which is bounded from above (and of course, from below), is always less steep than a quadratic potential of the same effective mass. In fact, for a given mass $m_{\cal A}$, choosing a large ${\cal F}$ corresponds to maximal steepness. In this limit, its behaviour mimics that of the quadratic potential, discussed in Sec.~\ref{sec:PhBrane_Quad}.  However, in the opposite limit, where ${\cal F}$ is as small as possible, the potential is much flatter. Phantom crossing occurs more slowly and it is less effective in describing the DESI DR2 data. 
Our plots in Figs.~\ref{fig:DE_PhBrane_Axion_Hubble}--\ref{fig:DE_PhBrane_Axion_EoS}, are generated for rather intermediate values of ${\cal F}$, namely, ${\cal F}=1.1$, and $1.21$. A more systematic parameter estimation using MCMC is discussed in Sec.~\ref{sec:MCMC}.

\section{Observational Constraints from DESI DR2}
\label{sec:MCMC}  

\subsection{Data combination}

In order to further gauge the viability of the models in the context of observed data, we perform a Markov Chain Monte Carlo (MCMC) analysis using a combination of the following most recent cosmological datasets\,:

\begin{itemize}

\item[] \textit{Type Ia Supernovae (SNe Ia)}: We make use of the Union~3 compilation \cite{Rubin:2023ovl}, which is based on a collection of 2087 Type~Ia supernovae in the redshift range $0.05 < z < 2.26$. The data comprise 22 distance modulus values, determined through the implementation of the Unity~1.5 Bayesian hierarchical framework, which allows for constraint of the expansion history through the luminosity distance $D_L(z)$. In our analysis we marginalize over the B-band absolute magnitude, $M_B$. Though there are differences between the methodology and results from the three latest supernovae compilations (Union-3, Pantheon+ \cite{Brout:2022vxf} and DES 5-year \cite{DES:2024tys}), they have been shown to display a degree of consistency such that we do not expect our results to change if we were to use one of the other ones instead \cite{Matthewson_2025}. Here we select Union-3 because of the relative computational ease that arises by virtue of the smaller associated covariance matrix.

\item[] \textit{Baryon Acoustic Oscillations (BAO)}: 
Using the DESI DR2 \cite{DESI:2025zgx}, we gain access to further constraints on the expansion history through the Hubble distance $D_H(z)$, the angular diameter distance $D_M(z)$ and the angle-averaged distance $D_V(z)$. The first two kinds of distance measurements are collected in six redshifts in the range $0.51 < z < 2.33$, while the angle-averaged measurement has an effective redshift of $z = 0.295$. 

\item[] \textit{CMB (compressed)}: In order to ensure the viable high-redshift behaviour of the resulting model fits, we make use of a combination of \textit{Planck\/} and Atacama Cosmology Telescope (ACT) CMB data. In particular, for reasons of numerical efficacy,\footnote{Using the full CMB likelihood would require careful and lengthy modifications to the dynamics of cosmological perturbations inside a Boltzmann solver, which we reserve for a future work.} we use CMB distance priors $R, \ell_a, \omega_b$, first used for this purpose by \cite{Wang_2007}, and shown to be a robust compression of CMB data.   Since these require high redshift quantities to be computed, the differential equations are solved starting at much higher redshift ($z>1000$) in the MCMC than in the analysis presented so far. The data and covariance of the CMB distance priors are taken from \cite{bansal2025expansionhistorypreferencesdesidr2} where they are computed for the PR3 \textit{Planck\/} \textsc{plik} likelihood \cite{2020Planck} and the Data Release~6 of the Atacama Cosmology Telescope \cite{Madhavacheril_2024}.

\end{itemize}

\subsection{MCMC Results}
\enlargethispage{\baselineskip}

We make use of the publicly available likelihoods along with custom theory code implemented in \texttt{Cobaya}~\cite{Torrado:2020dgo} to perform the parameter estimation, using MCMC sampling. The best-fit parameter values for the various models are reported in Table~\ref{table:MCMC_CPL}, and compared with the results for the CPL parametrisation\,\footnote{Although we display the best-fit curves and contours for the CPL parametrisation  here, it is important to stress the fact that the phantom-crossing preference of the DESI DR2 dataset is not merely a consequence of CPL parametrisation . Rather, a wide variety of reconstructions of the DE EoS prefer the phantom-crossing behaviour~\cite{Keeley:2025stf,DESI:2025wyn,DESI:2025fii,Lee:2025pzo}.} under the same framework. What is striking is that the $\chi^2$ values of all of the models are very close to that of the CPL parametrisation   (\textit{i.e.\@} $\l| \Delta\chi^2 \r| \ll 1$), indicating that the fits to the actual data of the models we investigate are able to attain a level comparable with CPL.

\begin{table}[H]
\small
 \vspace{0.15in}
 \begin{tabular}{c || c | c | c | c | c | c | c |} 
\hline\Tstrut
\hspace{-1in} Parameters $\longrightarrow$ &  $\Omega_{0m}$ & $H_0$ & $\Omega_{0b}h^2$ & Model Par-I & Model Par-II  & $\Omega_{0 \ell}$ &  $\Delta\chi^2$  \\ [0.4ex] 
\hspace{-1in}  Models  $\downarrow$  &  & (${\rm km/s/Mpc}$) &  &  &  &  $\l( \, \pm \, 68 \%  \, \r)$ &  ($\equiv \chi^2 - \chi^2_{\rm CPL}$) \\ [2ex] 
 \hline\hline\Tstrut
\hspace{-1in} CPL  & 0.3276 & 66.13 & 0.02238 & $w_0 = -0.67$ & $w_a = -1.07$ & -- &  0 \\ [2ex] 
 \hline\Tstrut
\hspace{-1in}  Quadratic (GR) &  0.3209 &  66.32 &  0.02254 & $\f{m}{H_0} = 1.038$ & -- & -- &  7.99 \\  [2ex] 
 \hline\Tstrut
\hspace{-1in}   Quadratic & 0.3239 & 66.53 & 0.02236 & $\f{m}{H_0} = 1.26$ & -- & $0.0132^{+0.0098}_{-0.0082}$ & 0.06 \\ [2ex] 
 \hline\Tstrut
 \hspace{-1in}  Quartic  & 0.3241 & 66.52 &  0.02239 & $ \f{\lambda\, m_p^2}{H_0^2} =  0.0514$ & -- & $0.0136^{+0.0125}_{-0.0049}$ & 0.19 \\ [2ex] 
 \hline\Tstrut
\hspace{-1in}  SB Steep  & 0.3241 & 66.48 & 0.02238 & $\frac{m_\Phi}{H_0} = 1.17$  & $\bm{\f{\lambda\, m_p^2}{H_0^2}  = 0.0012}$ & $0.0127^{+0.01216}_{-0.0057}$ & 0.09 \\ [2ex] 
 \hline\Tstrut
\hspace{-1in}   SB Flat  & 0.3252 & 66.41 & 0.02237 & $\frac{m_\Phi}{H_0} = 2.68$ & $\bm{\f{\lambda\, m_p^2}{H_0^2}  = 0.64}$ & $0.0115^{+0.0081}_{-0.005}$ & -0.16 \\ [2ex] 
 \hline\Tstrut
 \hspace{-1in}  Axion  & 0.3243 & 66.48 & 0.02238 & $\frac{m_{\mathcal A}}{H_0} = 1.28 $ & $\bm{\frac{\mathcal F}{m_p} = 6.40}$ & $0.01302^{+0.0087}_{-0.0076}$ & 0.06 \\ [2ex] 
 \hline\Tstrut
 \hspace{-1in}  Exp  & 0.3227 & 66.65 & 0.02237 & $ \lambda =  0.997$ & $\bm{\f{V_0}{m_p^2H_0^2}  = 0.846}$ & $0.0142^{+0.0064}_{-0.0084}$ & 0.24 \\ [2ex] 
 \hline
\end{tabular}
\caption{Best-fit values of different parameters  corresponding to the medians of the MCMC chains  for various potentials  are tabulated here. Very small values of $\Delta\chi^2$  (except for the GR case) in the bottom-most row indicate that our models explain the latest observational results (combining DESI DR2, Union\,3 and CMB \textit{Planck}\,$+$\,ACT datasets) very well, at least as well as the CPL parametrisation.}
\label{table:MCMC_CPL}
\end{table}

\begin{figure}[htb]
\begin{center}
\includegraphics[width=0.85\textwidth]{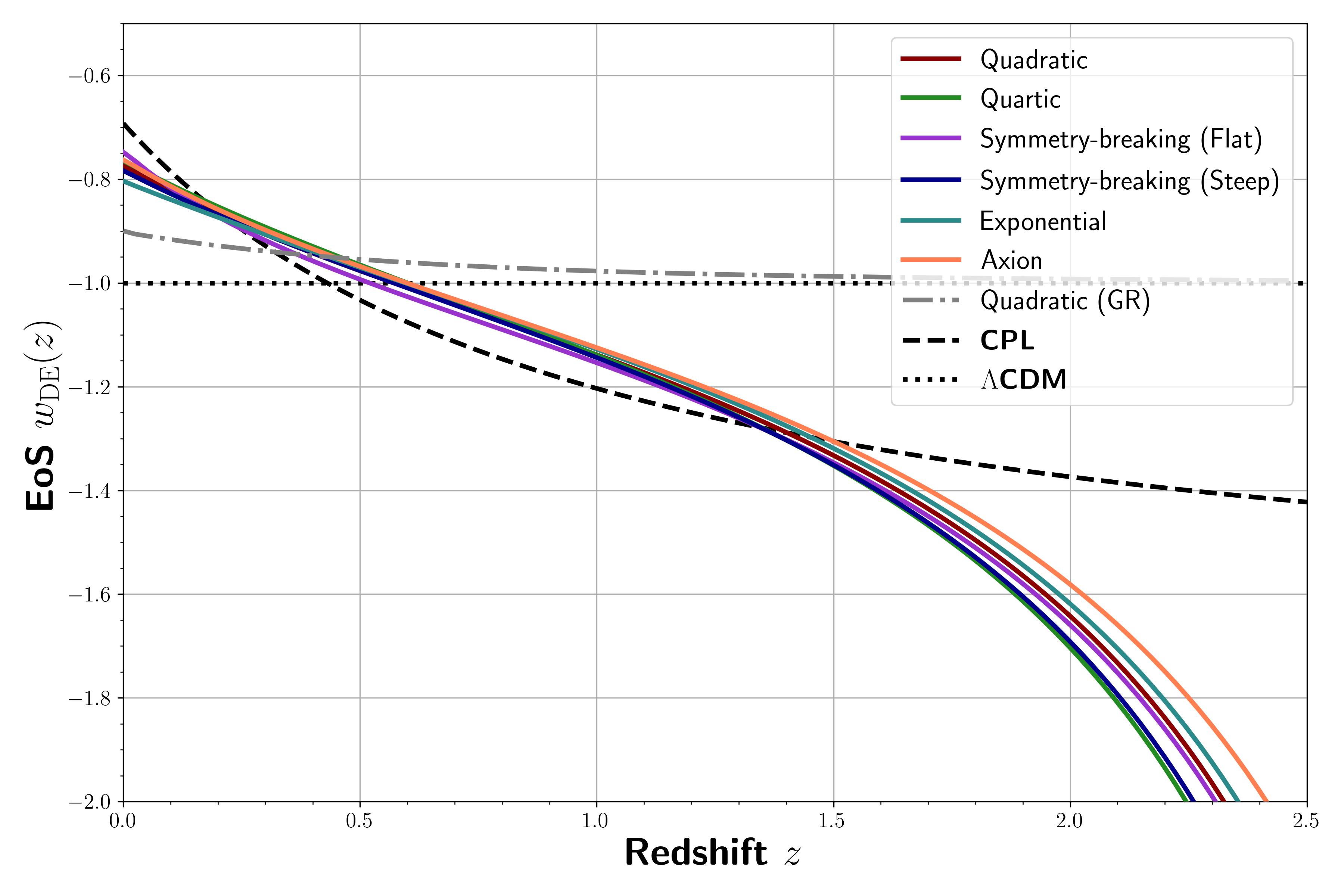}
\caption{Best-fit curves of the effective EoS of DE,  $w_{\rm DE}(z)$, corresponding to the medians of the MCMC chains are shown in  coloured solid curves, for various potentials considered in this work\,: quadratic potential (red),  quartic potential (green), steep wing (blue) and flat wing (purple)  of the symmetry-breaking potential, the exponential potential (cyan) and the Axion potential (orange). The corresponding curve for the CPL parametrisation  is shown as a  dashed black curve.  EoS for the quadratic potential in GR is shown in dot-dashed gray curve. Similarly, the constant EoS of DE in $\Lambda$CDM, namely, $w_{\rm DE} = -1$, is shown as a dotted black (horizontal) line.}
\label{fig:bestfits_EoS}
\end{center}
\end{figure}

 In fact, the median $w_{\rm DE}(z)$ curves of all our  braneworld  DE models display phantom behaviour for redshifts $z \gtrsim 0.5$, before crossing the phantom divide line at lower redshifts to exhibit $w_{\rm DE}(z) > -1$ at the present epoch,  as can be seen in Fig.~\ref{fig:bestfits_EoS}. It is also interesting to note that the low-redshift behaviour of the EoS of DE in all our models differs from that of the CPL parametrisation   in a similar way,  which is discussed in more detail in Sec.~\ref{sec:compare}.

\begin{figure}[htp]
  \vspace{-0.2in}
\begin{center}
\includegraphics[width=\textwidth]{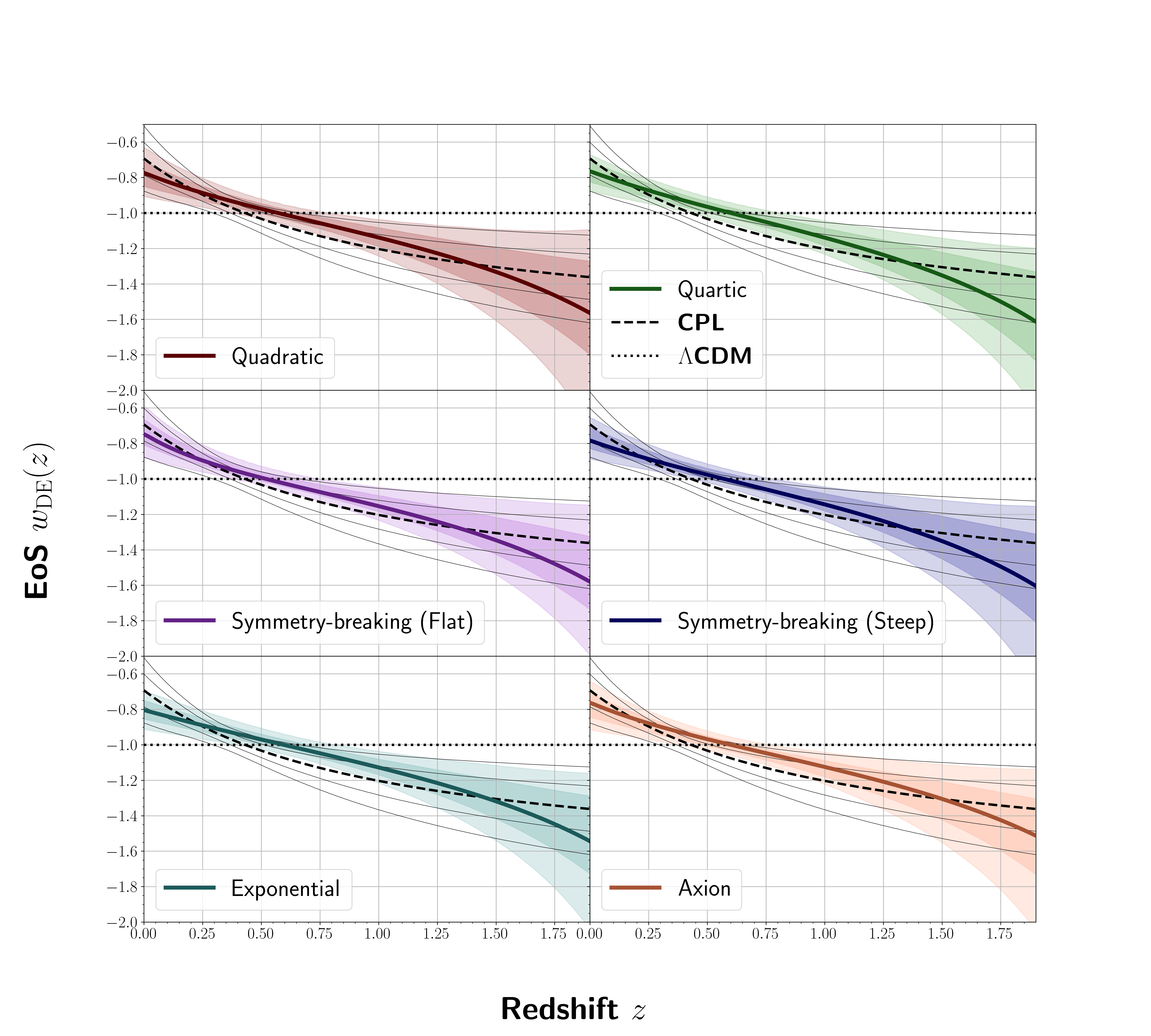}
\caption{1$\sigma$ and 2$\sigma$ contours for the effective EoS $w_{\rm DE}(z)$ of DE for various potentials considered in this work are shown, along with the  CPL contours. The medians of the samples from each model are shown in coloured solid curves, while the CPL EoS is shown as a dashed black curve, with finer solid black contours. The  $\Lambda$CDM expectation of a constant DE EoS at $w_{\rm DE}(z)=-1$ is shown with a dotted horizontal black line.}
\label{fig:mcmc_EoS}
\end{center}
 \vspace{-0.2in}
\end{figure}

\begin{figure}[htp]
 \vspace{-0.1in}
\begin{center}
\includegraphics[width=1.0\textwidth]{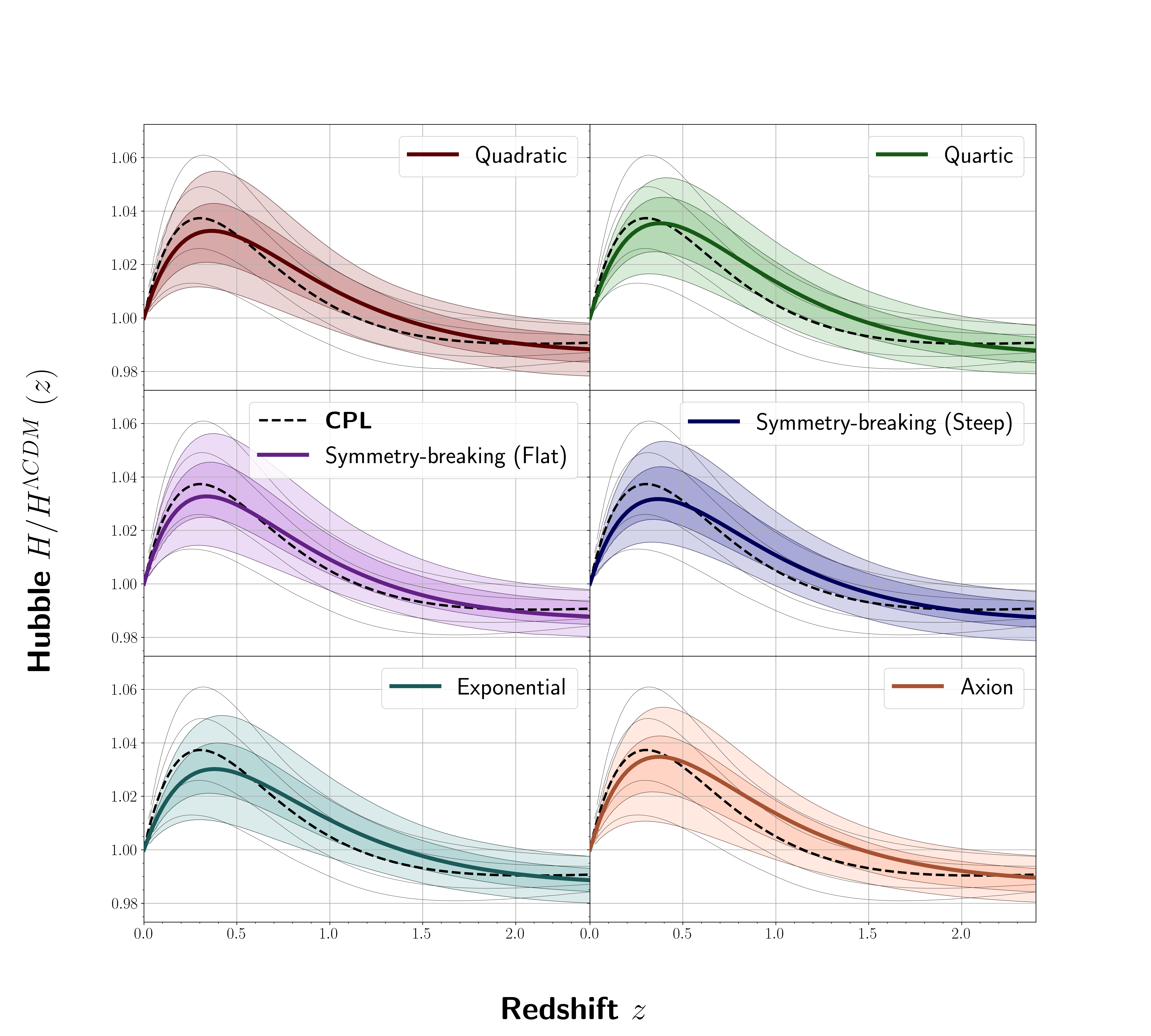}
\caption{$1\sigma$ and $2\sigma$ contours for the ratio of the Hubble factor to its $\Lambda$CDM counterpart are shown for each potential considered in this work. The medians of the samples from each model are shown in coloured solid curves. The corresponding median and contours for the CPL parametrisation  are shown as dashed black curve and fine solid black curves, respectively.}
\label{fig:mcmc_H}
\end{center}
\vspace{-0.2in}
\end{figure}

In figures \ref{fig:mcmc_EoS} and \ref{fig:mcmc_H} we show the median curves of $w_{\rm DE}(z)$ and $H(z)/H^{\Lambda \rm CDM}(z)$, for each potential considered in this work, along with the $1\sigma$ and $2\sigma$ contours calculated from the MCMC samples. For comparison, we plot also the corresponding curve and contours for the CPL parametrisation, demonstrating that the general dynamics of dark energy in the braneworld displays a  behaviour similar to that of CPL parametrisation,\footnote{ However, see Sec.~\ref{sec:compare} for a discussion on comparison between different  potentials in the braneworld scenario considered in this work, as well as contrasting them with the CPL parametrisation.} when constrained using the latest cosmological datasets.

 These results suggest that the physical models investigated here serve as viable explanations for the observed data, at least at the level of the CPL parametrisation  and within the precision achieved using the current datasets. More importantly, the phantom divide crossing behaviour, which is suggested by the CPL parametrisation   (as well as non-parametric methods, see Ref.~\cite{DESI:2025fii,DESI:2025wyn}) is  explained physically, within the framework of thawing scalar field potentials in the braneworld scenario. 

\begin{figure}[htp]
 \vspace{-0.3in}
\begin{center}
\vspace{-0.3in}
\includegraphics[width=1.0\textwidth]{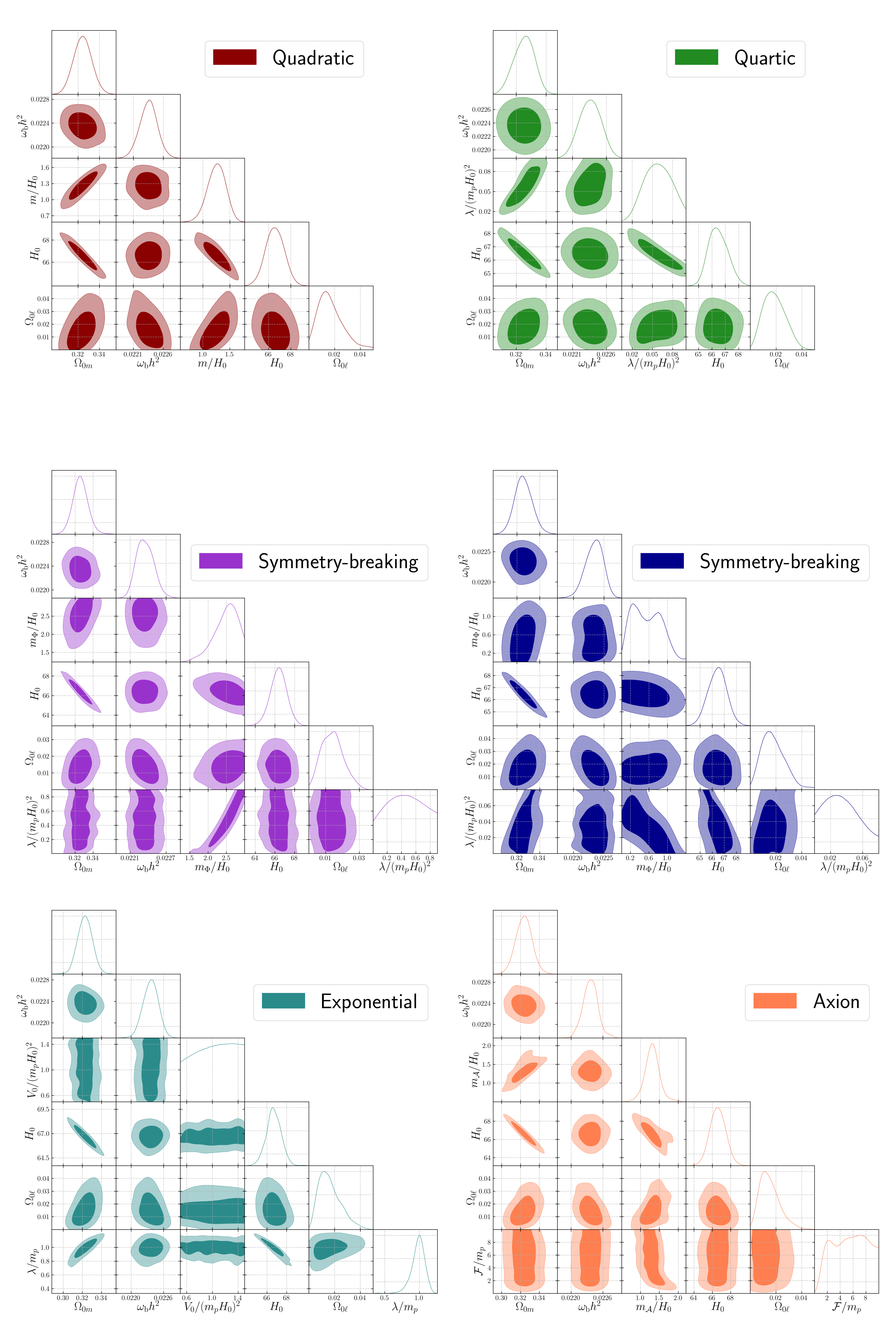}
\vspace{-0.3in}
\caption{A triangle plot displaying the posterior distribution of observables listed in table~\ref{table:MCMC_CPL} for various potentials considered in this work, namely, the quadratic potential in the {\bf top-left panel}, the quartic  potential in the {\bf top-right panel}, the symmetry breaking potential with flat wing in the {\bf middle-left  panel} and with steep wing in the {\bf middle-right  panel}, and finally the exponential potential in the {\bf bottom-left panel} and the axion potential in the {\bf bottom-right panel}.}
\label{fig:Triangle}
\end{center}
\end{figure}

 It is worth noting that in GR ($\Omega_{0\ell} = 0$), the DE EoS for these scalar field potentials does not exhibit phantom-divide crossing, as explicitly shown for the quadratic potential in the dot-dashed gray curve of Fig.~\ref{fig:bestfits_EoS}. In GR, the quadratic potential yields a significantly larger value of $\chi^2$ relative to CPL, with $\Delta\chi^2 \equiv \chi^2 - \chi^2_{\rm CPL} \simeq 7.99$, indicating a poor fit to the datasets. We find that this behaviour persists for all other potentials considered in this work. Moreover, $\Omega_{0\ell} = 0$ lies well outside the $1$--$2\,\sigma$ bounds on the best-fit values of $\Omega_{0\ell}$, as shown in Fig.~\ref{fig:Triangle} (the $1\,\sigma$ bounds on $\Omega_{0\ell}$ are explicitly listed in Table~\ref{table:MCMC_CPL}). Therefore, we conclude that scalar field DE in pure GR provides a poor fit to the data, whereas the braneworld scalar field models with $\Omega_{0\ell}\neq 0$ deliver a statistically significant improvement, offering a physically compelling realization of dynamical dark energy in light of the latest observations.

 Before moving forward, it is worth highlighting that for potentials with a single parameter, such as the quadratic and quartic potentials (along with the brane parameter $\Omega_{0\ell}$), all the parameters quoted in table~\ref{table:MCMC_CPL} are well constrained by the data. However, for two-parameter potentials, such as the symmetry-breaking, exponential and Axion potentials, the second parameter for each potential (marked in bold numbers) is not very well constrained by the data. Hence, there appears  to be a degeneracy that the considered datasets are unable to break, as can be seen more clearly from the bottom-most row of each panel of the triangle plots in Fig.~\ref{fig:Triangle}. The best-fit values of these parameters, as quoted in table~\ref{table:MCMC_CPL}, should therefore be taken not-too-literally. In fact, a range of values of the second parameter of each of these three potentials yields $\chi^2$ values that are comparable to each other.

\subsection{Comparison of results for different potentials}
\label{sec:compare}

 In the preceding analysis, the quadratic~(\ref{sec:PhBrane_Quad})~and~quartic~(\ref{sec:PhBrane_Quartic}) potentials in the braneworld scenario (with $\Omega_{0\ell}$) belong to two-parameter class of models, similar to the CPL parametrisation , while the symmetry-breaking~(\ref{sec:DE_PhBrane_SB_R}), exponential~(\ref{eq:pot_Exponential})~and~axion~(\ref{eq:pot_Axion}) potentials are three-parameter models. By comparing our results in Figs.~\ref{fig:DE_PhBrane_Quad_Hubble}--\ref{fig:DE_PhBrane_Quad_EoS} for the quadratic potential with Figs.~\ref{fig:DE_PhBrane_Quartic_Hubble}--\ref{fig:DE_PhBrane_Quartic_EoS} for the quartic potential, we observe that both potentials perform almost equally well\,---\,based purely on visual inspection (this can also be inferred more quantitatively from the values of $\Delta\chi^2$ displayed in table~\ref{table:MCMC_CPL}). This is somewhat surprising, as one might expect that the quartic potential would perform better due to its ability to produce a sharper phantom crossing.  Similarly, the results for the steep wing of the symmetry breaking potential in Figs.~\ref{fig:DE_PhBrane_SB_R_Hubble}--\ref{fig:DE_PhBrane_SB_R_EoS} are similar to those of the flat wing of the  symmetry breaking potential in Figs.~\ref{fig:DE_PhBrane_SB_L_Hubble}--\ref{fig:DE_PhBrane_SB_L_EoS}, as well as to those of the Axion potential in Figs.~\ref{fig:DE_PhBrane_Axion_Hubble}--\ref{fig:DE_PhBrane_Axion_EoS}.

\begin{figure}[H]
\begin{center}
\includegraphics[width=0.495\textwidth]{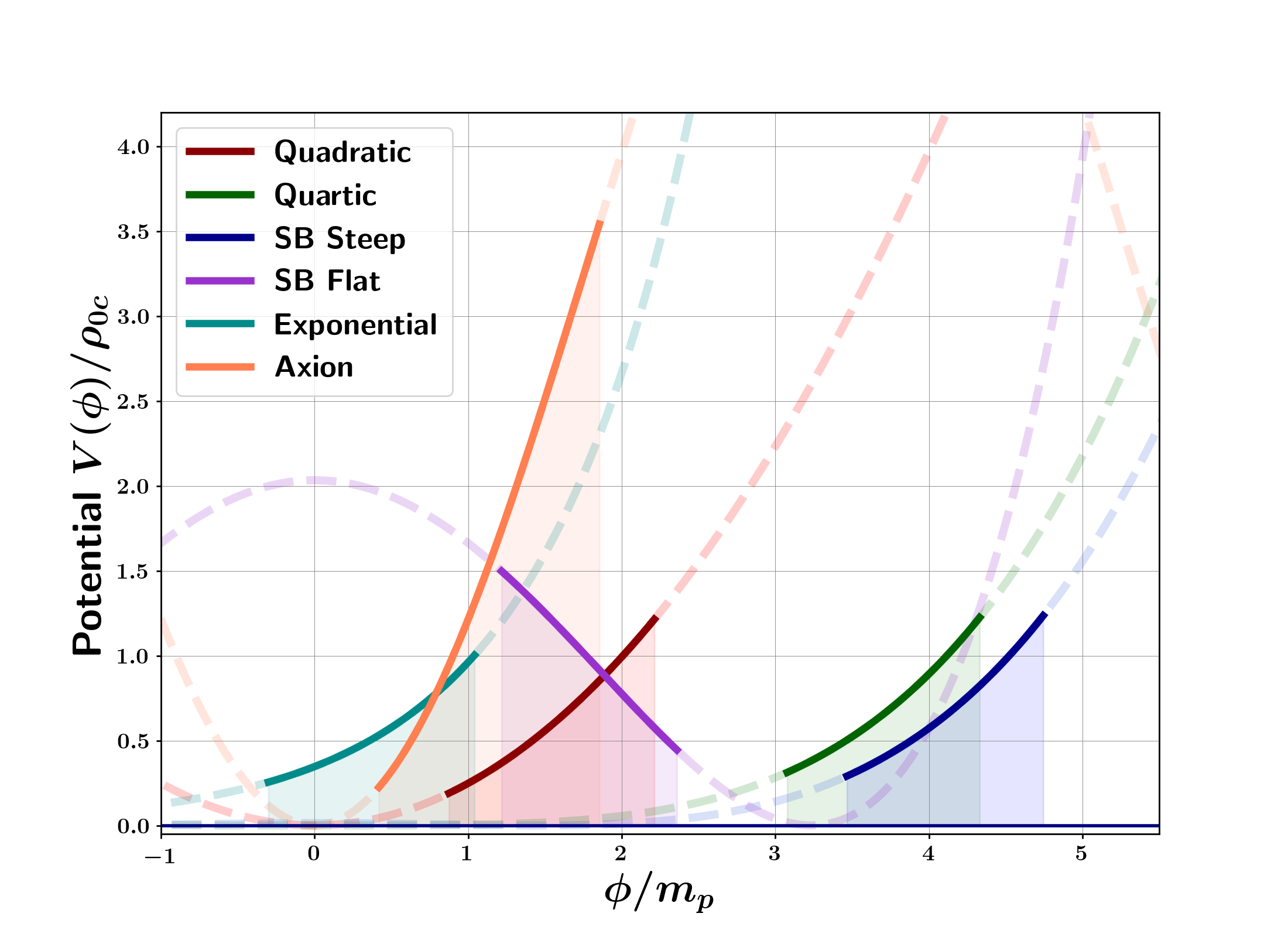} 
\includegraphics[width=0.495\textwidth]{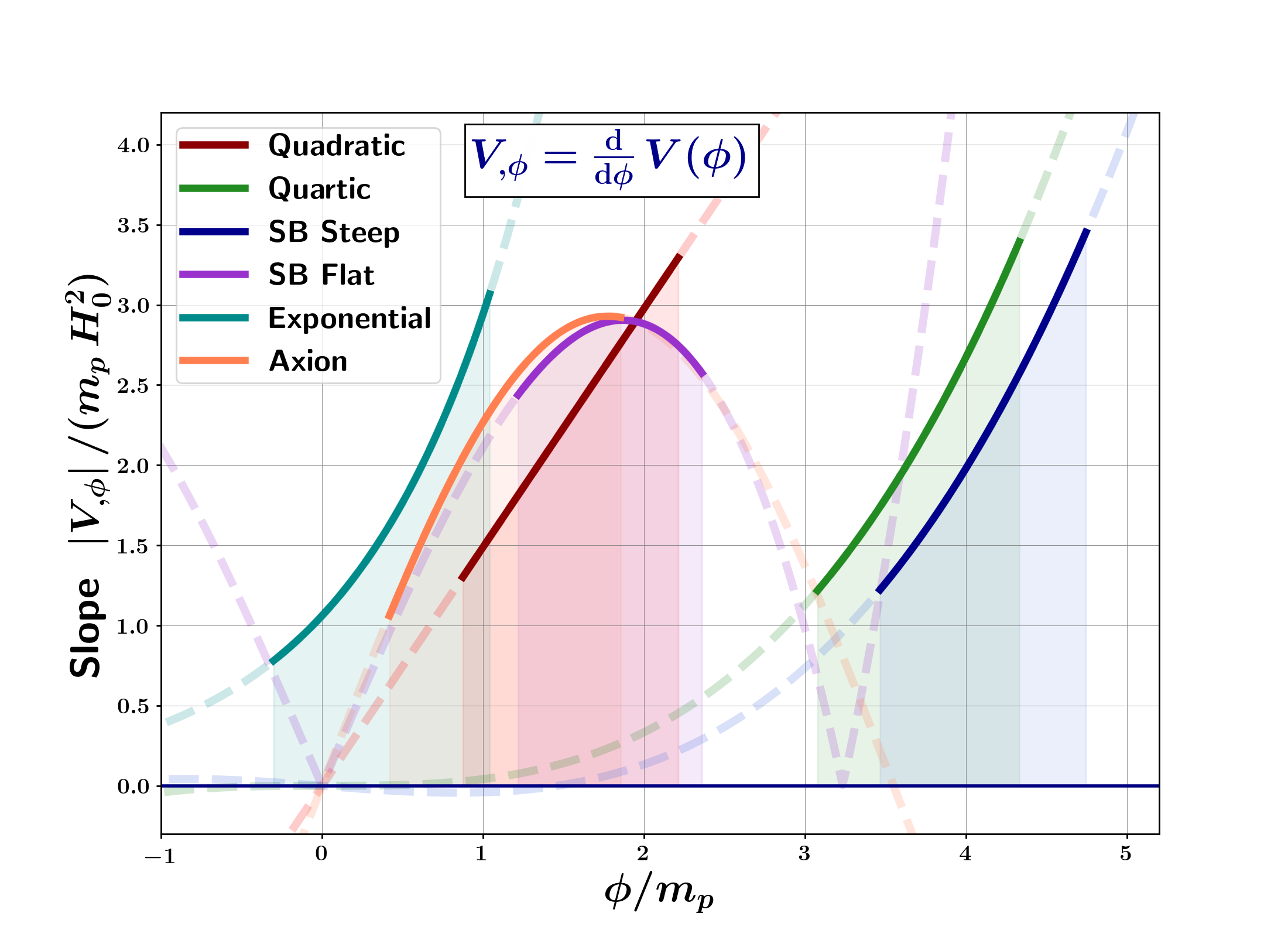}
\caption{{\bf Left panel} shows the quadratic potential (red),  quartic potential (green), steep wing (blue) and flat wing (purple)  of the symmetry-breaking potential, the exponential potential (teal) and the Axion potential (orange); while the {\bf right panel} shows the corresponding slopes of these potentials. The {\bf darker (solid) curves} with vertical shades correspond to the field ($\phi$) range explored during the full simulation time, \textit{i.e.\@} from $z=99$ to $z \lesssim -1$. The {\bf fainter (dashed) curves} correspond to plots of $V(\phi)$ and $V_{,\phi}$ for much larger range of $\phi$ values. We have plotted each potential and its derivative for parameters which correspond to $\Delta\chi^2$ values that are very close to the $\Delta\chi^2$ values quoted in table~\ref{table:MCMC_CPL} for the best-fit parameters.}
\label{fig:pot_compare}
\end{center}
\end{figure}

The reason for this similarity lies in the fact that the slopes of  these potentials\,---\,within the range of field values relevant to the system's full evolution (from 
$z=99$ to $z \lesssim -1$, as used in our simulations)\,---\,are of the same order of magnitude  for the chosen parameter values in the figures. This is illustrated clearly in Fig.~\ref{fig:pot_compare}, which directly compares these potentials.

 However, it is important to stress that for a given (effective) mass,  flat potentials, such as the Axion potential~(\ref{eq:pot_Axion}) and the flat wing of the symmetry-breaking potential~(\ref{eq:pot_SB}), are always less steep compared a quadratic potential~(\ref{eq:pot_Quad}) of the same mass. Since DESI DR2 prefers a relatively steeper change in the DE EoS around the phantom crossing, the flat potentials can explain the observed results for relatively higher values of the (effective) mass as compared to the quadratic potential, as observed in our simulations, also see table~\ref{table:MCMC_CPL}.

 There are a number of important points to highlight  here  in order to contrast between the DE dynamics for the CPL parametrisation  and the potentials considered in our braneworld scenario, such as,
\begin{enumerate}
\item At relatively higher redshifts (towards the right-hand side of  Fig.~\ref{fig:bestfits_EoS}), the EoS of DE for the CPL parametrisation  is different from our models. This is because the EoS in our models approach a pole in $w(z)$ as discussed in Appendix~\ref{app:Analytics}. Therefore,  our models exhibit  concave-shaped EoS curves, as opposed to the CPL curves which are necessarily convex, as can be inferred from Eq.~(\ref{eq:CPL}). 

Of course, at very high redshifts, $z\gg1$, the CPL EoS tends towards a constant, given by (from best-fit values in the table)
$$
w_{\rm {DE,\,CPL}} \to w_0 + w_a \simeq -1.74\,, \quad z \gg 1 \, ,
$$
and so do our EoS curves, which tend towards $w_{\rm DE} \to -0.5$ deep inside the  matter domination, because the deceleration parameter $q(z) \to 0.5$.
\item At lower redshifts, $z \ll 1$, Eq.~(\ref{eq:CPL}) demonstrates that the CPL EoS changes linearly with redshift displaying a constant slope, namely, 
$$
w_{\rm {DE,\,CPL}} = w_0 +w_a \times z \simeq -0.67 -1.07 \times z\,, \quad z \ll 1 \, .
$$ 
For the thawing potentials in our braneworld scenario, the behaviour is more complicated and depends on the slope of the potential used. However,  in our thawing DE models, since the potential is decreasing as we approach $z=0$, the scalar field is closer to its minimum and $\rho_\phi$ is much smaller than its frozen value at higher redshifts. Hence, the value of the EoS is primarily dictated by $\Omega_{0\ell}$, see Eq.~(13) in Ref.~\cite{Bag:2018jle}.  The best-fit values of $\Omega_{0\ell}$ (given DESI DR2) for all our models are therefore almost equal, and hence the EoS of DE in all these model behaves similarly around $z\simeq 0$. Small differences come from the fact that closer to $z=0$, $V(\phi)$ and $V_{,\phi}$ for these models are slightly different, as can be seen from Fig.~\ref{fig:pot_compare}. 

For a choice of model parameters sufficiently far away from the best-fit values quoted in table~\ref{table:MCMC_CPL}, the scalar field might remain frozen until the present epoch. In that case it will behave like a cosmological constant (on the brane) and the present EoS of DE will be phantom-like with $w_{\rm eff} < -1$. For such a choice of parameter space, the EoS of DE will not exhibit phantom-crossing by $z=0$.

\end{enumerate}

\section{Discussion}
\label{sec:discussion}

The recent DESI observations regarding the evolution of cosmic dark energy\,---\,particularly the indication of a phantom crossing\,---\,are both surprising and difficult to reconcile within the framework of standard single-component dynamical dark energy models. If interpreted literally, these results suggest a more complex underlying physics than previously anticipated.

To address this challenge and replicate the dark energy behaviour established by DESI, we explore a new cosmological model consisting of a scalar field propagating on  the ghost-free  braneworld with a single extra spatial dimension. Both scalar field and braneworld may arise quite naturally from a more fundamental superstring or higher-dimensional theoretical framework. In this framework, the scalar field serves as the only source of thawing dark/vacuum energy on the brane. Braneworld DE on its normal (ghost-free) branch is known to exhibit universal phantom-like features \cite{Sahni:2002dx,Bag_2021} which we utilise to model early-time evolution. At early times, an ultra-light scalar field remains frozen to its initial value resulting in an effective phantom-like EOS of dark energy with $w_{\rm DE} < -1$ at $z \geq 1$. As the universe expands, the dynamics of the scalar field becomes increasingly prominent, introducing quintessential characteristics at later times. This interplay naturally leads to a crossing of the phantom divide within our models so that the  EoS of DE exhibits $w_{\rm DE} > -1$ at $z \lesssim 1$\,.

 We find that scalar fields with a variety of thawing potentials are quite effective in realizing this behaviour. Notably, simple potentials\,---\,such as quadratic, quartic, symmetry-breaking, exponential and axion\,---\,yield excellent results, and offer a compelling quantitative fit to the DESI DR2 data when analysed using MCMC sampling. In fact, single-parameter potentials, such as the quadratic or quartic potential, are able to explain the observational data as well as the frequently employed CPL parametrisation . Our models therefore represent a physically well-motivated scenario for explaining the latest observational results from DESI. 

\medskip

It will be interesting to extend the present analysis, which has been carried out at the level of background cosmological dynamics, by investigating the evolution of linear perturbations in the braneworld scenario~\cite{Bag:2016tvc} in the presence of the scalar field. Such an analysis would provide a more comprehensive understanding of the model’s viability, particularly in the context of structure formation and the growth history of cosmic inhomogeneities. Previous studies have shown that the growth of perturbations in braneworld models can exhibit good agreement with large-scale structure observations~\cite{Viznyuk:2018eiz,Bag:2018jle}.  However, it has been noted in Ref.~\cite{Lombriser:2009xg}  that the ISW and galaxy cross-correlation can become negative in the phantom braneworld scenario in presence of a brane tension, which can be used as a key signature of the braneworld scenario. In this context, we note that our typical best-fit values of $\Omega_{0\ell} \simeq 0.013$ lie within the constraint $\Omega_{0\ell} \lesssim 0.02$ derived in Ref.~\cite{Lombriser:2009xg} using the parametrized post-Friedmann framework. Nonetheless, a rigorous and updated treatment that incorporates the most recent observational datasets\,---\,including those from DESI DR2, Type Ia supernovae, and CMB\,---\,would require a separate study which we intend to undertake  in a future work. In the meantime, the next DESI (DR3) data release is anticipated to deliver even tighter constraints on the evolution of dark energy.

\section*{Acknowledgements}

 We thank  Edmund Copeland for useful comments on the manuscript. SSM thanks Antonio Padilla for helpful discussions on the axion potential, and for comments on the braneworld model. We thank Rikpratik Sengupta for discussions during the early stages of this project.  SSM is  supported by a STFC Consolidated Grant [No. ST/T000732/1] as a Research Fellow at the University of Nottingham, UK\@. VS thanks the Anusandhan National Research Foundation (ANRF), India,
for the National Science Chair Professorship which provided partial funding for this work. YS is supported by the National Academy of Sciences of Ukraine under project 0121U109612 and by a grant from Simons Foundation International SFI-PD-Ukraine-00014573, PI~LB.  SSM thanks IUCAA for the hospitality.

 \medskip

For the purpose of open access, the authors have applied a CC BY public copyright license to any Author Accepted Manuscript version arising.

\section*{Appendices}
\appendix
\section{Analytical estimate of the pole location}
\label{app:Analytics}
\begin{figure}[H]
\begin{center}
\vspace{-0.1in}
\includegraphics[width=0.495\textwidth]{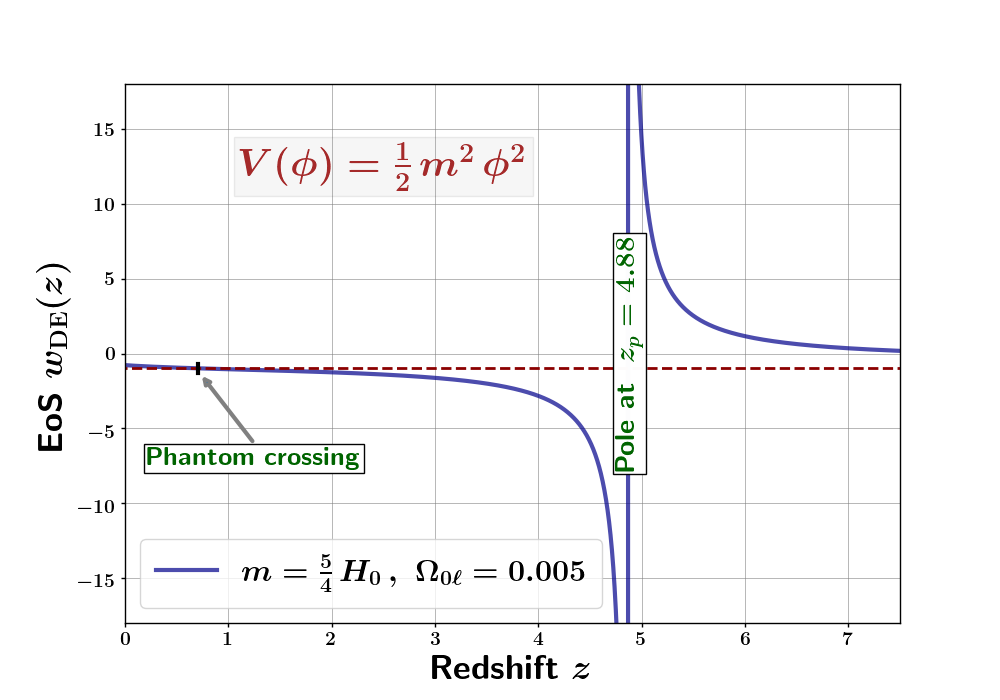} 
\includegraphics[width=0.495\textwidth]{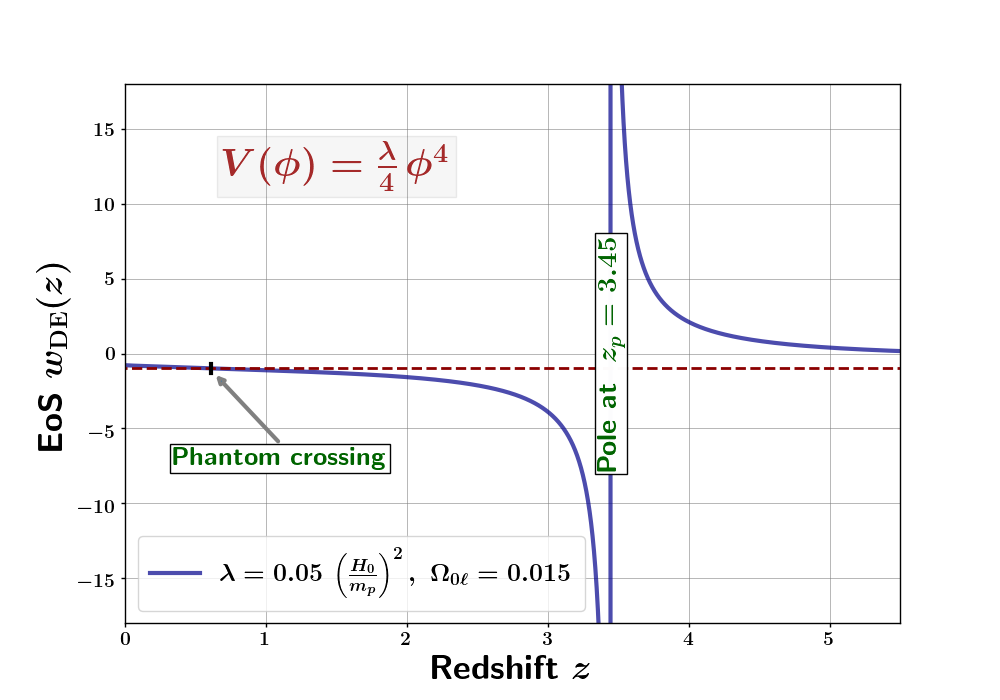}
\vspace{-0.2in}
\caption{Figure displays the redshift location of the pole in the effective EoS of DE on the phantom brane at a relatively higher redshift, for the quadratic potential~(\ref{eq:pot_Quad}) in the {\bf left panel}  and for the quartic potential~(\ref{eq:pot_Quartic}) in the {\bf right panel}.} 
\label{fig:EoS_pole}
\end{center}
\vspace{-0.1in}
\end{figure}

It is well-known that the effective EoS of DE on the phantom brane exhibits a pole~\cite{Sahni:2002dx} at a redshift where $\Omega_m(z) =1$, as can be inferred from Eq.~(\ref{eq:Brane_wDE}). Location of the pole depends on $\Omega_{0\ell}$, and for relatively small values of $\Omega_{0\ell}$, the pole is usually located at a higher redshift.   For the choice of parameters appearing in the figures of Sec.~\ref{sec:Dynamics_Models_Brane}, the corresponding pole is located at $z_p > 3$, which is outside the observational range of the DESI mission. Fig.~\ref{fig:EoS_pole} highlights the location of the pole for the quadratic (left panel) and quartic (right panel) potentials. In this appendix, we derive an analytical expression  for determining the (approximate) location of the pole.

We can determine the redshift location of the pole as follows. Eq.~(\ref{eq:h_PhBrane}) can be written as
\beq
1 + \sqrt{\Omega_{\ell}} = \sqrt{\Omega_m + \Omega_\phi + \Omega_{\ell}} \, ,
 \label{eq:Omega_1}
\eeq
where
\beq
\Omega_m = \f{\Omega_{0m}}{h^2} \, , \qquad \Omega_\phi = \f{\rho_\phi}{3m_p^2H_0^2}\, \f{1}{h^2} \, , \qquad \Omega_{\ell} = \f{\Omega_{0\ell}}{h^2} \, .
\label{eq:Omegas_def}
\eeq
Upon squaring both sides of Eq.~(\ref{eq:Omega_1}), we obtain
\beq
 1 =  \Omega_m + \Omega_\phi -  2\,\sqrt{\Omega_{\ell}} \, .
\eeq
Interpreting the above equation in the framework of GR as a Friedmann equation of the form 
\beq
1 = \Omega_m + \Omega_{\rm DE} \, ,
\eeq
yields
\beq
\Omega_{\rm DE} = \Omega_\phi - 2\,\sqrt{\Omega_{\ell}}  \, .
\label{eq:Omega_DE}
\eeq

The location $z_p$ of the pole in the effective EoS of DE can be inferred from Eq.~(\ref{eq:Brane_wDE}), by considering 
$$\Omega_m = 1 ~ \Rightarrow ~ \Omega_{\rm DE} = 0 \, ,$$
which, using Eq.~(\ref{eq:Omega_DE}), leads to
\beq
\Omega_\phi = 2 \, \sqrt{\Omega_{\ell}} \, ,
\eeq
or, equivalently, 
$$ h  = \f{1}{2} \, \l( \f{\rho_\phi}{3m_p^2H_0^2} \r) \, \f{1}{\sqrt{\Omega_{0\ell}}} \, .$$
Since, at the location of the pole, $\Omega_{\rm DE} = 0$, and $\Omega_m=1$, we have $h^2 = \Omega_{0m}\,(1+z_p)^3$, which results in
 \beq
\l(1+z_p\r)^{3/2} = \f{1}{2} \l( \f{\rho_\phi}{3m_p^2H_0^2} \r) \l( \f{1}{\sqrt{\Omega_{0\ell} \,\Omega_{0m}}} \r) \, ,
\label{eq:pole_z_gen}
 \eeq
 leading to the final expression for the redshift location of the pole\,:
 \beq
z_p =  \l[ \f{1}{4} \l(\f{\rho_\phi}{3m_p^2H_0^2}\r)^2 \f{1}{\Omega_{0\ell} \,\Omega_{0m}} \r]^{1/3} -1 \, .
\label{eq:pole_z_gen1}
 \eeq 
If we constrain ourselves to the specific range of parameter space of the scalar field potential for which $\phi$ is still frozen at the epoch when the pole appears, \textit{i.e.\@} $\phi(z_p) \simeq \phi_i$, then we can approximate $\rho_\phi$ as $\rho_\phi(z_p) = \rho_{\phi,i}$. In this case, the redshift location of the pole can be approximated as
 \beq
z_p \simeq  \l[ \f{1}{4} \, \f{\Omega_{\phi,i}^2}{\Omega_{0\ell} \,\Omega_{0m}} \r]^{1/3} -1 \, .
\label{eq:pole_z_initial}
 \eeq 
 Moreover, since $\rho_\phi$ will eventually start decreasing with time, we can write $\rho_{0\phi} = \xi \, \rho_{\phi, i}$ where $\xi \leq 1$ is a fraction. In terms of $\xi$, Eq.~(\ref{eq:pole_z_initial}) becomes
  \beq
z_p \simeq  \l[ \f{1}{4\,\xi^2} \, \f{\Omega_{0\phi}^2}{\Omega_{0\ell} \,\Omega_{0m}} \r]^{1/3} -1 \, ,
\label{eq:pole_z_initial_xi}
 \eeq 
which demonstrates that, for higher values of $\Omega_{0\ell}$ (with all other parameters being the same), the pole appears at lower redshifts. The pole disappears in the GR limit $\Omega_{0l}\to 0$. For the choice of parameters appearing in the figures of Sec.~\ref{sec:Dynamics_Models_Brane}, the above formula is a good approximation, and yields a pole at $z_p > 3$, which is outside the observational range of the DESI mission. Note also, that for $\xi = 1$, the above Eq.~(\ref{eq:pole_z_initial_xi}) agrees with Eq.~(15) of 
Ref.~\cite{Alam:2016wpf}.

 However, for the choice of parameters where  the scalar field has already started rolling down its potential by the redshift $z = z_p$ (or equivalently, when $\Omega_m = 1$), the approximation Eq.~(\ref{eq:pole_z_initial_xi}) becomes inaccurate, and one needs to determine  $\rho_\phi(z)$ numerically in order to obtain the value of $z_p$ more accurately, using Eq.~(\ref{eq:pole_z_gen1}). Nevertheless, Eq.~(\ref{eq:pole_z_gen1}) demonstrates that if $\rho_\phi$ is smaller than $\rho_{\phi,i}$, then the pole appears at a lower redshift than what one would infer from Eq.~(\ref{eq:pole_z_initial_xi}).

 Before concluding this section, it is worth mentioning that, depending upon the functional form of the potential, a second pole in the EoS of DE might appear in the future, if $\Omega_{\rm DE}(z_p) = 0$ at some $z_p < 0$. Or equivalently, from Eq.~(\ref{eq:de}),  if $\rho_\phi(z_p) = 3m_p^2H_0^2\sqrt{\Omega_{0\ell}}\, h(z_p)$ at some $z_p < 0$. Incidentally, this occurs for all the thawing potentials studied in this work. As mentioned before, two poles in the effective EoS of DE also appear in the coupled matter--quintessence model in GR studied in Ref.~\cite{Andriot:2025los}. However, since the second pole is located in the future in our scenario, it does not carry much significance from the observational perspective, and hence we do not discuss it further in this work.

\section{Evolution of other physical quantities in our models}
\label{app:PhBrane_others}
In Sec.~\ref{sec:Dynamics_Models_Brane}, we displayed plots for the Hubble parameter and the effective EoS of DE for various models considered in this work. In this appendix, we include plots for the evolution of the DE fraction~(\ref{eq:Brane_fDE}), deceleration parameter~(\ref{eq:Brane_q}) and $Om$ diagnostic~(\ref{eq:Om})  corresponding to each potential. 
\subsection{Quadratic potential}
\label{app:Quadratic}
\begin{figure}[H]
\vspace{-0.1in}
\begin{center}
\vspace{-0.1in}
\includegraphics[width=0.495\textwidth]{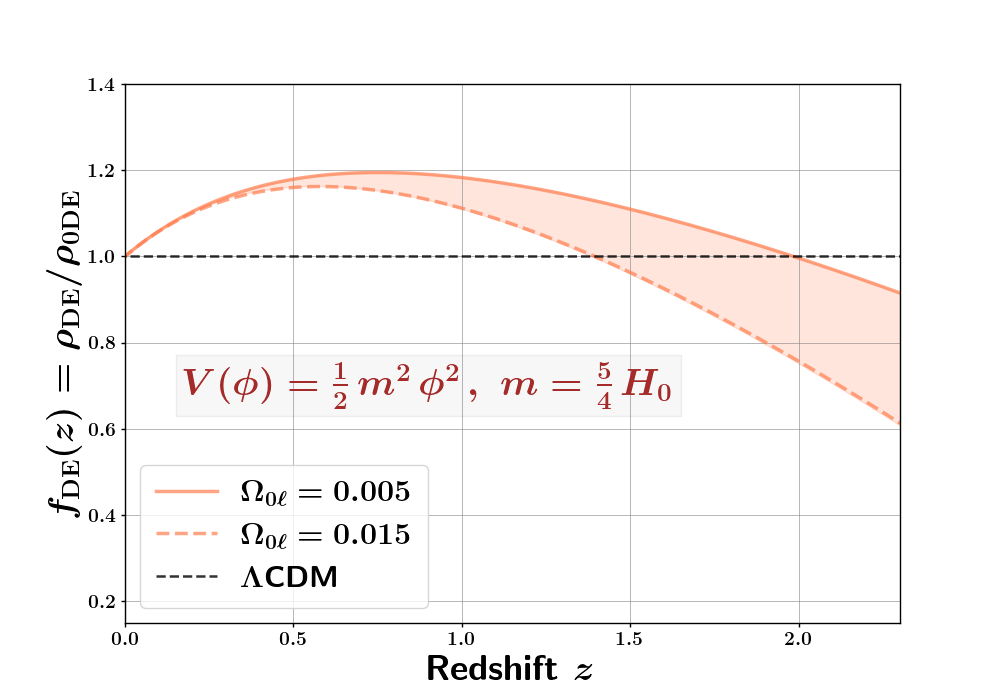}
\includegraphics[width=0.495\textwidth]{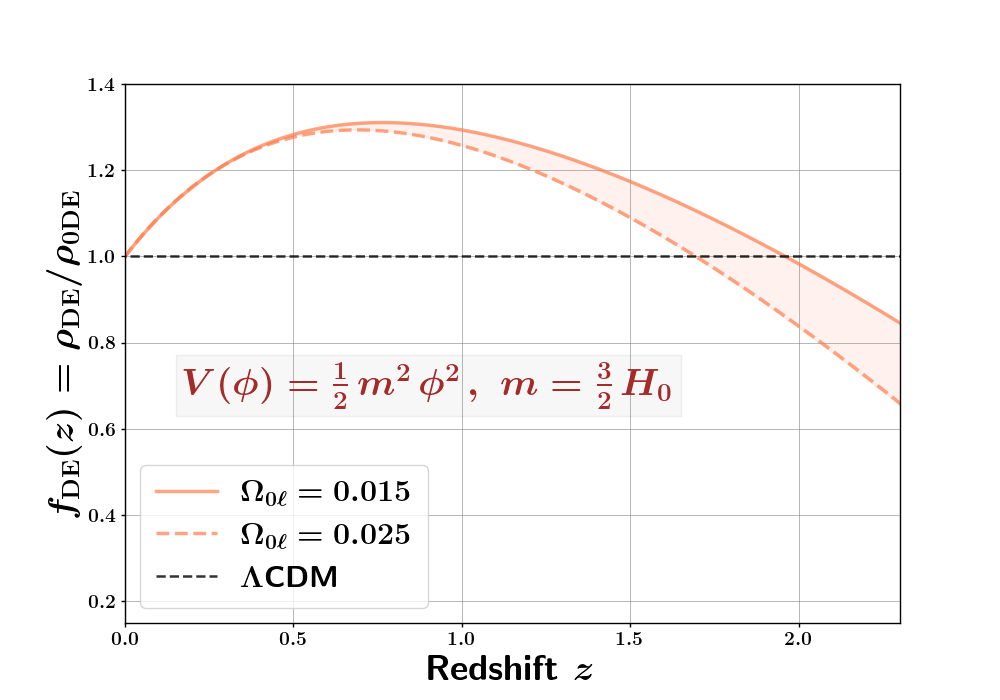}
\vspace{-0.2in}
\caption{The DE density relative to its present-epoch value  corresponding to the quadratic potential (\ref{eq:pot_Quad}) is shown for $m=\frac54 H_0$  ({\bf left panel}), and  for  $m=\frac32 H_0$ ({\bf right panel}).}
\label{fig:DE_PhBrane_Quad_Density}
\end{center}
\end{figure}
\begin{figure}[H]
\vspace{-0.1in}
\begin{center}
\vspace{-0.0in}
\includegraphics[width=0.495\textwidth]{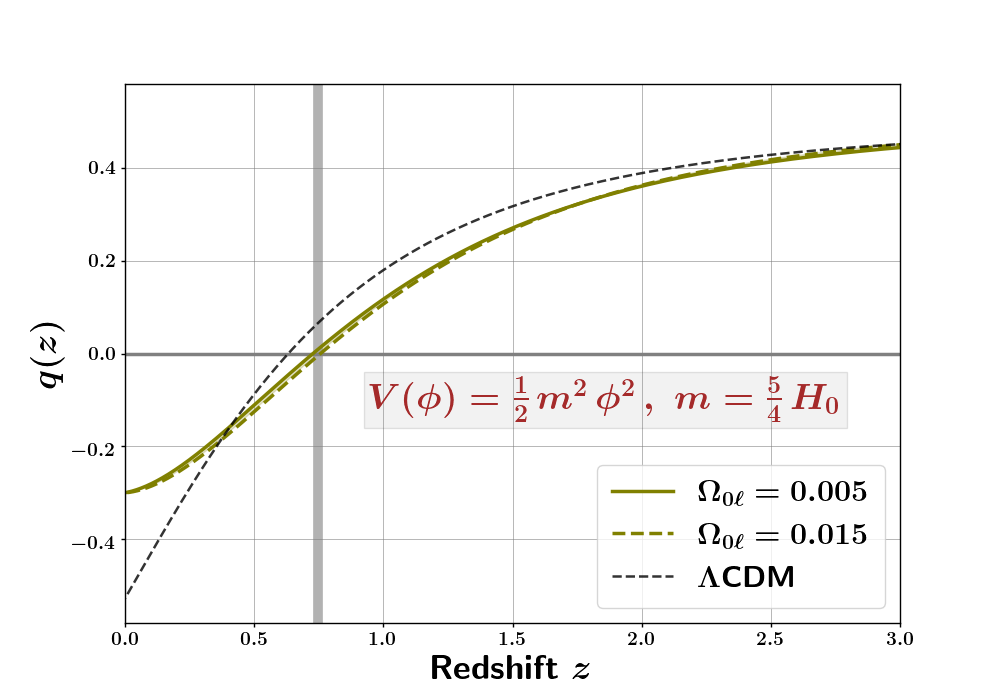}
\includegraphics[width=0.495\textwidth]{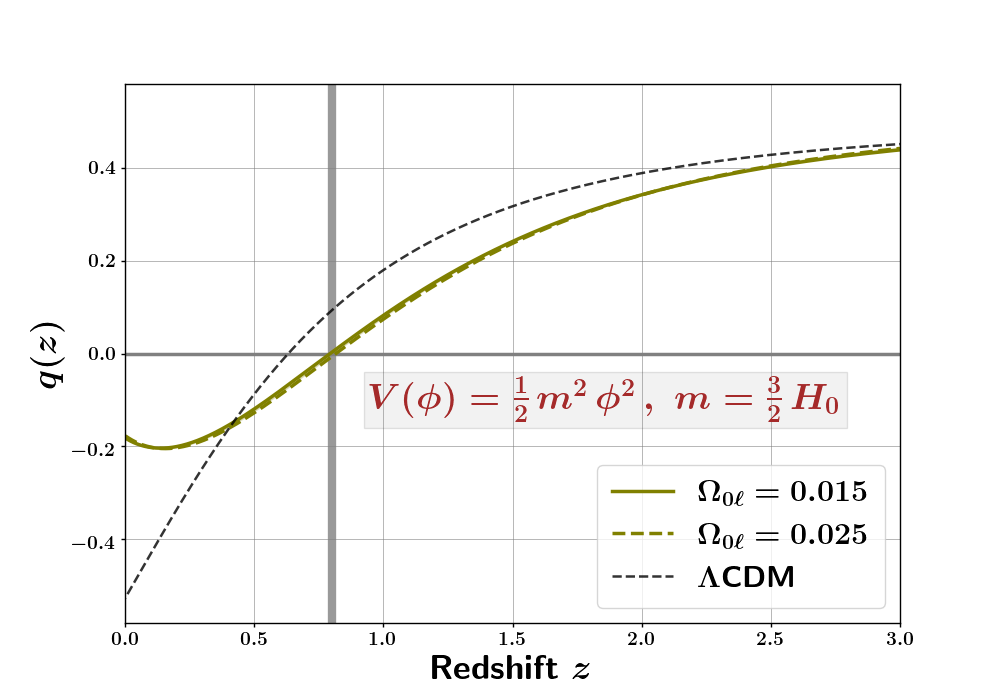}
\vspace{-0.2in}
\caption{The deceleration parameter corresponding to the quadratic potential (\ref{eq:pot_Quad}) is shown for $m = \frac54 H_0$ ({\bf left panel}), and  for  $m = \frac32 H_0$ ({\bf right panel}). }
\label{fig:DE_PhBrane_Quad_q}
\end{center}
\end{figure}

\begin{figure}[H]
\vspace{-0.0in}
\begin{center}
\vspace{-0.2in}
\includegraphics[width=0.495\textwidth]{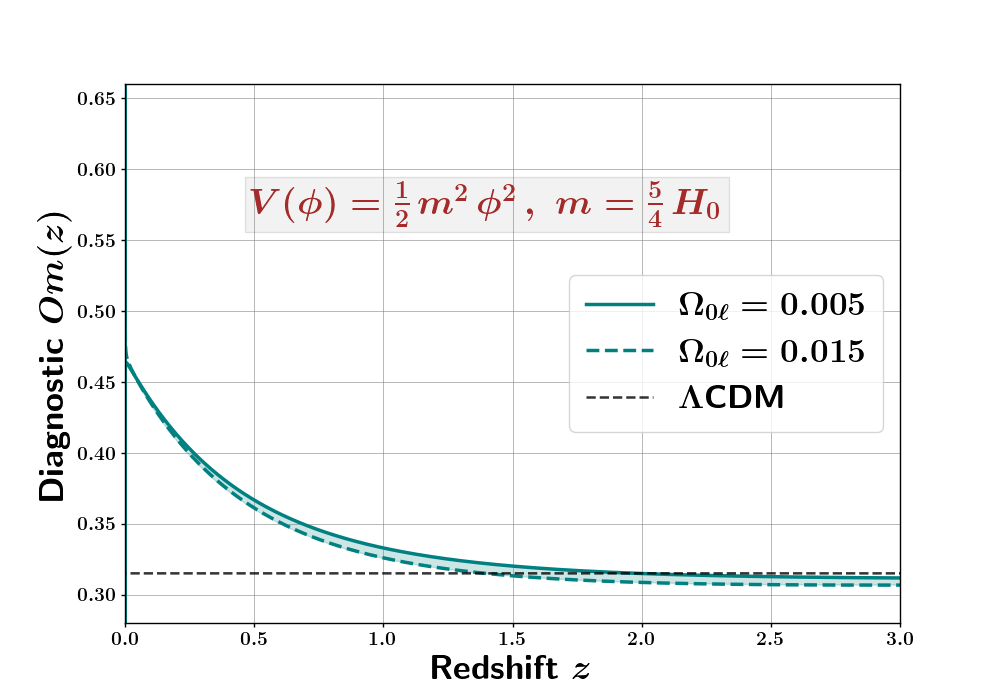}
\includegraphics[width=0.495\textwidth]{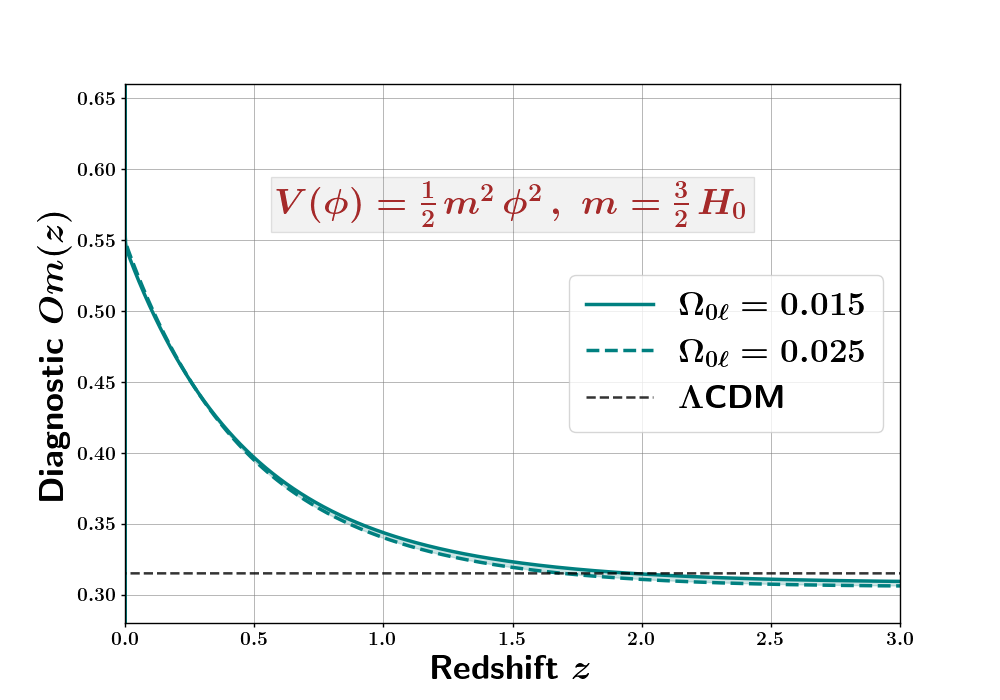}
\vspace{-0.2in}
\caption{The $Om$ diagnostic parameter corresponding to the quadratic potential (\ref{eq:pot_Quad}) is shown for $m = \frac54 H_0$ ({\bf left panel}), and  for  $m = \frac32 H_0$ ({\bf right panel}).}
\label{fig:DE_PhBrane_Quad_Om}
\end{center}
\vspace{-0.2in}
\end{figure}

\subsection{Quartic potential}
\label{app:Quartic}
\begin{figure}[H]
\vspace{-0.1in}
\begin{center}
\vspace{-0.0in}
\includegraphics[width=0.495\textwidth]{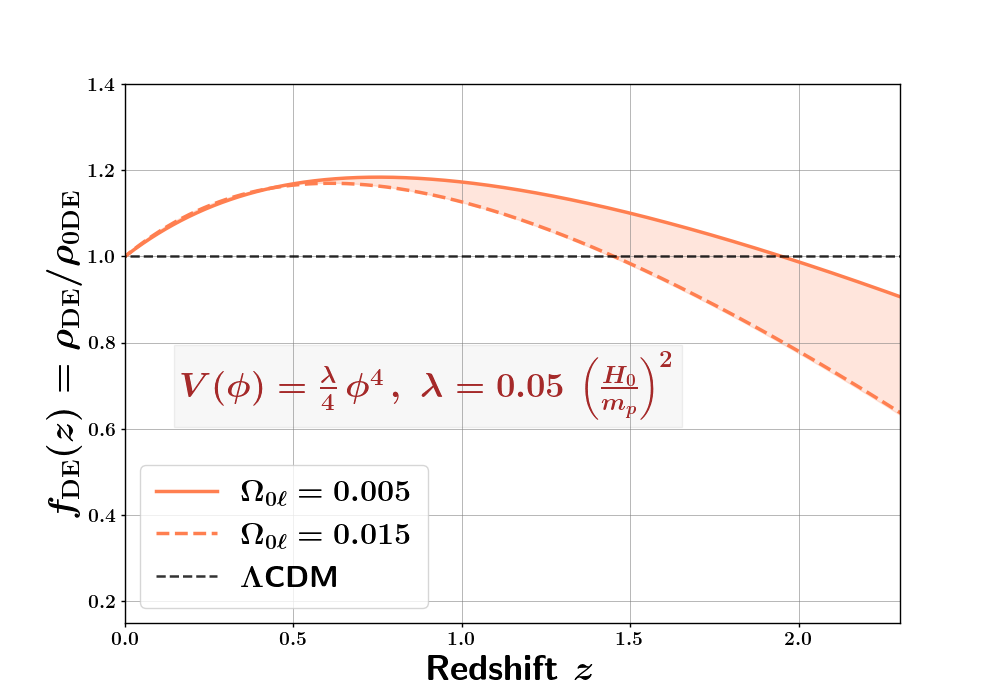}
\includegraphics[width=0.495\textwidth]{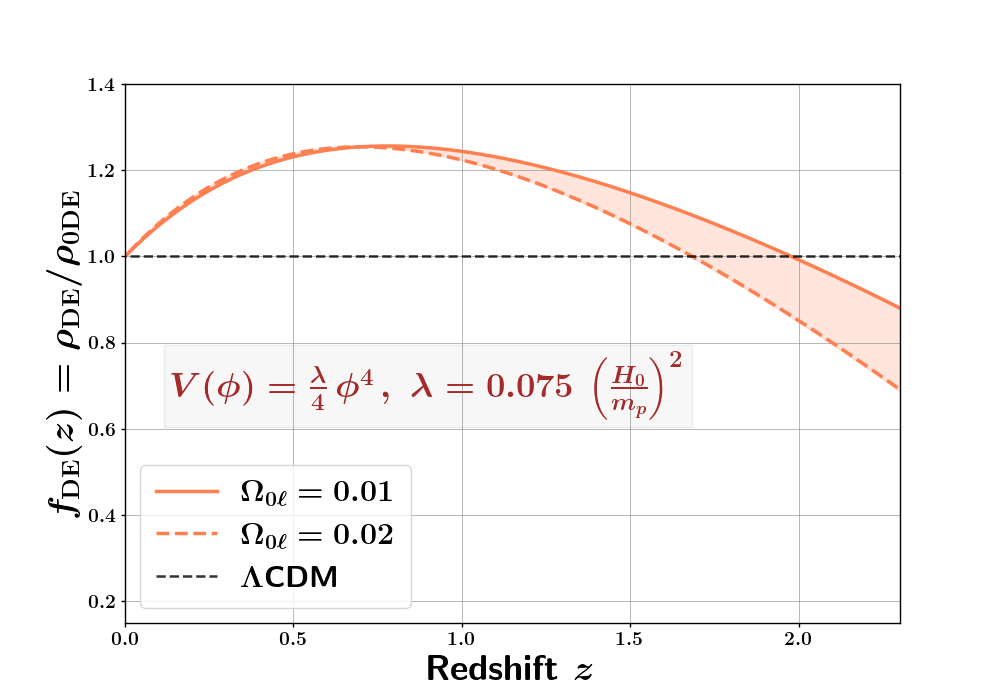}
\vspace{-0.2in}
\caption{The DE density relative to its present-epoch value for the quartic potential (\ref{eq:pot_Quartic}) are shown for $\lambda=0.05 \left( H_0/m_p \right)^2$ ({\bf left panel}), and  for  $\lambda = 0.075 \left( H_0/m_p \right)^2$ ({\bf right panel}).}
\label{fig:DE_PhBrane_Quartic_Density}
\end{center}
\end{figure}
\begin{figure}[H]
\vspace{-0.2in}
\begin{center}
\vspace{-0.2in}
\includegraphics[width=0.495\textwidth]{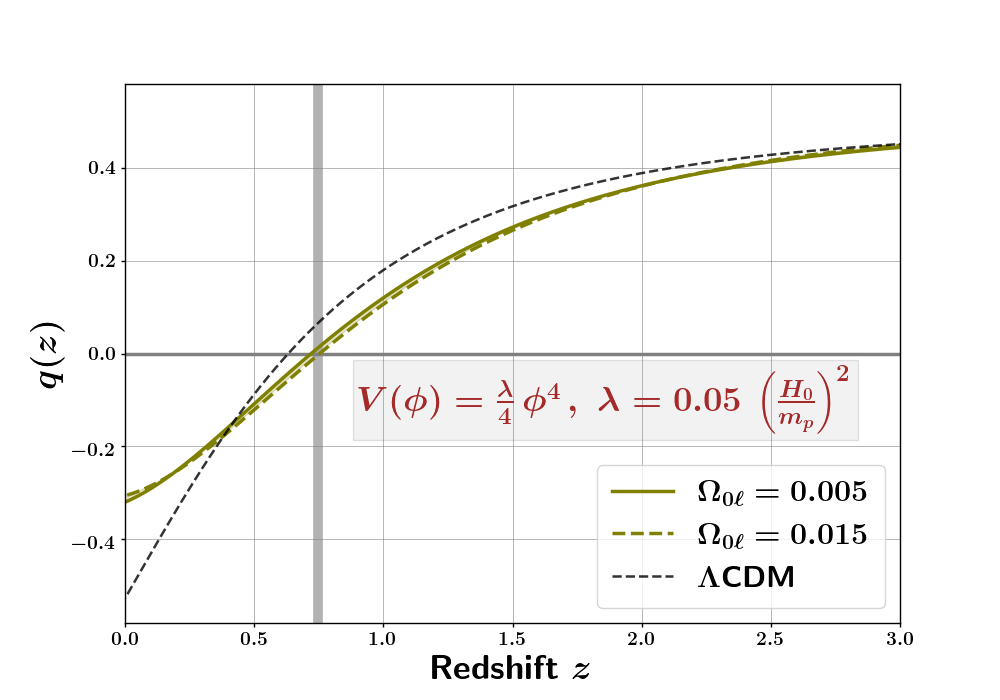}
\includegraphics[width=0.495\textwidth]{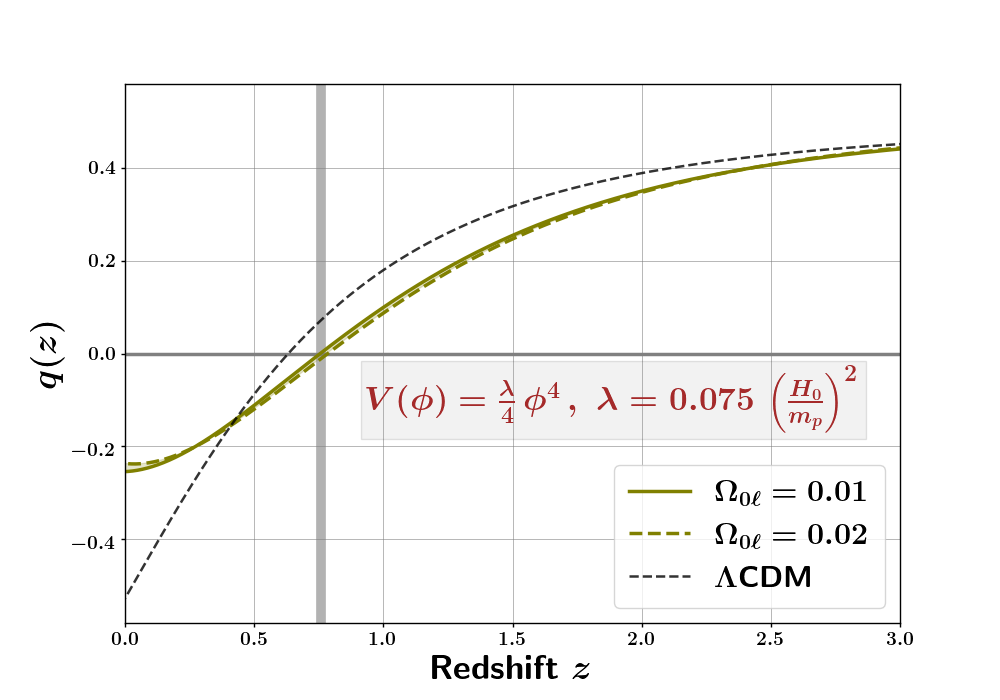}
\vspace{-0.2in}
\caption{The deceleration parameter  corresponding to the quartic potential (\ref{eq:pot_Quartic}) is shown for $\lambda=0.05 \left( H_0/m_p \right)^2$ ({\bf left panel}), and  for  $\lambda = 0.075 \left( H_0/m_p \right)^2$ ({\bf right panel}).}
\label{fig:DE_PhBrane_Quartic_q}
\end{center}
\end{figure}
\begin{figure}[H]
\vspace{-0.2in}
\begin{center}
\vspace{-0.2in}
\includegraphics[width=0.495\textwidth]{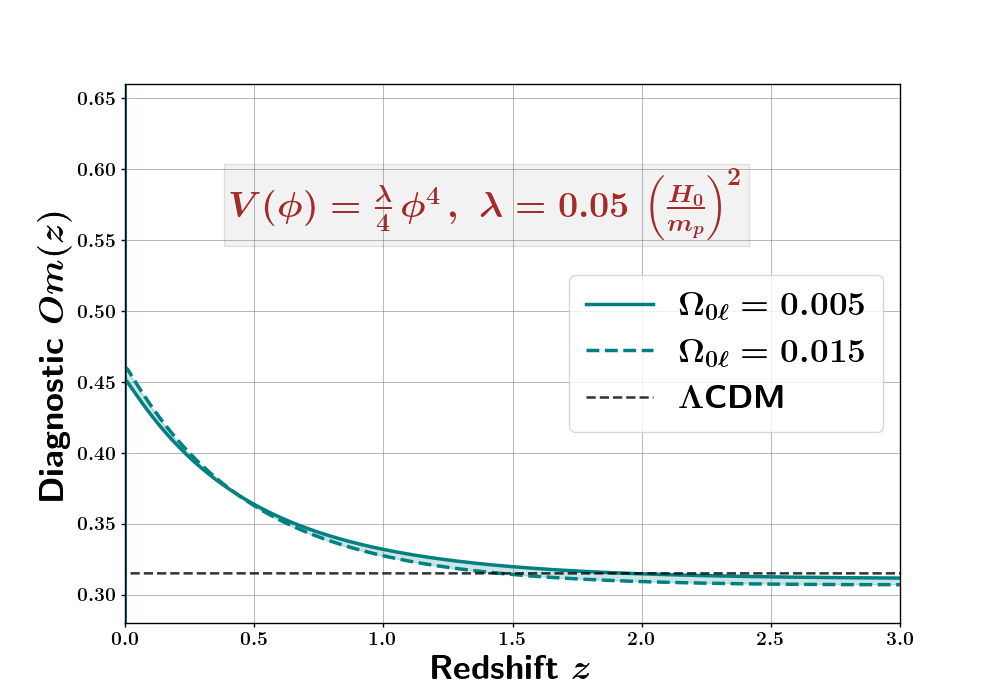}
\includegraphics[width=0.495\textwidth]{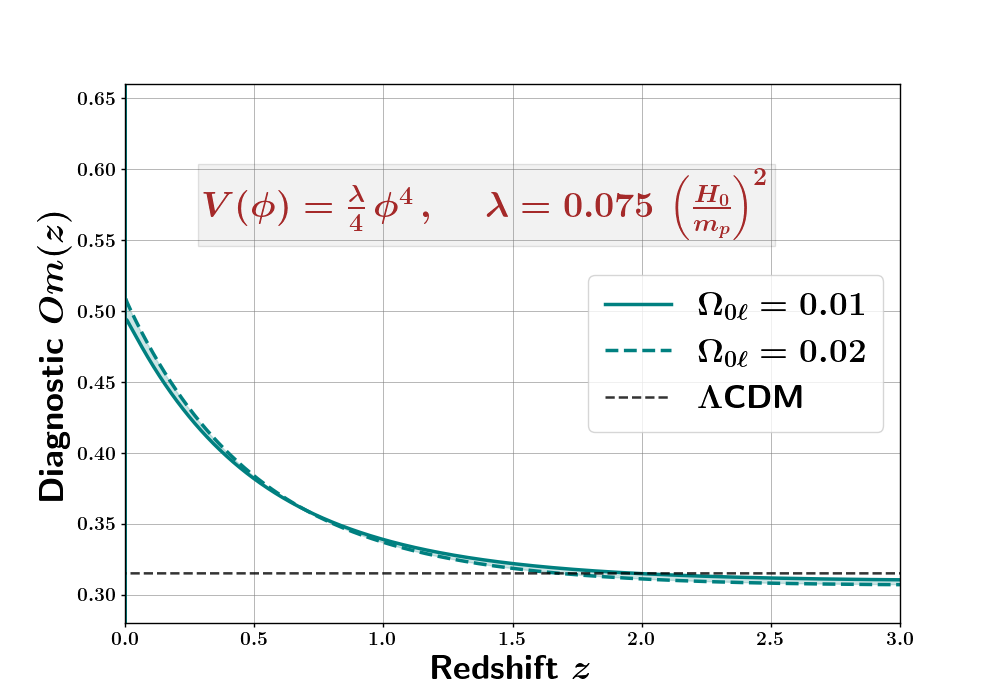}
\vspace{-0.2in}
\caption{The  $Om$ diagnostic parameter corresponding to the quartic potential (\ref{eq:pot_Quartic}) is shown for $\lambda=0.05 \left( H_0/m_p \right)^2$ ({\bf left panel}), and  for  $\lambda = 0.075 \left( H_0/m_p \right)^2$ ({\bf right panel}).}
\label{fig:DE_PhBrane_Quartic_Om}
\end{center}
\end{figure}
\subsection{Symmetry-breaking potential\,: steep wing}
\label{app:DE_PhBrane_SB_R}
\begin{figure}[H]
\vspace{-0.05in}
\begin{center}
\vspace{-0.2in}
\includegraphics[width=0.495\textwidth]{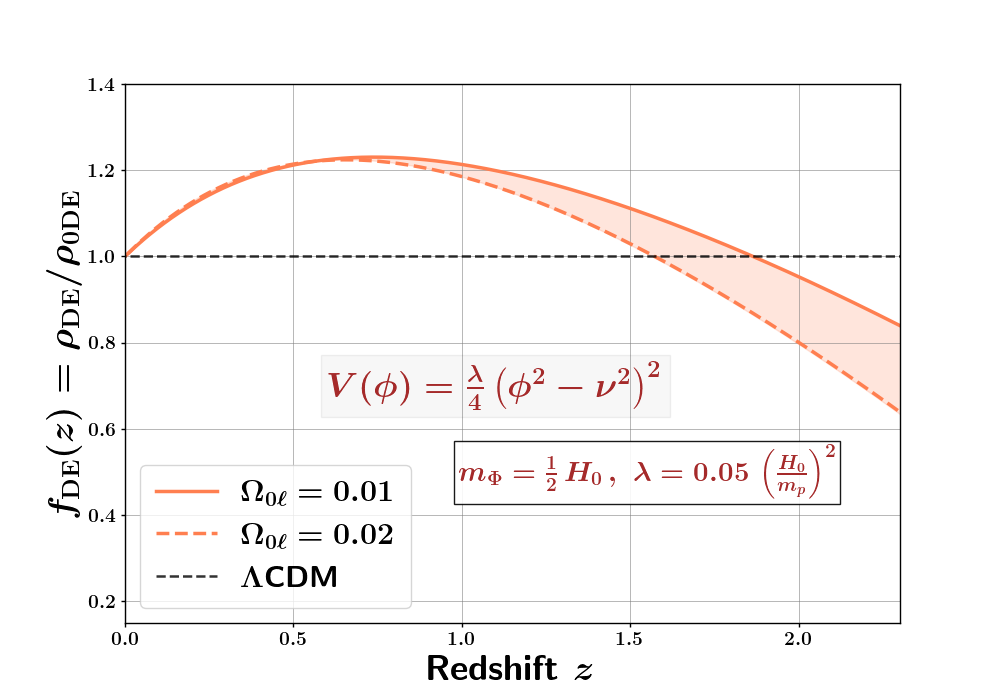}
\includegraphics[width=0.495\textwidth]{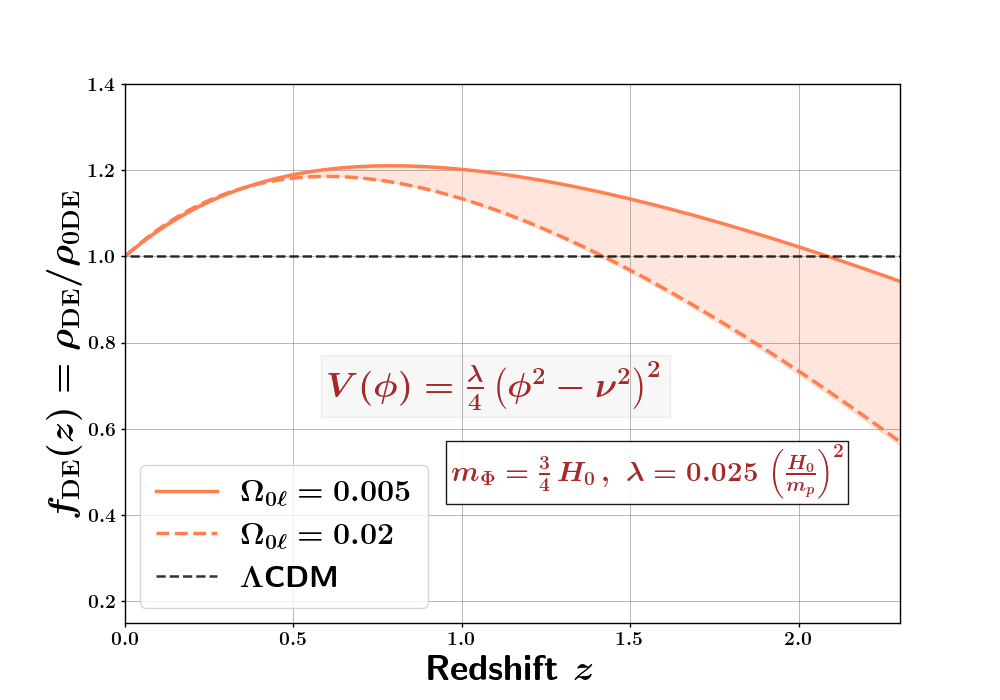}
\vspace{-0.2in}
\caption{The  DE density relative to its present-epoch value corresponding to the steep right wing of the symmetry-breaking  potential~(\ref{eq:pot_SB}) are shown for $m_\Phi=\frac{1}{2}\,H_0\,,~\lambda = 0.05 \left( H_0/m_p \right)^2$ ({\bf left panel}), and  for  $m_\Phi=\frac{3}{4}\,H_0\,,~\lambda = 0.025 \left( H_0/m_p \right)^2$ ({\bf right panel}).}
\label{fig:DE_PhBrane_SB_R_Density}
\end{center}
\vspace{-0.1in}
\end{figure}
\begin{figure}[H]
\vspace{-0.2in}
\begin{center}
\vspace{-0.2in}
\includegraphics[width=0.495\textwidth]{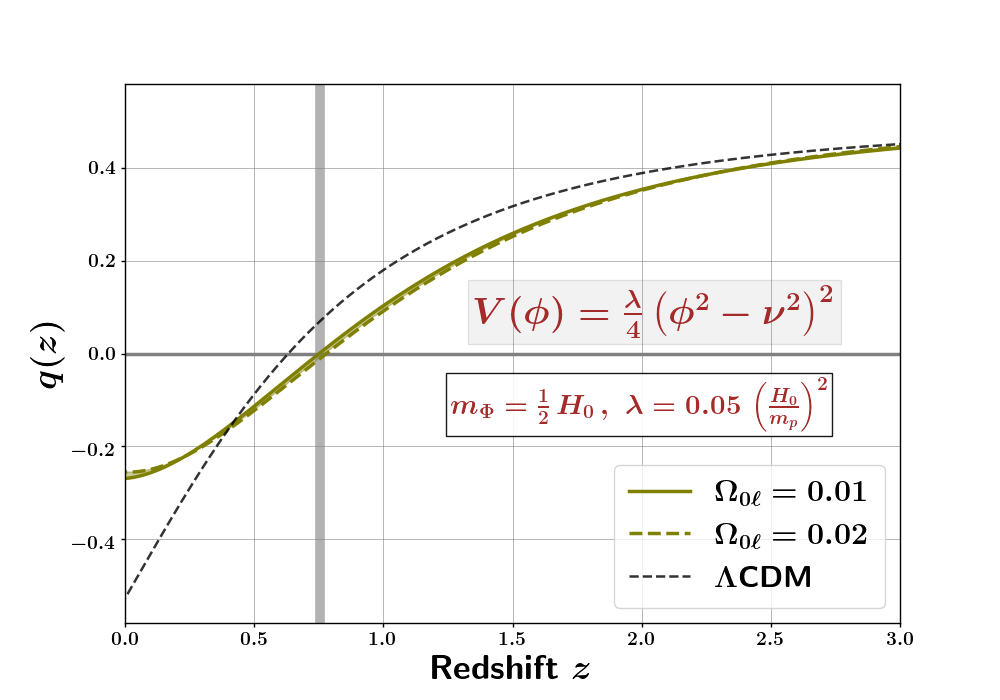}
\includegraphics[width=0.495\textwidth]{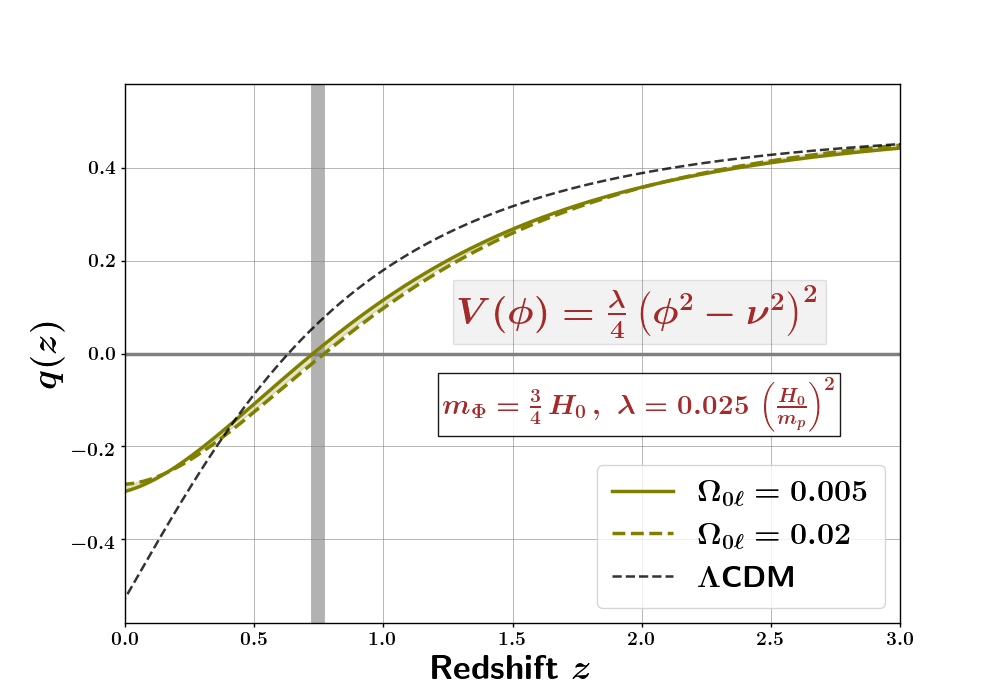}
\vspace{-0.2in}
\caption{The deceleration parameter corresponding to the steep right wing of the symmetry-breaking  potential~(\ref{eq:pot_SB}) is shown for $m_\Phi=\frac{1}{2}\,H_0\,,~\lambda = 0.05 \left( H_0/m_p \right)^2$ ({\bf left panel}), and  for  $m_\Phi=\frac{3}{4}\,H_0\,,~\lambda = 0.025 \left( H_0/m_p \right)^2$ ({\bf right panel}).} 
\label{fig:DE_PhBrane_SB_R_q}
\end{center}
\vspace{-0.1in}
\end{figure}
\begin{figure}[H]
\vspace{-0.2in}
\begin{center}
\vspace{-0.2in}
\includegraphics[width=0.495\textwidth]{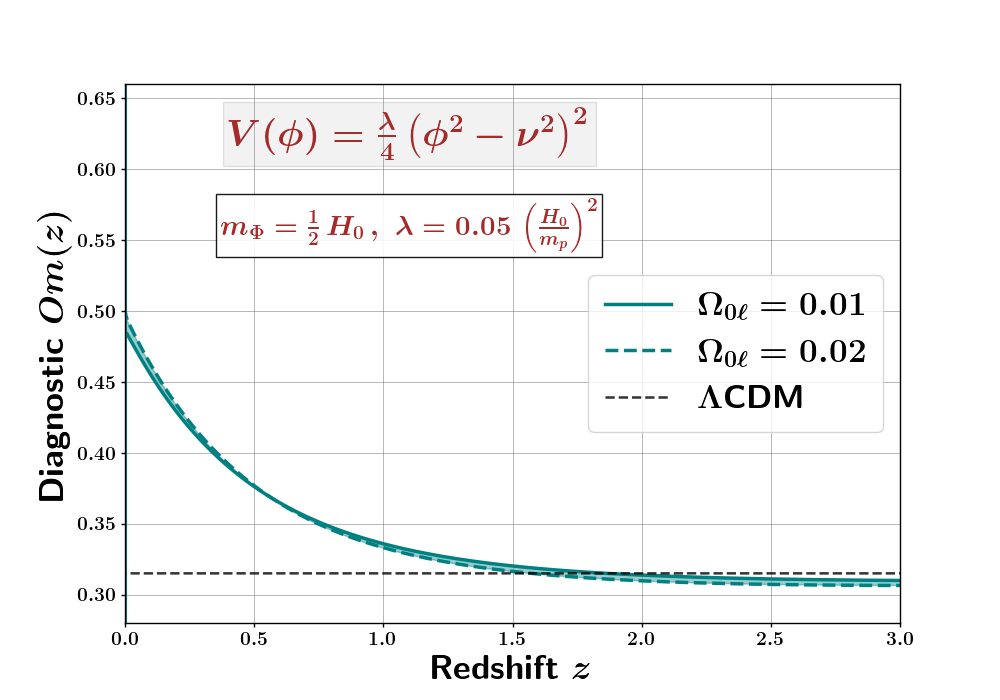}
\includegraphics[width=0.495\textwidth]{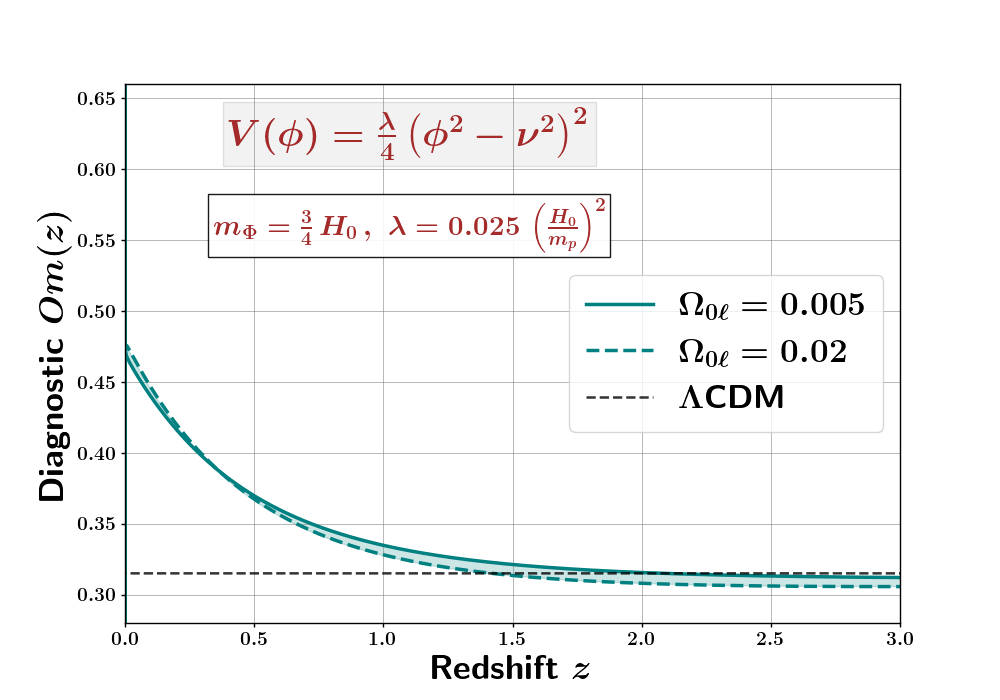}
\vspace{-0.2in}
\caption{The  $Om$ diagnostic parameter corresponding to the steep right wing of the symmetry-breaking  potential~(\ref{eq:pot_SB}) is shown for $m_\Phi=\frac{1}{2}\,H_0\,,~\lambda = 0.05 \left( H_0/m_p \right)^2$ ({\bf left panel}), and  for  $m_\Phi=\frac{3}{4}\,H_0\,,~\lambda = 0.025 \left( H_0/m_p \right)^2$ ({\bf right panel}).} 
\label{fig:DE_PhBrane_SB_R_Om}
\end{center}
\end{figure}
\subsection{Symmetry-breaking potential\,: flat wing}
\label{app:DE_PhBrane_SB_L}
\begin{figure}[H]
\vspace{-0.05in}
\begin{center}
\vspace{-0.2in}
\includegraphics[width=0.495\textwidth]{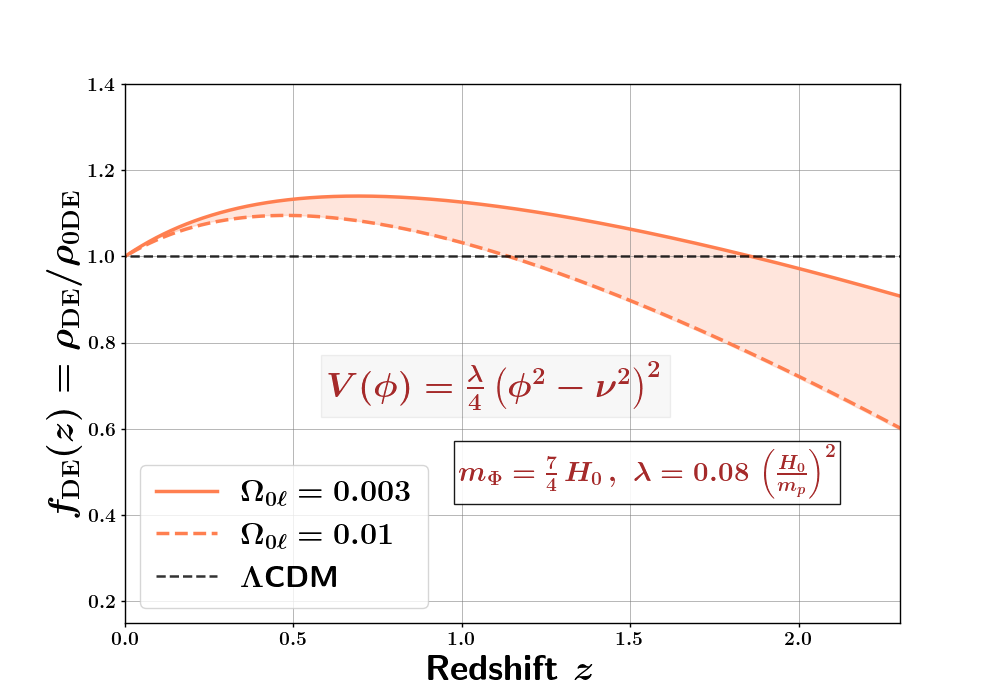}
\includegraphics[width=0.495\textwidth]{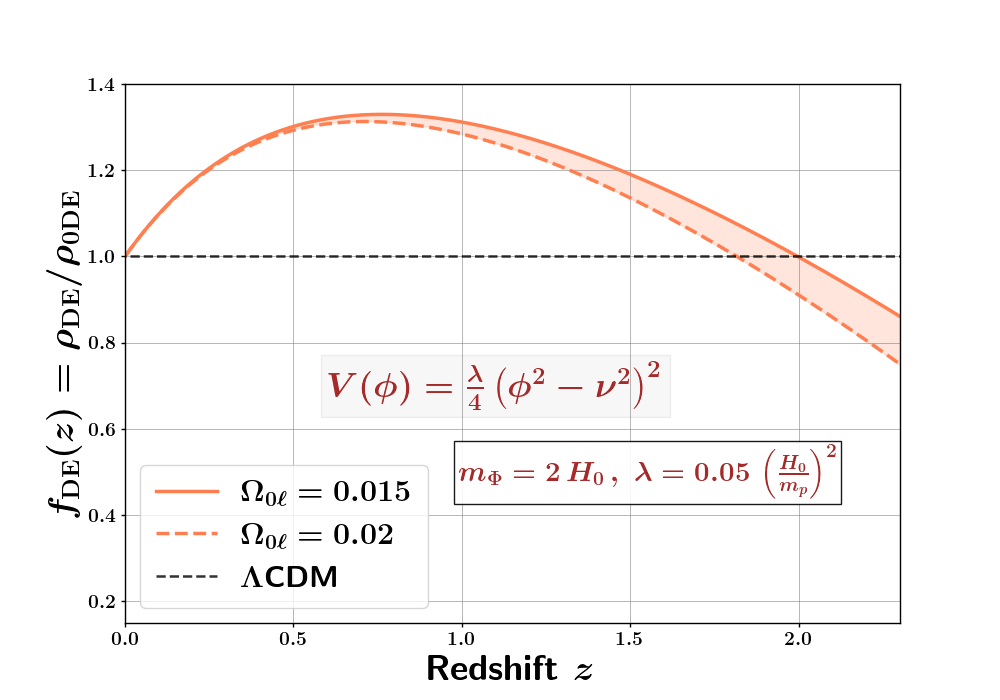}
\vspace{-0.2in}
\caption{The  DE density relative to its present-epoch value corresponding to the flat left wing of the symmetry-breaking  potential~(\ref{eq:pot_SB}) is shown for $m_\Phi=\frac{7}{4}\,H_0\,$, $\lambda = 0.08 \left( H_0/m_p \right)^2$ ({\bf left panel}), and  for  $m_\Phi=2\,H_0\,$, $\lambda = 0.05 \left( H_0/m_p \right)^2$ ({\bf right panel}). }
\label{fig:DE_PhBrane_SB_L_Density}
\end{center}
\vspace{-0.1in}
\end{figure}
\begin{figure}[H]
\vspace{-0.1in}
\begin{center}
\vspace{-0.2in}
\includegraphics[width=0.495\textwidth]{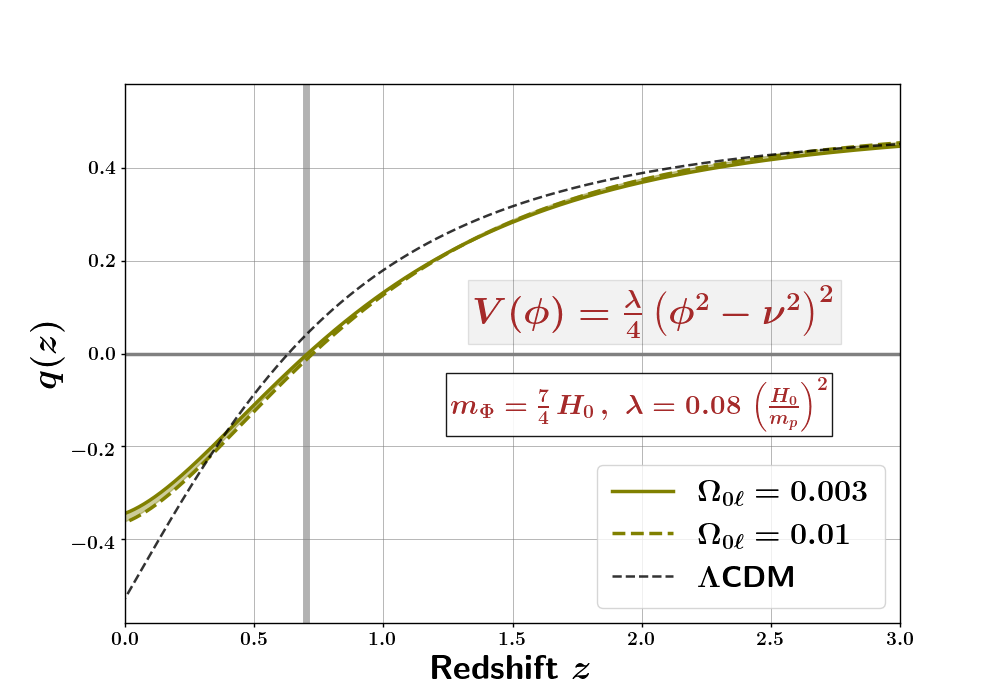}
\includegraphics[width=0.495\textwidth]{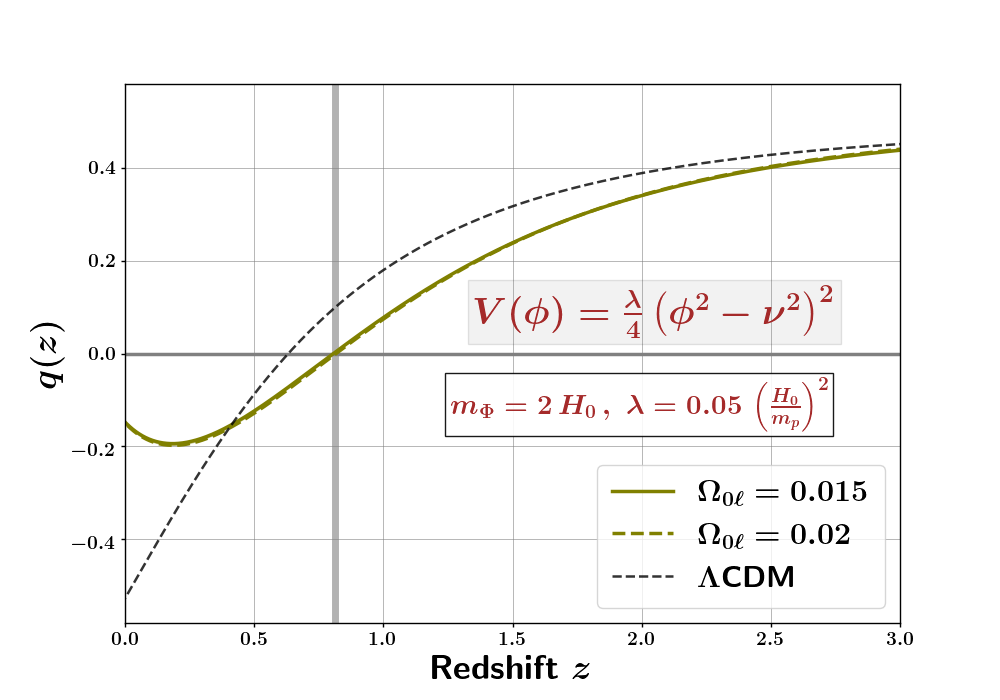}
\vspace{-0.2in}
\caption{The deceleration parameter corresponding to the flat left wing of the symmetry-breaking  potential (\ref{eq:pot_SB}) is shown for $m_\Phi=\frac{7}{4}\,H_0\,$, $\lambda = 0.08 \left( H_0/m_p \right)^2$ ({\bf left panel}), and  for  $m_\Phi=2\,H_0\,$, $\lambda = 0.05 \left( H_0/m_p \right)^2$ ({\bf right panel}).} 
\label{fig:DE_PhBrane_SB_L_q}
\end{center}
\vspace{-0.1in}
\end{figure}
\begin{figure}[H]
\vspace{-0.1in}
\begin{center}
\vspace{-0.2in}
\includegraphics[width=0.495\textwidth]{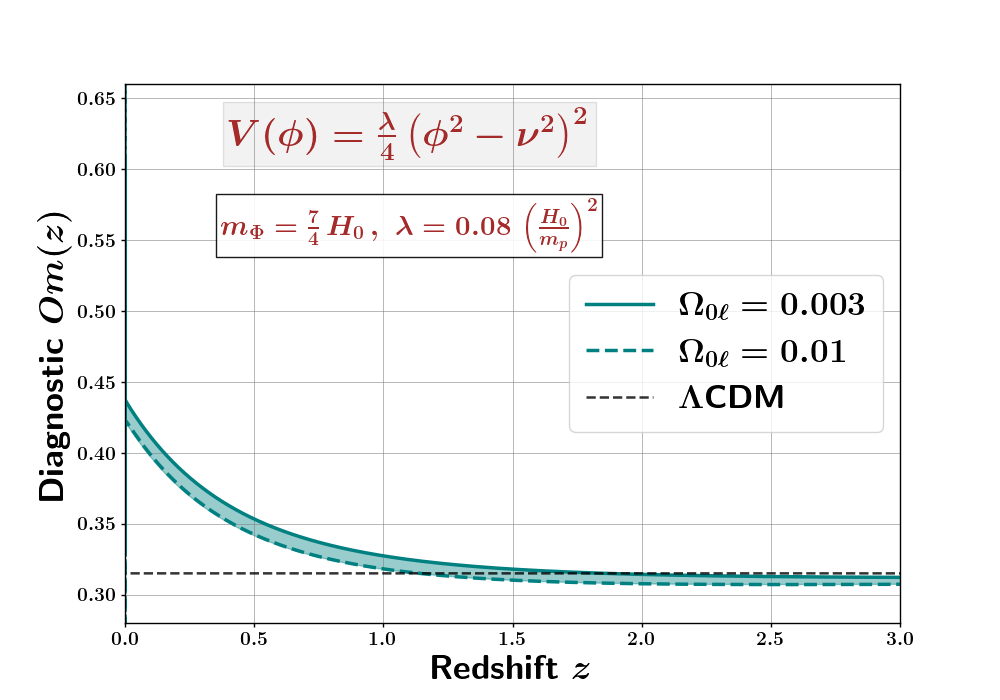}
\includegraphics[width=0.495\textwidth]{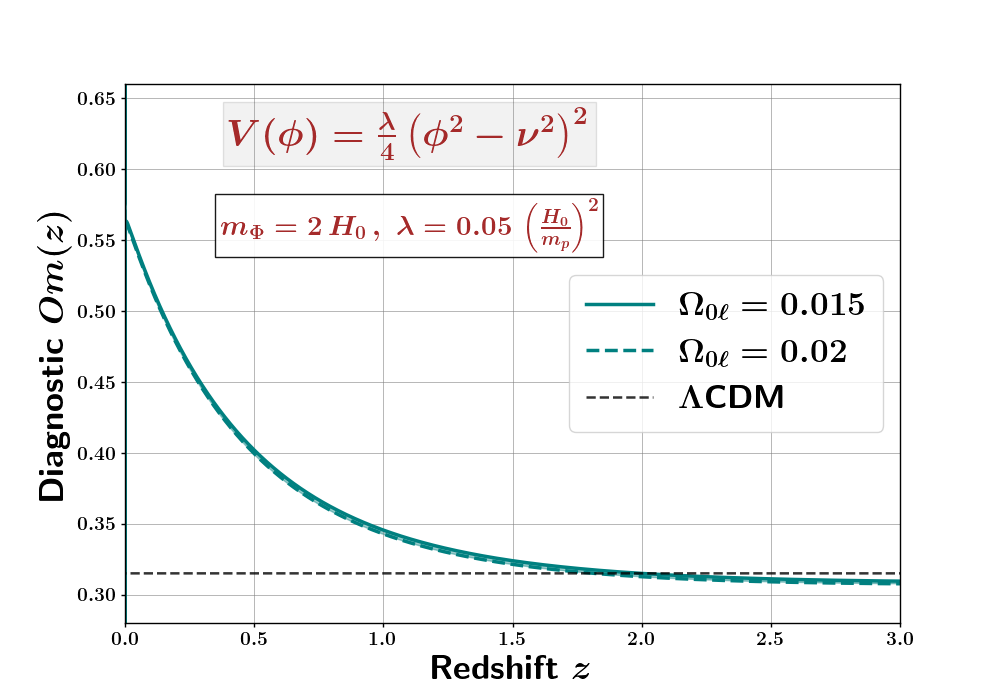}
\vspace{-0.2in}
\caption{The  $Om$ diagnostic parameter corresponding to the flat left wing of the symmetry-breaking  potential (\ref{eq:pot_SB}) is shown for $m_\Phi=\frac{7}{4}\,H_0\,$, $\lambda = 0.08 \left( H_0/m_p \right)^2$ ({\bf left panel}), and  for  $m_\Phi=2\,H_0\,$, $\lambda = 0.05 \left( H_0/m_p \right)^2$ ({\bf right panel}).} 
\label{fig:DE_PhBrane_SB_L_Om}
\end{center}
\vspace{-0.1in}
\end{figure}

\subsection{Exponential potential}
\label{app:Exp}
\begin{figure}[H]
\begin{center}
\includegraphics[width=0.495\textwidth]{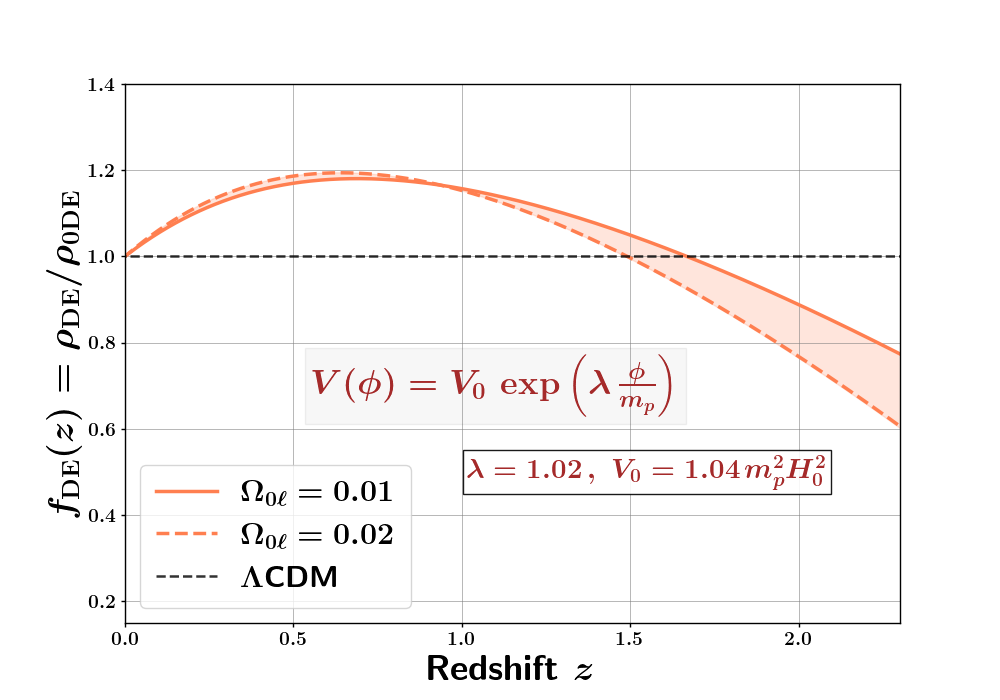}
\includegraphics[width=0.495\textwidth]{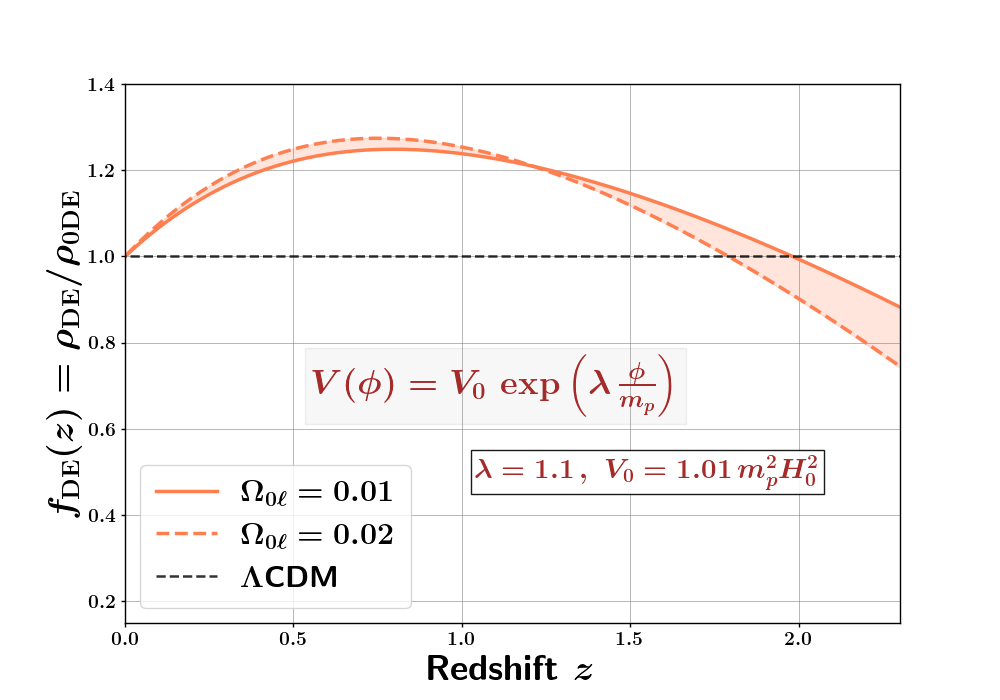}
\vspace{-0.2in}
\caption{The  DE density relative to its present-epoch value  for the exponential  potential~(\ref{eq:pot_Exponential}) is shown for $\lambda=1.02, \, V_0 =1.04\, m_p^2H_0^2$ ({\bf left panel}), and  for  $\lambda= 1.1, \, V_0 =1.01\, m_p^2H_0^2$ ({\bf right panel}).}
\label{fig:DE_PhBrane_Exp_Density}
\end{center}
\end{figure}
\begin{figure}[H]
\vspace{-0.2in}
\begin{center}
\vspace{-0.2in}
\includegraphics[width=0.495\textwidth]{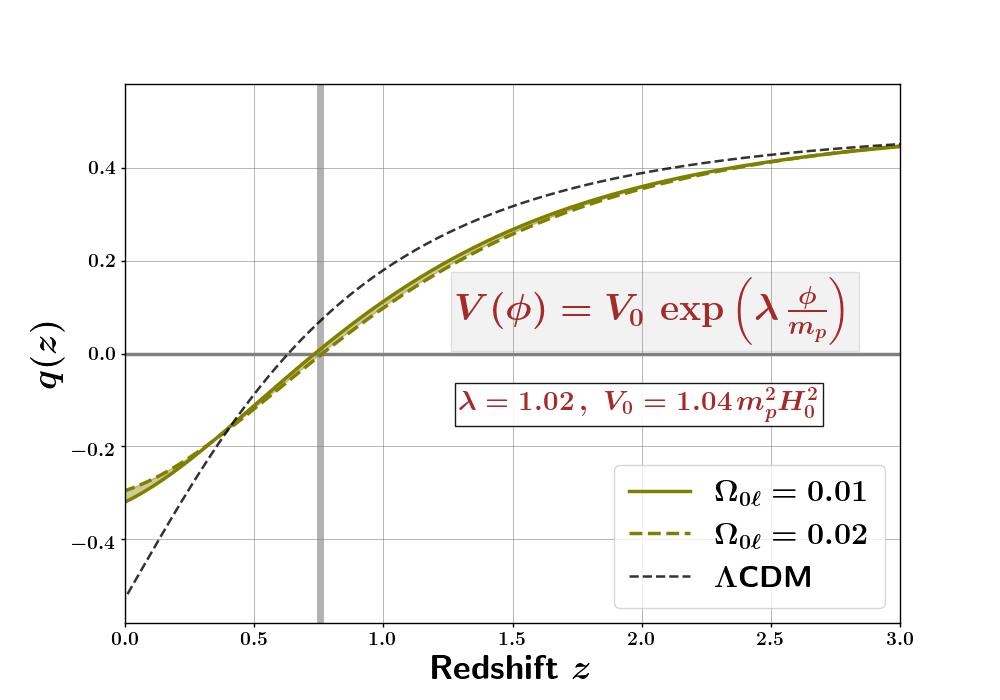}
\includegraphics[width=0.495\textwidth]{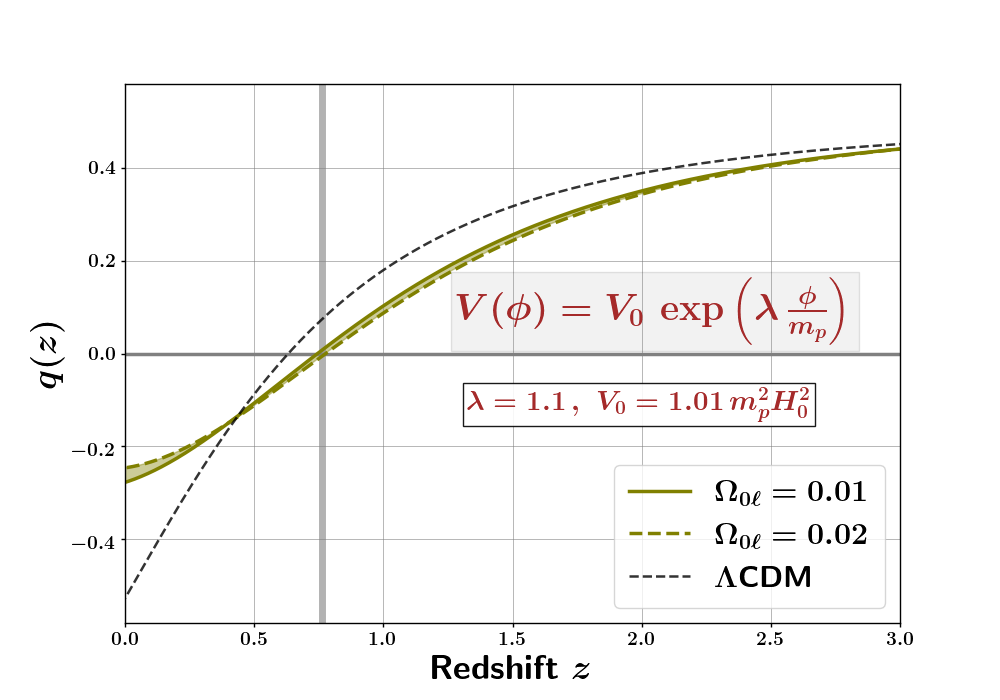}
\vspace{-0.2in}
\caption{The deceleration parameter  for the exponential  potential~(\ref{eq:pot_Exponential}) is shown for $\lambda=1.02, \, V_0 =1.04\, m_p^2H_0^2$ ({\bf left panel}), and  for  $\lambda= 1.1, \, V_0 =1.01\, m_p^2H_0^2$ ({\bf right panel}).} 
\label{fig:DE_PhBrane_Exp_q}
\end{center}
\end{figure}
\begin{figure}[H]
\vspace{-0.2in}
\begin{center}
\vspace{-0.2in}
\includegraphics[width=0.495\textwidth]{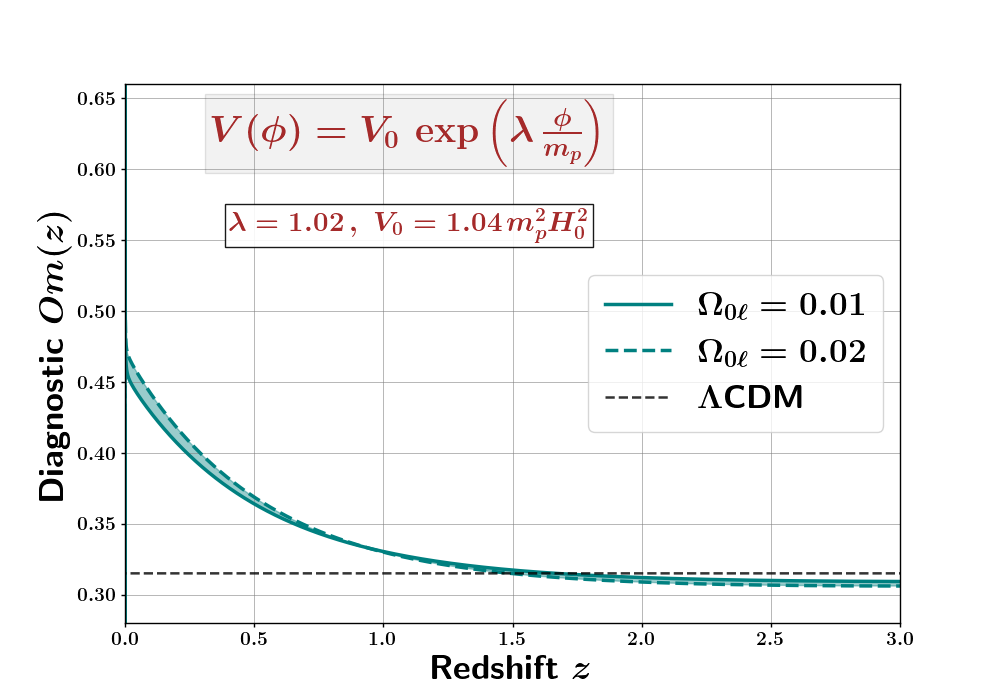}
\includegraphics[width=0.495\textwidth]{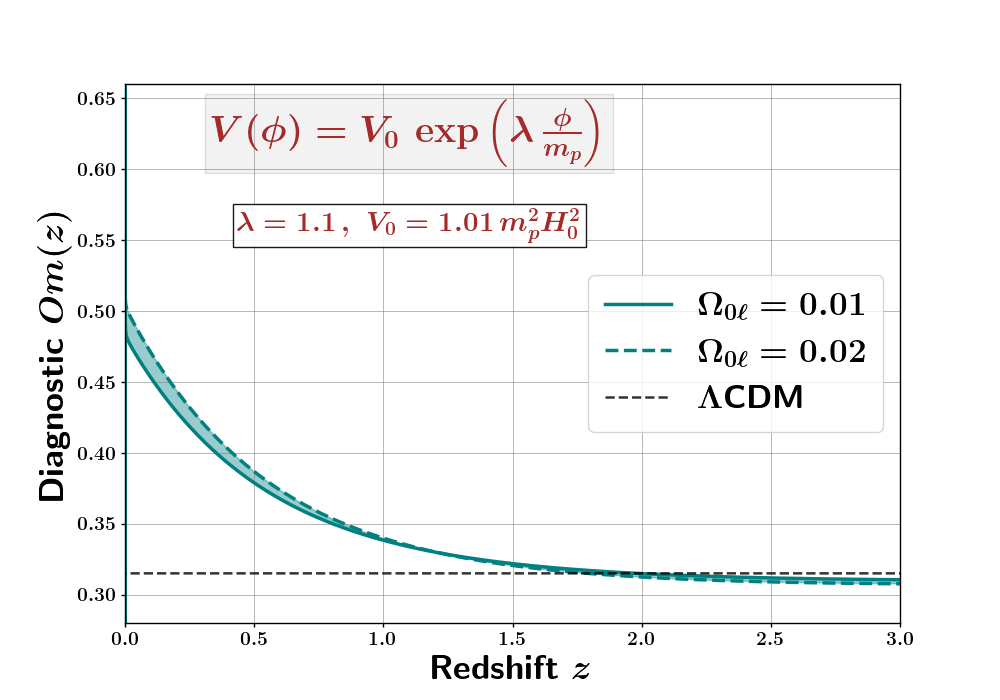}
\vspace{-0.2in}
\caption{The  $Om$ diagnostic parameter  for the exponential  potential~(\ref{eq:pot_Exponential}) is shown for $\lambda=1.02, \, V_0 =1.04\, m_p^2H_0^2$ ({\bf left panel}), and  for  $\lambda= 1.1, \, V_0 =1.01\, m_p^2H_0^2$ ({\bf right panel}).} 
\label{fig:DE_PhBrane_Exp_Om}
\end{center}
\end{figure}

\subsection{Axion potential}
\label{app:DE_PhBrane_Axion}

\begin{figure}[H]
\begin{center}
\includegraphics[width=0.495\textwidth]{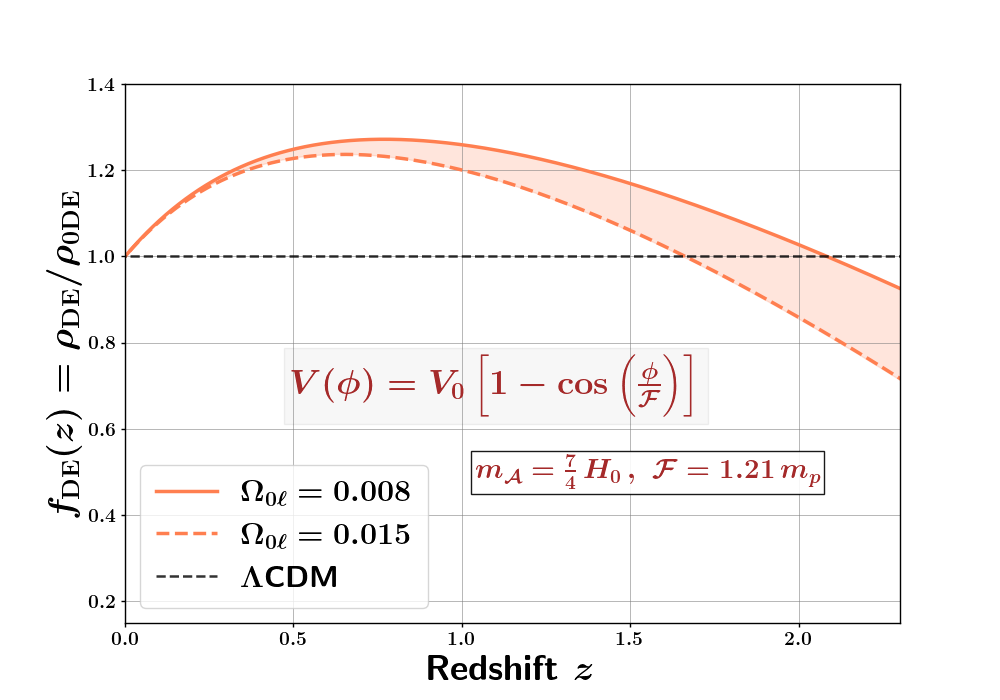}
\includegraphics[width=0.495\textwidth]{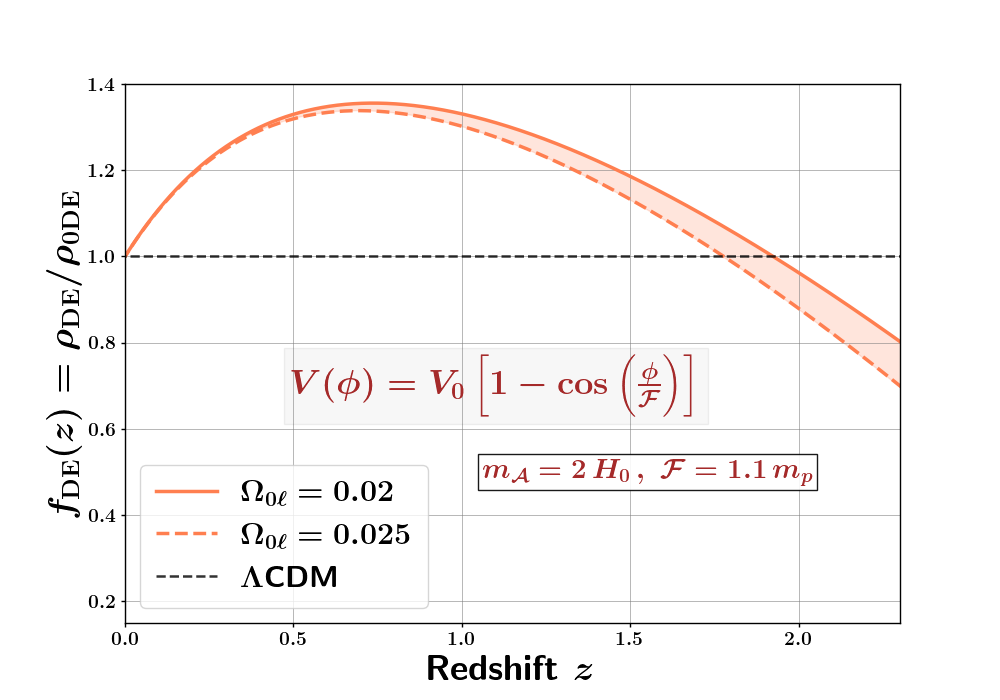}
\caption{The  DE density relative to its present-epoch value  for the Axion  potential~(\ref{eq:pot_Axion}) is shown for $m_{\cal A}=\frac{7}{4}\,H_0\,,~{\cal F}=1.21\,m_p$ ({\bf left panel}), and  for  $m_{\cal A}=2\,H_0\,,~{\cal F}=1.1\,m_p$ ({\bf right panel}).}
\label{fig:DE_PhBrane_Axion_Density}
\end{center}
\end{figure}
\begin{figure}[H]
\vspace{-0.2in}
\begin{center}
\vspace{-0.2in}
\includegraphics[width=0.495\textwidth]{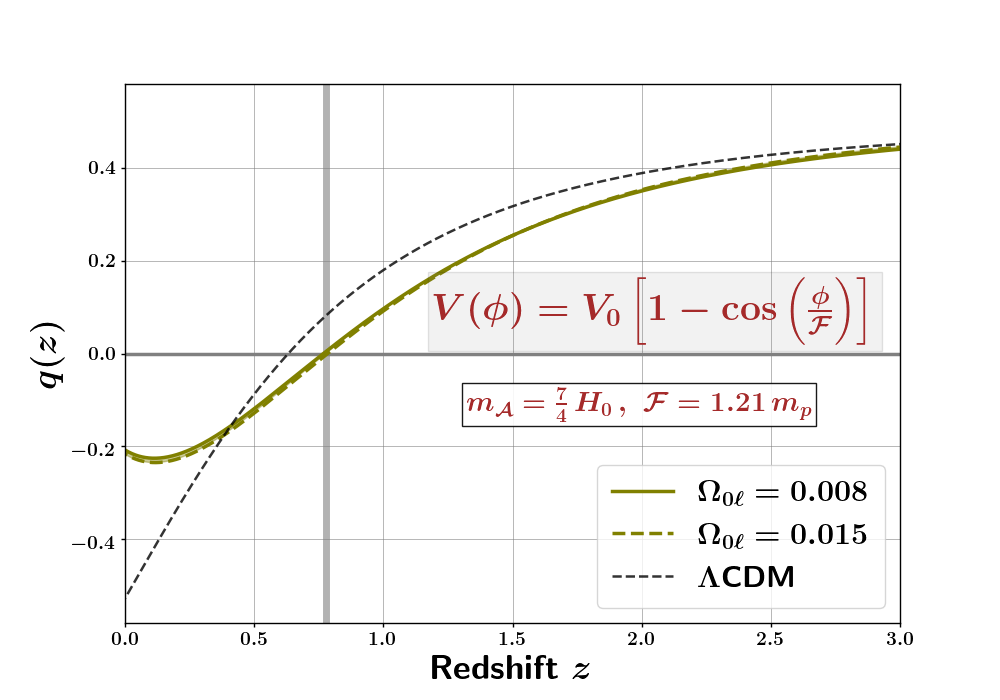}
\includegraphics[width=0.495\textwidth]{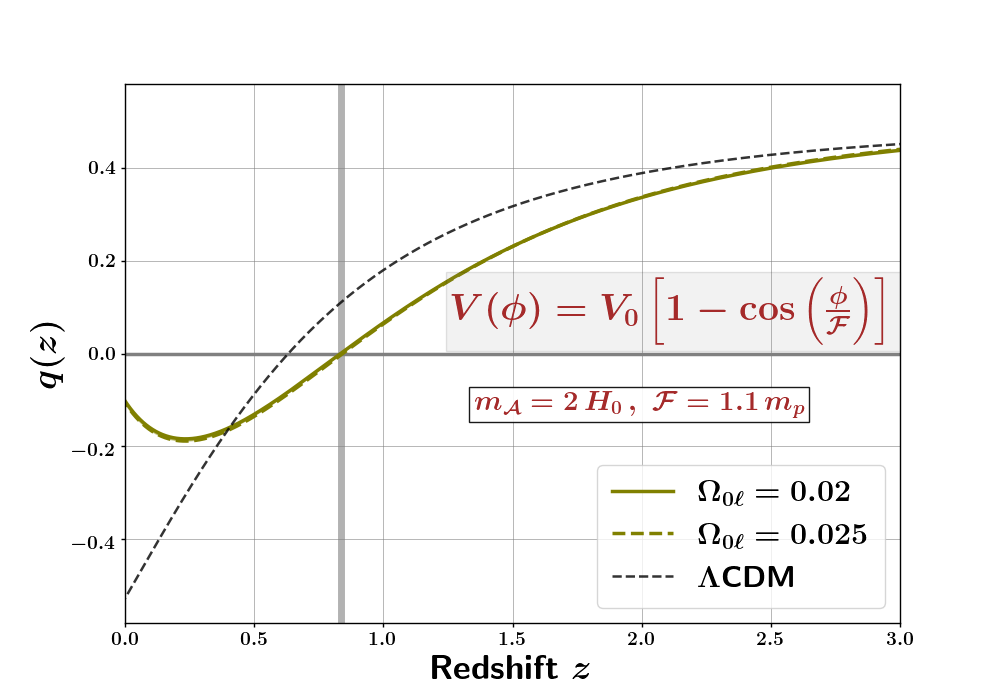}
\vspace{-0.2in}
\caption{The deceleration parameter corresponding to the Axion  potential~(\ref{eq:pot_Axion}) is shown for $m_{\cal A}=\frac{7}{4}\,H_0\,,~{\cal F}=1.21\,m_p$ ({\bf left panel}), and  for  $m_{\cal A}=2\,H_0\,,~{\cal F}=1.1\,m_p$ ({\bf right panel}).} 
\label{fig:DE_PhBrane_Axion_q}
\end{center}
\end{figure}
\begin{figure}[H]
\vspace{-0.2in}
\begin{center}
\vspace{-0.2in}
\includegraphics[width=0.495\textwidth]{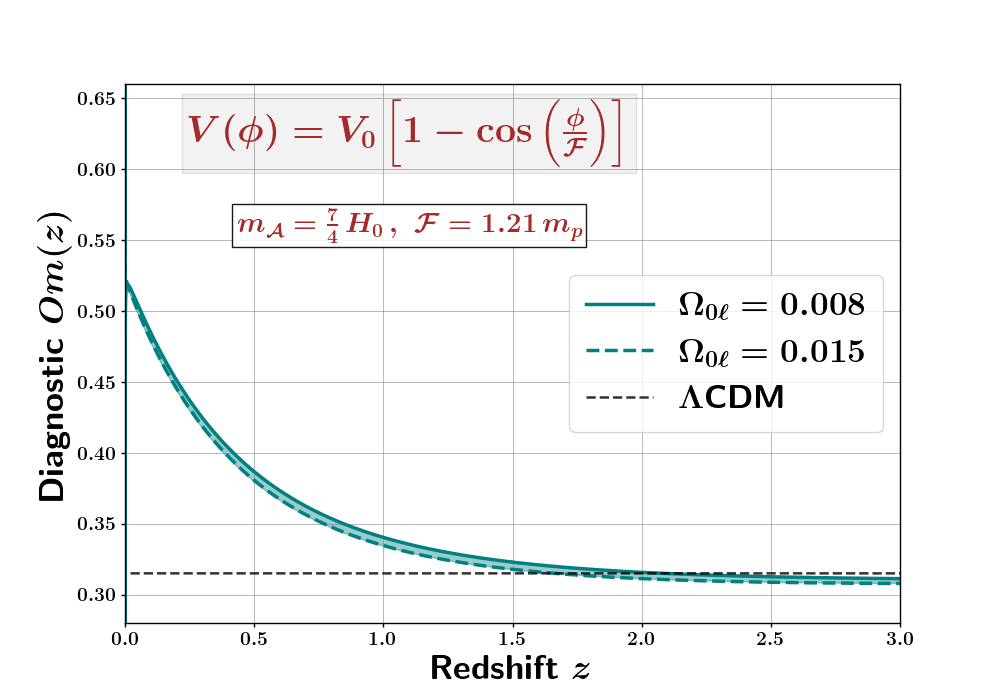}
\includegraphics[width=0.495\textwidth]{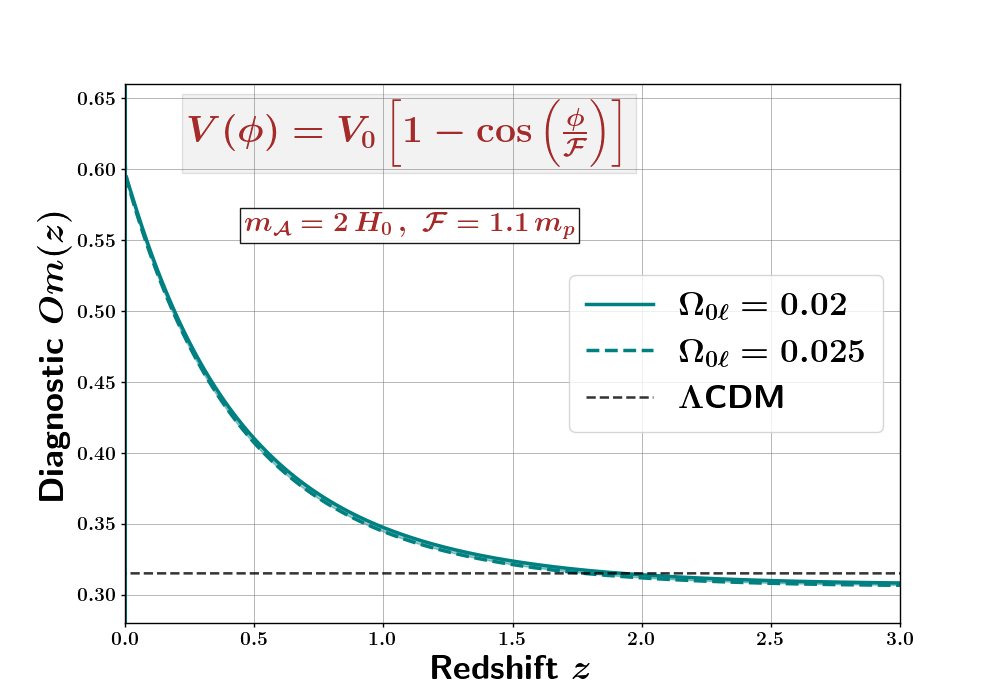}
\vspace{-0.2in}
\caption{The  $Om$ diagnostic parameter for the Axion  potential~(\ref{eq:pot_Axion}) is shown for $m_{\cal A}=\frac{7}{4}\,H_0\,,~{\cal F}=1.21\,m_p$ ({\bf left panel}), and  for  $m_{\cal A}=2\,H_0\,,~{\cal F}=1.1\,m_p$ ({\bf right panel}).} 
\label{fig:DE_PhBrane_Axion_Om}
\end{center}
\end{figure}

\section{Plateau potential}
\label{sec:PhBrane_Tmodel}

\begin{figure}[htb]
\centering
\includegraphics[width = 0.495\textwidth]{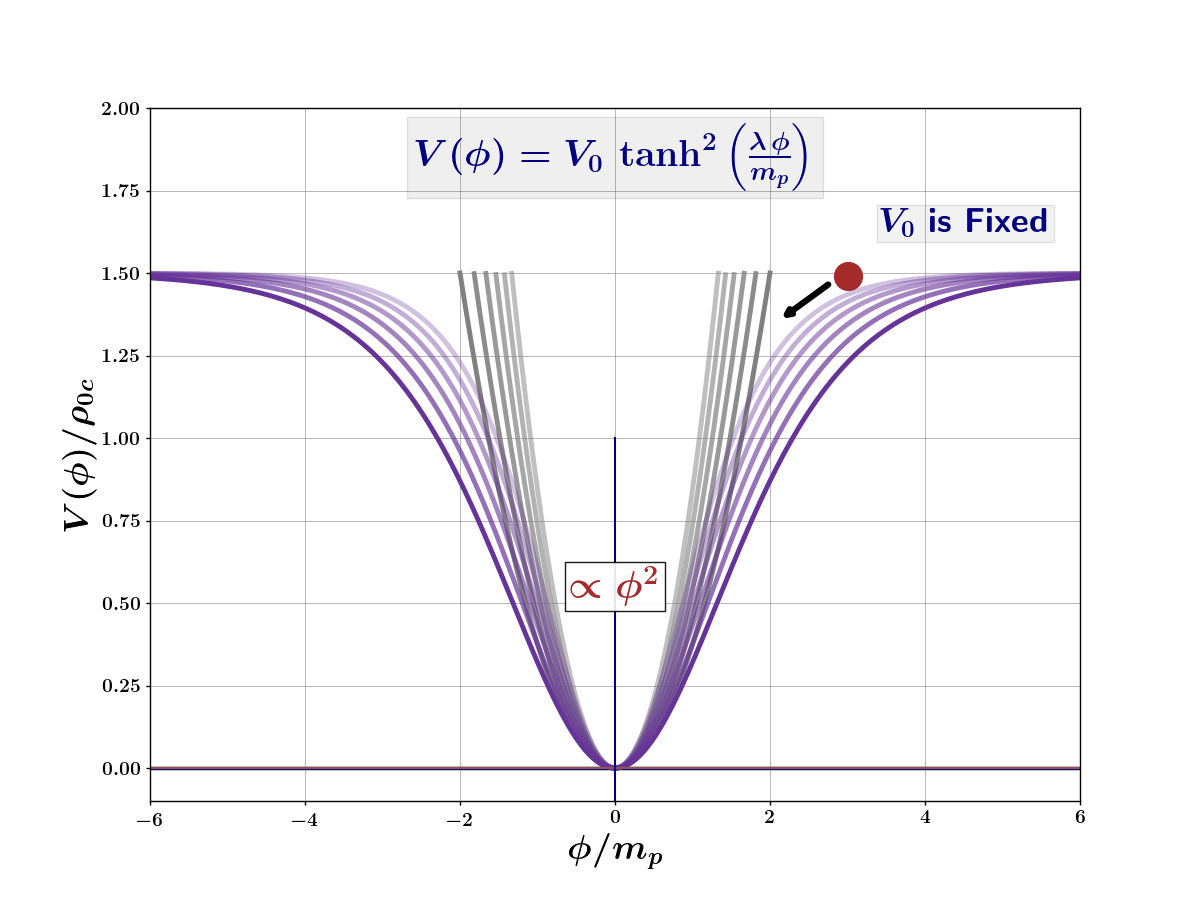}
\includegraphics[width = 0.495\textwidth]{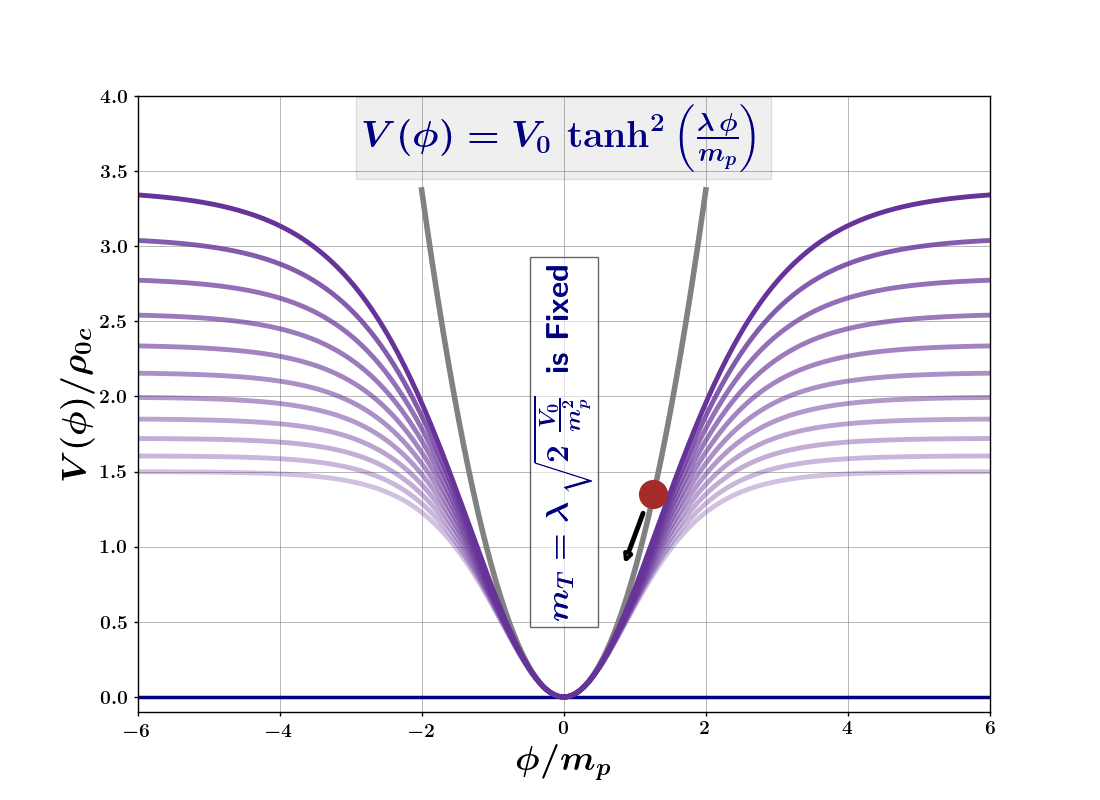}
\caption{A schematic plot of the plateau potential, given in Eq.~(\ref{eq:pot_Tmodel}), is shown by the purple curves: the {\bf left panel} illustrates the potential for a fixed value of 
$V_0$ and varying $\lambda$, while the {\bf right panel} shows it for a fixed $m_T$ and varying $V_0$. The quadratic approximation [Eq.~(\ref{eq:pot_Tmodel_quad})] is indicated by the light-gray curves.}
    \label{fig:DE_pot_Tmodel}
\end{figure}

Consider the plateau potential\,\footnote{The {\em plateau} potential is popularly known as the T-Model $\alpha$-attractor potential \cite{Kallosh:2013hoa} in the context of inflation.} given by
\begin{equation}
V(\phi) = V_0 \, \tanh^2{\l(\lambda \, \frac{\phi}{m_p}\r)}  \, ,
\label{eq:pot_Tmodel}
\end{equation}
which is naturally stabilised at the minimum, and exhibits asymptotically flat behaviour for $\phi \gg \lambda/m_p$, see Fig.~\ref{fig:DE_pot_Tmodel}. 

Note that, for $\phi \ll m_p/\lambda$, the potential can be approximated as a quadratic potential up to leading order in $\lambda\,\phi/m_p$, that is,
\begin{equation}
V(\phi) \big\vert_{\,\phi \, \ll \, m_p/\lambda} \, \simeq \, \frac{1}{2}\,m_T^2\,\phi^2 - {\cal O}\l(\frac{\lambda\,\phi}{m_p}\r)^4  \, ,
\label{eq:pot_Tmodel_quad}
\end{equation}
with 
\begin{equation}
m_T = \lambda \, \sqrt{\frac{2\,V_0}{m_p^2}} \quad \Rightarrow \quad \frac{m_T}{H_0} = \lambda \, \sqrt{\frac{6\,V_0}{3\,m_p^2\,H_0^2}} \, .
\label{eq:pot_Tmodel_mass}
\end{equation}

\begin{figure}[H]
\vspace{-0.0in}
\begin{center}
\vspace{-0.1in}
\includegraphics[width=0.495\textwidth]{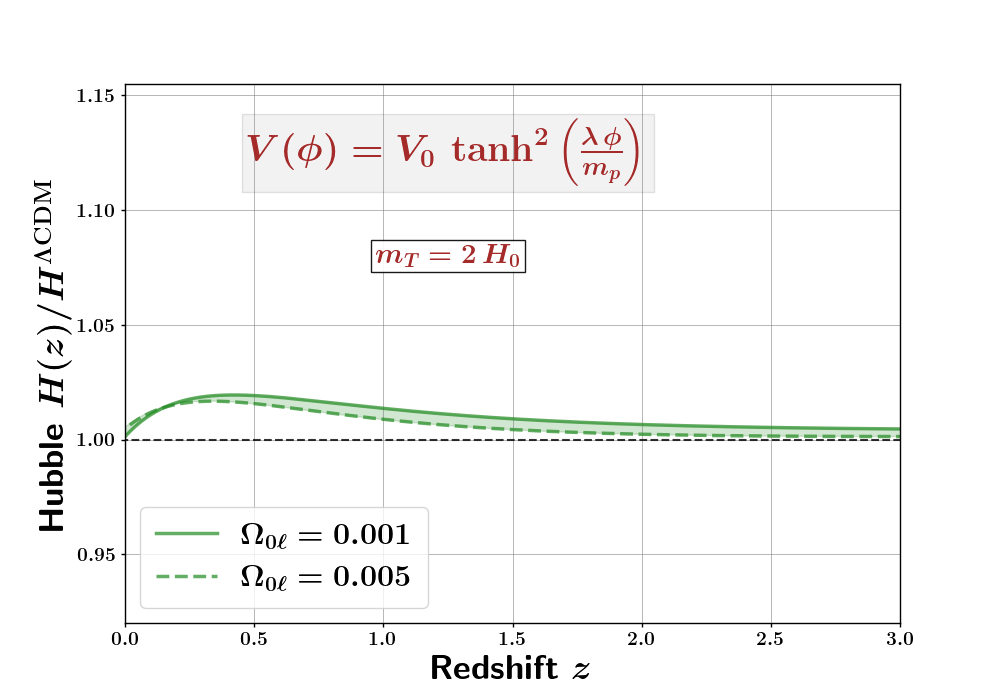}
\includegraphics[width=0.495\textwidth]{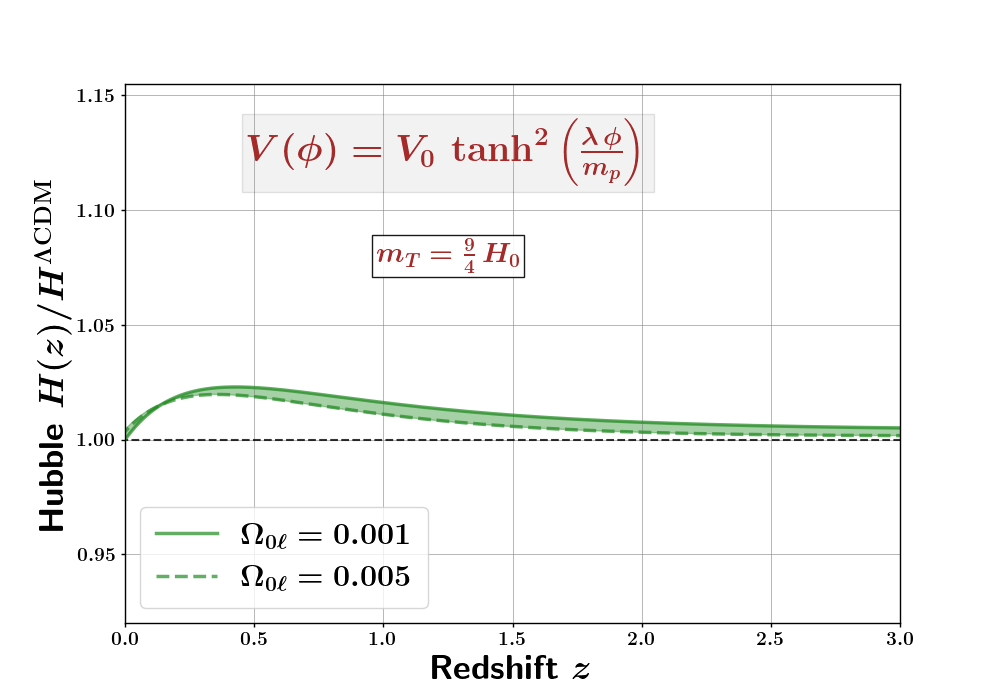}
\vspace{-0.2in}
\caption{The Hubble parameter corresponding to the plateau   potential~(\ref{eq:pot_Tmodel}) is shown for $m_{T}=2\,H_0$ ({\bf left panel}), and $m_{T}=\f{9}{4}\,H_0$ ({\bf right panel}).}
\label{fig:DE_PhBrane_Tmodel_Hubble}
\end{center}
\end{figure}

\begin{figure}[H]
\vspace{-0.2in}
\begin{center}
\vspace{-0.2in}
\includegraphics[width=0.495\textwidth]{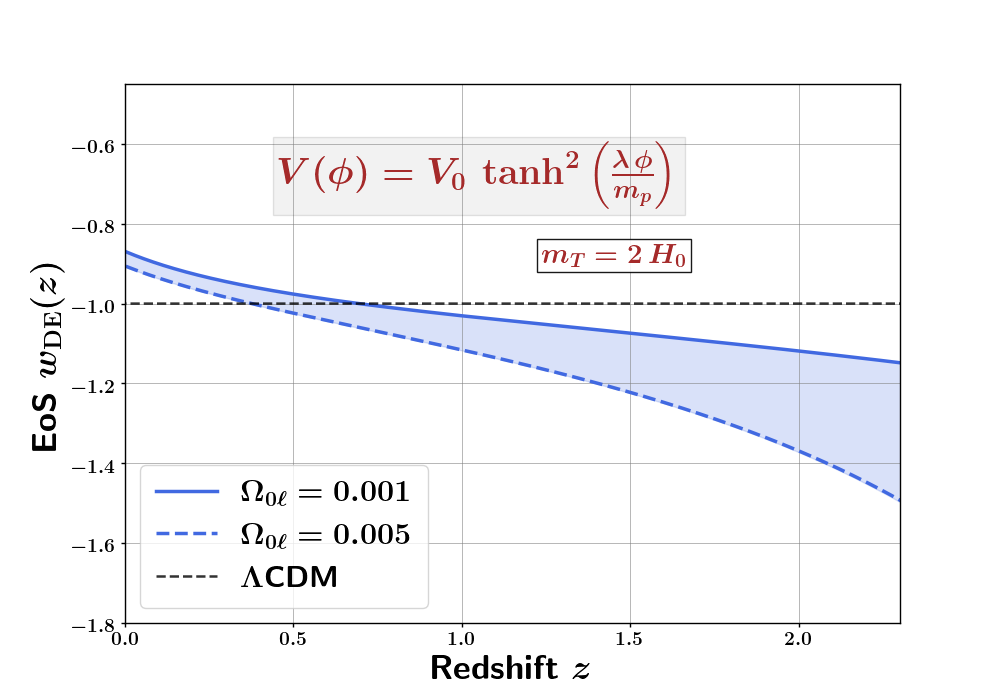}
\includegraphics[width=0.495\textwidth]{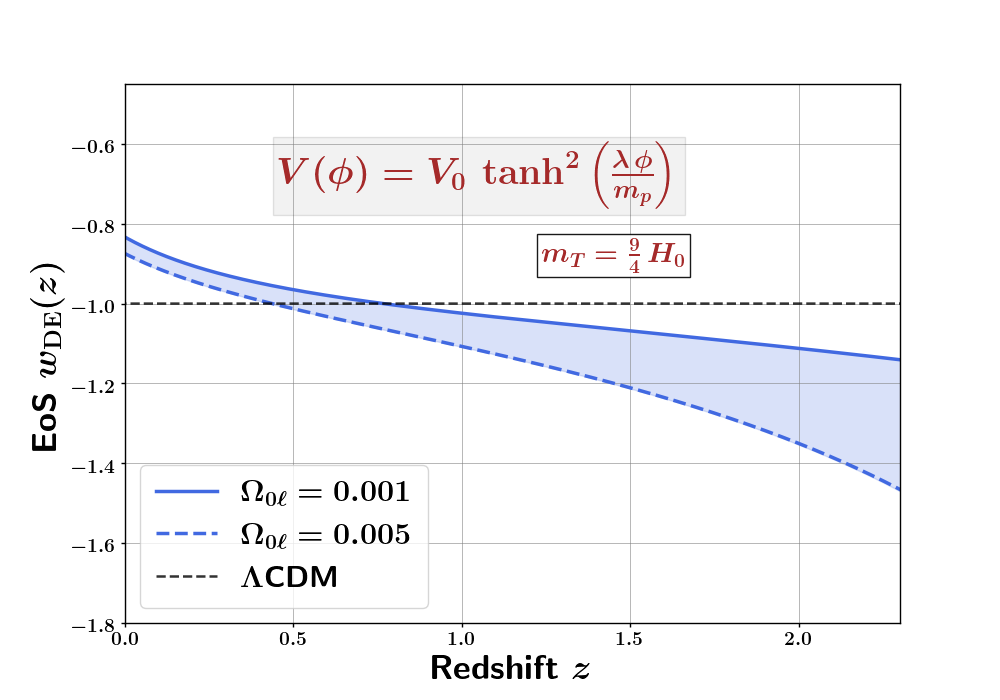}
\vspace{-0.2in}
\caption{The DE equation-of-state parameter corresponding to the plateau   potential~(\ref{eq:pot_Tmodel}) is shown for $m_{T}=2\,H_0$ ({\bf left panel}), and  for  $m_{T}=\f{9}{4}\,H_0$ ({\bf right panel}).}
\label{fig:DE_PhBrane_Tmodel_EoS}
\end{center}
\end{figure}

\begin{figure}[H]
\vspace{-0.1in}
\begin{center}
\vspace{-0.1in}
\includegraphics[width=0.495\textwidth]{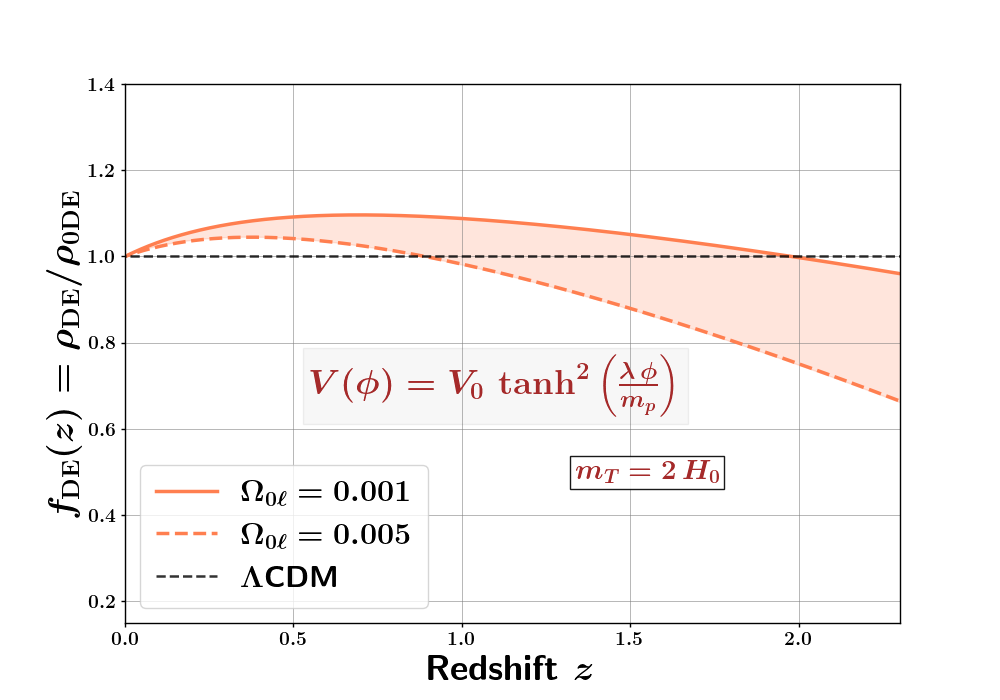}
\includegraphics[width=0.495\textwidth]{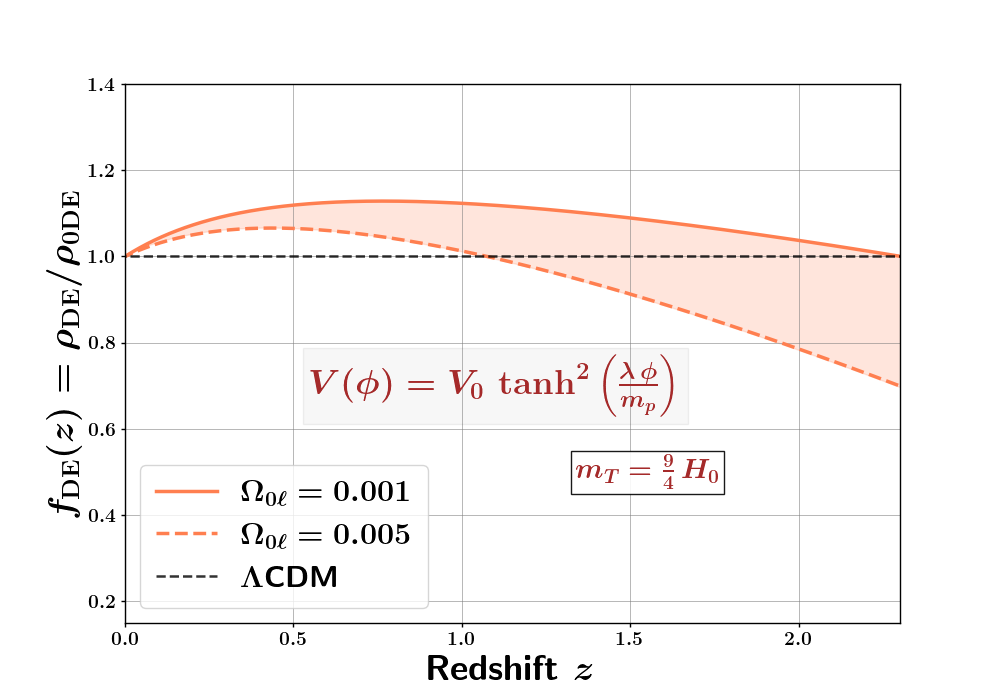}
\vspace{-0.2in}
\caption{The  DE density relative to its present-epoch value  corresponding to the plateau   potential~(\ref{eq:pot_Tmodel}) is shown for $m_{T}=2\,H_0$ ({\bf left panel}), and  for  $m_{T}=\f{9}{4}\,H_0$ ({\bf right panel}).}
\label{fig:DE_PhBrane_Tmodel_Density}
\end{center}
\end{figure}

\begin{figure}[H]
\vspace{-0.3in}
\begin{center}
\vspace{-0.2in}
\includegraphics[width=0.495\textwidth]{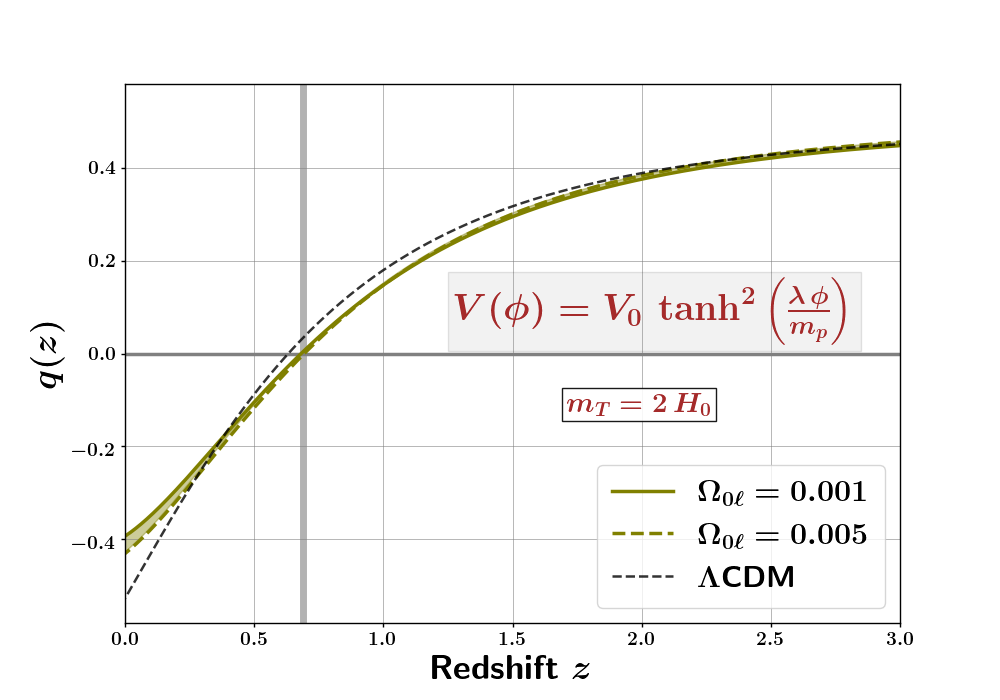}
\includegraphics[width=0.495\textwidth]{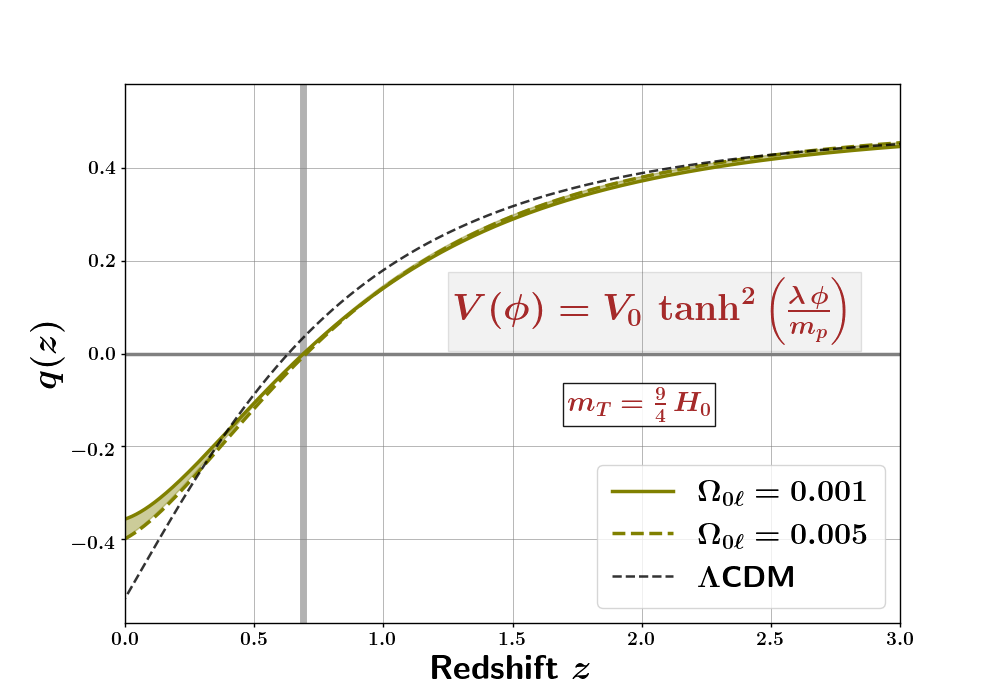}
\vspace{-0.2in}
\caption{The deceleration parameter corresponding to the plateau  potential~(\ref{eq:pot_Tmodel}) is shown for $m_{T}=2\,H_0$ ({\bf left panel}), and  for  $m_{T}=\frac{9}{4}\,H_0$ ({\bf right panel}).} 
\label{fig:DE_PhBrane_Tmodel_q}
\end{center}
\end{figure}

\begin{figure}[H]
\vspace{-0.2in}
\begin{center}
\vspace{-0.1in}
\includegraphics[width=0.495\textwidth]{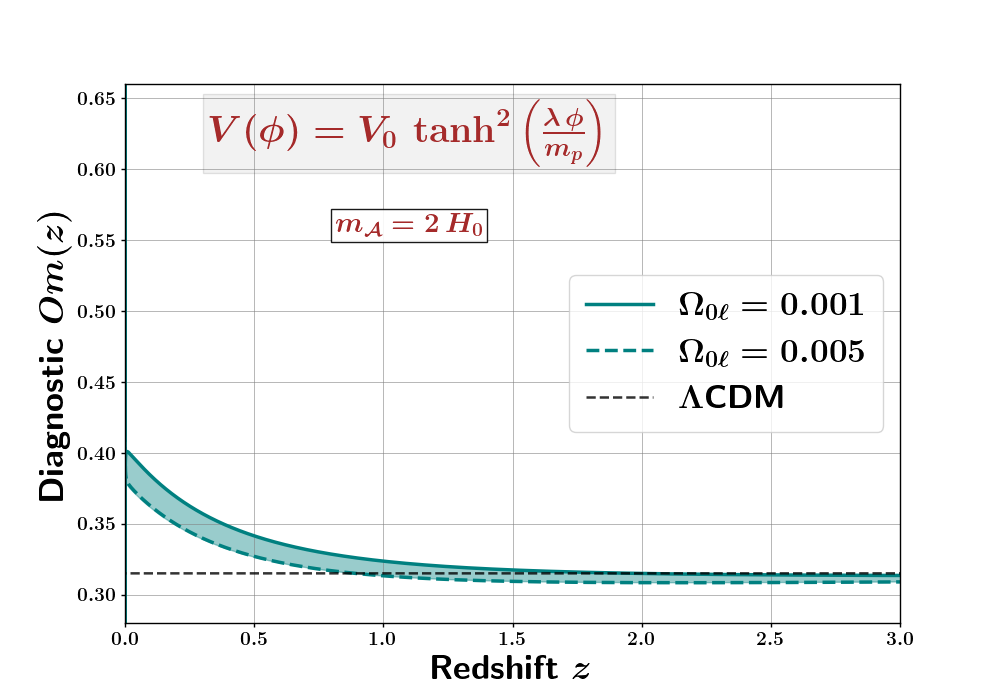}
\includegraphics[width=0.495\textwidth]{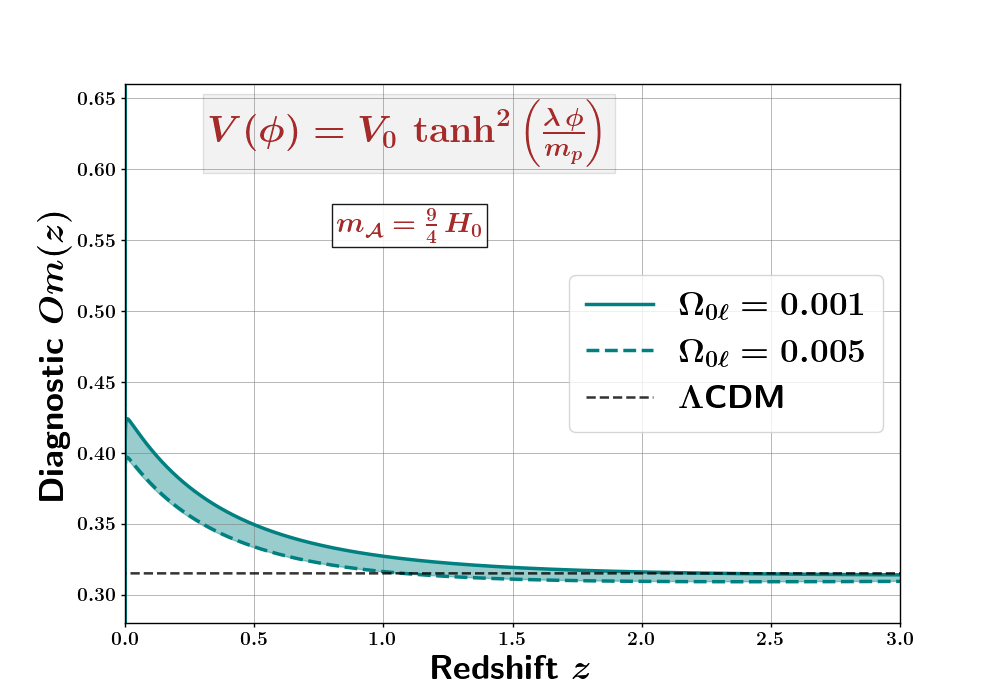}
\vspace{-0.2in}
\caption{The $Om$ diagnostic parameter corresponding to the plateau  potential~(\ref{eq:pot_Tmodel}) is shown for $m_{T}=2\,H_0$ ({\bf left panel}), and  for  $m_{T}=\frac{9}{4}\,H_0$ ({\bf right panel}).} 
\label{fig:DE_PhBrane_Tmodel_Om}
\end{center}
\end{figure}

We perform numerical simulations for a range of (fixed) values of $m_T$. Our results are illustrated in Figs.~\Ref{fig:DE_PhBrane_Tmodel_Hubble}--\ref{fig:DE_PhBrane_Tmodel_Om}.  The plots are generated for two distinct values of the scalar field mass, namely $m_T = 2 H_0$, and $\frac{9}{4} H_0$, while keeping the value of $V_0$ fixed.

Note that the value of 
$m_T$ is a function of $V_0$ and $\lambda$, as given by Eq.~(\ref{eq:pot_Tmodel_mass}). To ensure a minimum required amount of dark energy at present, $V_0$ must exceed a certain critical value. For a fixed mass, choosing a very large $V_0$ necessitates a correspondingly small $\lambda$. In this limit, the dark energy dynamics of the plateau potential closely resembles that of a purely quadratic potential; see the right panel of Fig.~\ref{fig:DE_pot_Tmodel}.

Therefore, we consider the opposite limit by setting $V_0 = \frac{9}{2} m_p^2 H_0^2$,  which is sufficiently small to allow a given scalar field mass to be achieved with a correspondingly large value of $\lambda$. In this regime, the dynamics is expected to exhibit a significant deviation from the purely quadratic case\,---\,which is precisely what we observe.

A visual comparison of our plots with those of Ref.~\cite{DESI:2025fii} indicate that the DESI DR2 constraints appear to be  somewhat consistent with a plateau potential on the phantom brane for the parameters chosen in Figs.~\Ref{fig:DE_PhBrane_Tmodel_Hubble}--\ref{fig:DE_PhBrane_Tmodel_Om}. 

This potential appears to perform less effectively than the ones discussed in Sec.~\ref{sec:Dynamics_Models_Brane} because the results  are displayed for $V_0 = \frac{9}{2} m_p^2 H_0^2$, as highlighted in the previous paragraph. If we had chosen higher values of $V_0$, then our results would have resembled those of the quadratic potential, which more consistent with the DESI DR2 data.  


\begin{figure}[H]
\vspace{-0.3in}
\begin{center}
\vspace{-0.2in}
\includegraphics[width=0.495\textwidth]{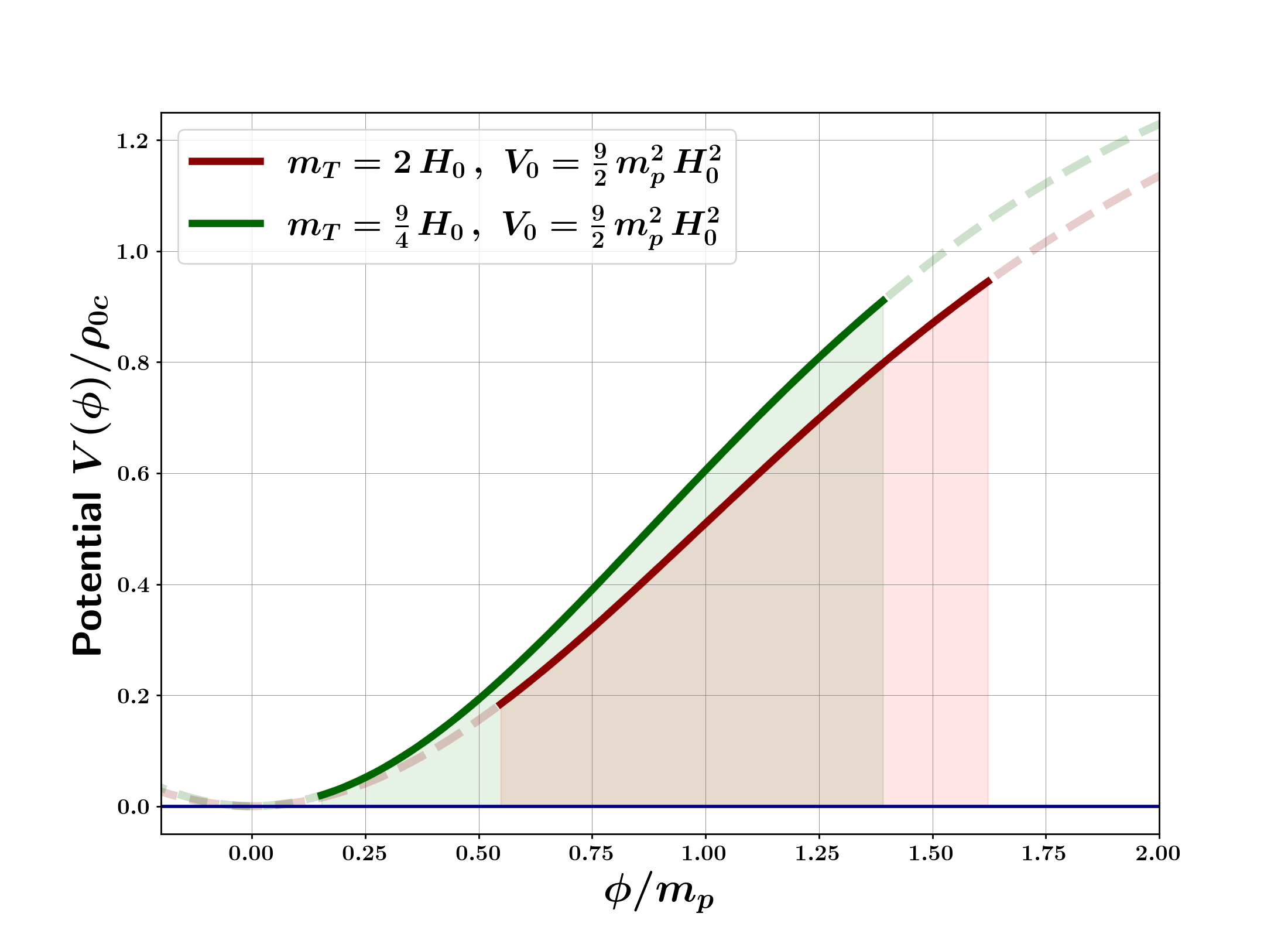} 
\includegraphics[width=0.495\textwidth]{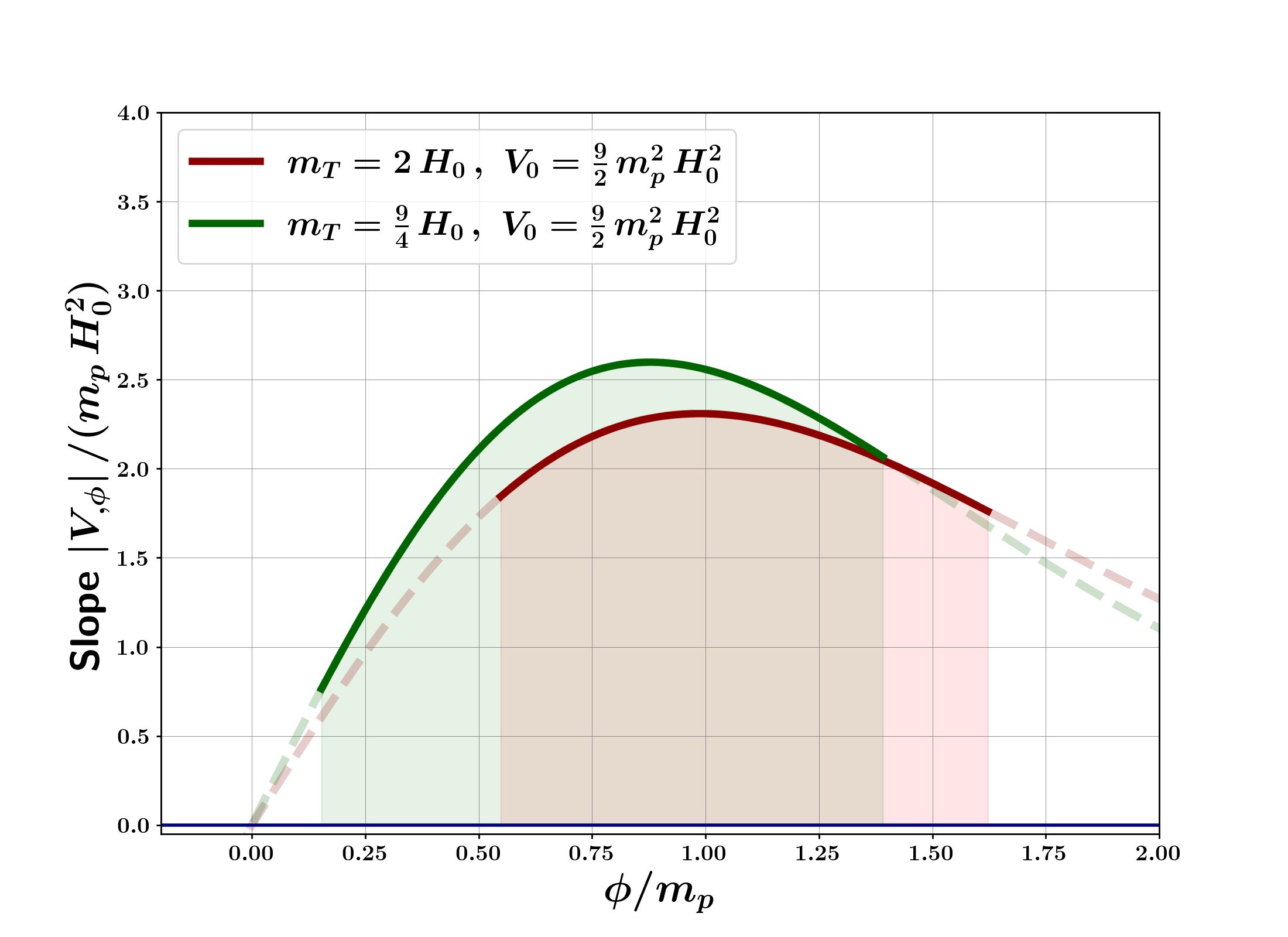}
\vspace{-0.2in}
\caption{ {\bf Left panel} shows the plateau potential~(\ref{eq:pot_Tmodel}) with $V_0 = \f{9}{2}\,m_p^2\,H_0^2$, along with $m_T=2\,H_0$ (red)  and $m_T=\f{9}{4}\,H_0$ (green). The {\bf right panel} shows the corresponding slopes. The {\bf darker (solid) curves} with vertical shades correspond to the field ($\phi$) range explored during the full simulation time, \textit{i.e.\@} from $z=99$ to $z \lesssim -1$. The {\bf fainter (dashed) curves} correspond to a much wider range of values of $\phi$.} 
\label{fig:pot_compare_TModel}
\end{center}
\vspace{-0.2in}
\end{figure}

\printbibliography

\end{document}